\documentclass[]{pasj01}
\Received{$\langle$reception date$\rangle$}
\Accepted{$\langle$acception date$\rangle$}
\Published{$\langle$publication date$\rangle$}

\usepackage{xspace}
\usepackage{graphicx}
\usepackage{bm}
\usepackage{ulem}
\usepackage{lscape}

\renewcommand\thefootnote{\roman{footnote}}

\newcommand{\molline}[3]{#1($J$=#2--#3)}
\newcommand{\chandra}{\textit{Chandra}\xspace}
\newcommand{\nustar}{\textit{NuSTAR}\xspace}

\newcommand{\ergs}{erg s$^{-1}$}

\newcommand{\rega}{CNR-E\xspace} 
\newcommand{\regb}{CNR-SE\xspace}

\begin{document}


\maketitle

\title{
A \chandra and ALMA Study of X-ray-irradiated Gas in the Central $\sim$100 pc of the Circinus Galaxy
}


\author{Taiki Kawamuro$^{\ast,1,\dagger}$}
\altaffiltext{1}{National Astronomical Observatory of Japan, Osawa, Mitaka, Tokyo 181-8588, Japan}

\email{taiki.kawamuro@nao.ac.jp}

\author{Takuma Izumi$^{1,\ddagger}$}

\author{Masatoshi Imanishi$^{1,2}$}
\altaffiltext{2}{Department of Astronomy, School of Science, Graduate University for Advanced Studies (SOKENDAI), Mitaka, Tokyo 181-8588}           
\altaffiltext{$\dagger$}{JSPS fellow (PD)}
\altaffiltext{$\ddagger$}{NAOJ fellow}


\begin{abstract}

  We report a study of X-ray-irradiated gas in the central $\sim$100 pc of the Circinus galaxy,
  hosting a Compton-thick active galactic nucleus (AGN), at 10-pc resolution using \chandra and ALMA. Based
  on $\sim$200 ksec \chandra/ACIS-S data, we created an image of the Fe K$\alpha$ line at 6.4 keV,
  tracing X-ray-irradiated dense gas. The ALMA data in Bands 6 ($\sim$270 GHz) and 7 ($\sim$350
  GHz) cover five molecular lines: \molline{CO}{3}{2}, \molline{HCN}{3}{2}, \molline{HCN}{4}{3},
  \molline{HCO$^{+}$}{3}{2}, and \molline{HCO$^{+}$}{4}{3}. The detailed spatial distribution of
  dense molecular gas was revealed, and compared to the iron line image. The molecular gas emission
  appeared faint in regions with bright iron emission. Motivated by this, we quantitatively discuss
  the possibility that the molecular gas is efficiently dissociated by AGN X-ray irradiation (i.e.,
  creating an X-ray-dominated region). Based on a non-local thermodynamic equilibrium model, we
  constrained the molecular gas densities and found that they are as low as interpreted by 
  X-ray dissociation. Furthermore, judging from inactive star formation (SF) reported in the
  literature, we suggest that the X-ray emission has potential to suppress SF, particularly in
  the proximity of the AGN. 
 
\end{abstract}


\section{INTRODUCTION}\label{sec:int} 

Active galactic nuclei (AGN) liberate the enormous gravitational energy of the 
mass accreted by supermassive black holes (SMBHs) into their host galaxies. 
As an extreme case, theoretical works have suggested that AGN are energetic enough
to terminate galaxy evolution, or star formation (SF), by blowing out the surrounding
gas and heating the halo to prevent cooling flows (e.g., \cite{Sil98}; \cite{DiM05};
\cite{Cro06}; \cite{Fab12}). Then, many observational investigations on how the AGN interacts 
with the host galaxy have been conducted (e.g., \cite{Stu11}; \cite{Pag12};
\cite{Har12}; \cite{Tom15}, \yearcite{Tom17}; \cite{Fio17}; \cite{Woo17}). 
Among such studies, spatially resolved spectroscopy has played an important 
role in mapping and/or quantifying AGN effects such as ionized gas and outflows 
(e.g., \cite{Fer10}; \cite{Liu13a}, \yearcite{Liu13b}; \cite{Cic14}; \cite{Har14}; 
\cite{Bae16}; \cite{Bae17}; \cite{Kaw17}; \cite{Oos17}). Furthermore, 
the spatial associations with the SF have been examined, and the results have
showed that AGN may suppress and/or promote SF (e.g.,
\cite{Can12}; \cite{Cre15a}, \yearcite{Cre15b}; \cite{Fer15}; \cite{Mai17}). 

AGN are usually luminous in the X-ray regime, and thus an unavoidable AGN effect is 
X-ray irradiation, which can ionize and heat the surrounding gas. This
region, referred to as the X-ray-dominated region (XDR; e.g., \cite{Mal99}), is
more expansive than the so-called photo-dissociation region by ultraviolet (UV) photons 
due to the high column-penetrating power. Thus far, great efforts have been made 
to understand the physical (i.e., size, density, and temperature; \cite{Pro14}) 
and chemical conditions (e.g., \cite{Lep96}; \cite{Mal96}, \yearcite{Mal99}; \cite{Use04};
\cite{Mei05}; \cite{Mei07}; \cite{Vit14}) in the XDR. Among processes suggested,
an interesting one is that photoelectrons produced by X-rays dissociate
molecules. This may consequently suppress SF activity, given the  
positive correlation between the molecular gas and SF rate (SFR) surface densities
(e.g., \cite{Ken07}; \cite{Big08}), suggesting a causal link between the chemical state of 
interstellar matter and the ability to form stars. 

Mapping the Fe K$\alpha$ fluorescent line at 6.4 keV is a simple way to identify
X-ray-irradiated gas. This line is produced through ionization by X-rays with energies 
above 7.1 keV, corresponding to the K-edge of neutral iron. A medium becomes optically thick
for the 7.1 keV photon around $\log N_{\rm H}/{\rm cm}^{-2} \sim 23.9$, given the low
cross-section and the abundance ($\sigma_\textrm{Fe} \approx 3.5\times10^{-20}$ cm$^{2}$
and $A_\textrm{Fe} \approx 3.3\times10^{-5}$; \cite{Mor83}). As such, the Fe K$\alpha$ 
line has some advantages to investigate regions of X-ray-irradiated dense gas, such as the galaxy center.
First, the low cross-section is convenient for tracing the entire region of dense gas; otherwise, 
only the surface could be probed. Second, the resultant Fe K$\alpha$ photons still retain
good penetrating power, and therefore AGN X-ray-affected regions can be unveiled with a 
small bias against the absorption. Finally, X-rays at energies above the Fe K edge are 
little contaminated by optically thin thermal emission from the SF (e.g., \cite{LaM12}; \cite{Kaw16}).
So far, \chandra, which achieves the highest angular resolution in the X-ray band, has
played a leading role in the investigation of morphological X-ray properties, and has successfully 
unveiled the extended Fe K$\alpha$ emission of some nearby AGN host galaxies
(e.g., NGC 4945, NGC 1068, the Circinus galaxy, and ESO 428$-$G014; \cite{You01}; \cite{Mar12}, 
\yearcite{Mar13}, \yearcite{Mar17}; \cite{Fab17}). 

In this study, we investigated the physical and chemical conditions of X-ray-irradiated gas
at high spatial resolutions, to confirm whether X-rays dissociate molecular gas. For this purpose, 
we studied the central $\sim$100 pc of the Circinus galaxy (hereafter, Circinus), which hosts a 
Compton-thick AGN \citep{Mat99}, using \chandra and ALMA. Both observatories can achieve spatial
resolutions down to sub-arcsec. The redshift, systematic velocity, and distance of the galaxy
are respectively $z$ = 0.001448, $v = 434$ km s$^{-1}$ \citep{Kor04}, and 4.2 Mpc (e.g., \cite{Kar13}),
where 1 arcsec corresponds to 20 pc. The high spatial resolution, achievable given
the proximity, is crucial in discussion of the AGN effect, with the SF effect being discriminated
as much as possible (e.g., \cite{Izu16}). In addition, suppression of the transmitted 
emission by the Compton-thick material makes it easy to detect the faint extended X-ray emission.
Thus, Circinus  is one of the best laboratories for studying X-ray-irradiated gas. 
The AGN center, or the SMBH position, is RA, Decl. (J2000) = 14h13m09.953s, $-$65d20m21.047s, 
determined by the 22-GHz H$_2$O maser observation \citep{Gre03}. 

The remainder of this paper is organized as follows. Section~\ref{sec:circ}  
briefly summarizes what have been observed in Circinus. Section~\ref{sec:obs}
presents an overview of \chandra and ALMA data. \chandra X-ray images and spectra
are analyzed in Section~\ref{sec:cxo_ana}, and three subregions of interest are
defined. Molecular gas properties in the subregions are derived using ALMA data in Section~\ref{sec:alm_ana}.
Our discussion and summary are given in Sections~\ref{sec:dis} and \ref{sec:sum}, respectively. Unless
otherwise noted, errors are quoted at the 1$\sigma$ confidence level for a single parameter of interest.

\section{CIRCINUS}\label{sec:circ} 

Circinus has been observed at various wavelengths, and good quality and/or spatially well-resolved data, 
obtained thanks to the proximity, indicate the presence of an AGN in the center. Dust lanes with an $\sim$100
pc scale are located in the galaxy plane and contribute to the nuclear obscuration \citep{Wil00}, but optical
and near-infrared (IR) spectroscopies have captured AGN signatures such as strong coronal emission lines
(\cite{Oli94}; \cite{Moo96}). One-side ionization cones in the northwest direction, likely originating from the 
AGN, were also imaged at H$\alpha$ and [O~III] (\cite{Mar94}; \cite{Wil00}). Further, a near-IR [Si~VII] line
observation unveiled a cone in the opposite direction \citep{Pri05}. As a signature more closely related
to the AGN, the broad H$\alpha$ emission line was observed in polarized light \citep{Oli98}. X-ray data
have also given beneficial information. Soft X-ray observations detected flat continuum together with a
strong iron line, indicating Compton scattering and fluorescent  emission from gas illuminated by an X-ray
source (\cite{Mat96}; \cite{Smi01}). Then, hard X-ray observations confirmed the central engine heavily obscured
by a large column density ($\log N_{\rm H}/{\rm cm}^{-2} \sim 24$--$25$; \cite{Mat99}; \cite{Sol05}; \cite{Yan09};
\cite{Are14}). 

The SF is also a topic extensively discussed so far. \citet{Mar94} reported $\sim$200 pc scale star 
forming rings (see also \cite{Elm98b}), and molecular gas distributions at similar spatial scales
were also confirmed at various emission lines (e.g., \cite{Elm98c}; \cite{Cur98}, \yearcite{Cur99}; \cite{Izu18}).
The SFR in the rings was estimated to be a few $M_{\rm sun}$~yr$^{-1}$, but it decreases at scales $<$ 100 pc to 
be $\sim$0.01--0.1 $M_{\rm sun}$ yr$^{-1}$ (\cite{Mai00}; \cite{Esq14}).

\section{OBSERVATION DATA}\label{sec:obs} 

\subsection{\chandra Observations}

\begin{table}
\caption{\chandra Data List\label{tab:x_dat_list}} 
\begin{center}
\begin{tabular}{cccccc}
\hline \hline 
 ObsID & Obs. date (UT) & Grating &  Exp.    \\
       &                &         &  (ksec)  \\ 
   (1) &          (2)   &    (3)  &  (4)     \\ \hline 
 12823 &     2010/12/17 & NO      &  147     \\
 12824 &     2010/12/24 & NO      &  38      \\ 
       &                &         (Total & 185)    \\ \hline
 62877 & 2000/06/16     & YES     &      48  \\
 4770  & 2004/06/02     & YES     &      48  \\ 
 4771  & 2004/11/28     & YES     &      52  \\
 &                &  (Total & 148) \\
\hline
\multicolumn{1}{@{}l@{}}{\hbox to 0pt{\parbox{68mm}
{\footnotesize
  \textbf{Notes.}\\
  (1) Observation ID. 
  (2) Observation start date.
  (3) Check on the grating observation.
  (4) Exposure after data reduction.
}
\hss}}
\end{tabular}
\end{center}
\end{table}



\begin{table*}
  \caption{ALMA Data List\label{tab:alm_dat_lis}}
\begin{center}
\begin{tabular}{lllllll}
\hline \hline 
 Tag & Program Info. & Obs. date (UT) & Exp. & Molecules \\
 & & & (min) & \\ 
 (1) & (2)  & (3)  & (4) & (5) \\ \hline 
(a) & \#2015.1.01286.S (PI: F. Costagliola) & 2015/12/31 & 3 & \molline{HCO$^+$}{4}{3}, \molline{HCN}{4}{3}, \molline{CO}{3}{2}\\ 
(a) & \#2015.1.01286.S (PI: F. Costagliola) & 2015/12/31 & 5 & \molline{HCO$^+$}{3}{2}, \molline{HCN}{3}{2} \\ 
(b) & \#2016.1.01613.S (PI: T. Izumi) & 2016/11/24 & 125 & \molline{HCO$^+$}{4}{3} \\ 
\hline
\multicolumn{1}{@{}l@{}}{\hbox to 0pt{\parbox{170mm}
{\footnotesize
  \textbf{Notes.}\\
  (1)~(a) The dataset was used mainly to constrain and discuss the physical and chemical properties of the molecular gas
  (Sections~\ref{sec:radex}, \ref{sec:mol_dia}, and \ref{sec:xdr}).
  ~(b) The dataset was used to reveal the dense molecular gas distribution to be compared to the Fe K$\alpha$ line image
  (Section~\ref{sec:ima_com}). 
  (2) Project identification and principal investigator. 
  (3) Observation start date.
  (4) Total on-source exposure time. 
  (5) Molecular emission line(s) taken from each data. 
}
\hss}}
\end{tabular}
\end{center}
\end{table*}


\chandra observations towards Circinus were made 10 times with the \chandra/Advanced CCD 
  Imaging Spectrometer in the 1$\times$6 array CCD configuration (ACIS-S; \cite{Gar03}). Their focal
  points, where the highest spatial resolution can be achieved, were closely set to the nucleus 
  ($\approx $0.3 arcmin). Out of the ten, the High-Energy Transmission Grating (HETG; \cite{Can05}),
  consisting of two sets of gratings: the High-Energy Grating (HEG) and the Medium-Energy Grating,
  was used four times.

  As summarized in Table~\ref{tab:x_dat_list}, we utilized two data obtained without the HETG grating  
  (ObsID = 12823 and 12824) to discuss spatial distributions of X-ray emission. The remaining four 
  among the six non-grating data were ignored, because of small contributions to the total exposure ($\sim$200 ksec).
  \citet{Mar13} already reported X-ray spectral and image analyses using the same ObsID = 12823
  and 12824 data, but we made additional efforts to discriminate between the nuclear and extended
  emission (Sections~\ref{sec:xray_map} and ~\ref{sec:x_ana}).

  Moreover, among the four grating data of all the ten, we used three data (ObsID = 4770,  4771, and 62877),
  each obtained with a larger exposure of $\approx$60 ksec than the remaining one ($\approx$7 ksec). A spectrum of the nuclear 
  point source was constrained from them, and subsequently was used to determine how largely the
  nuclear X-ray emission contaminates spectra extracted from outer regions. That was also utilized
  to estimate fractions of the pile-up, in which more than one photon is counted as one photon within
  a single readout. Particularly around the nucleus, the pile-up effect cannot be ignored. Although
  the 0th order of the grating data is equivalent to the imaging, we did not take account of it for
  our main imaging analysis because of possible calibration uncertainties \citep{Are14}. In 
  Appendix~\ref{sec:app}, we however present a result that includes the 0th order data and find it 
  consistent with our main result. Note that in this supplemental analysis, we additionally 
  take account of usable data obtained with an aim to study a nearby object of SN 1996cr, $\sim$20 arcsec 
  away from Circinus. Most of them were taken in the grating mode.




\subsection{ALMA Observations} 

We analyzed Band 6 ($\sim$270 GHz) and 7 ($\sim$350 GHz) ALMA datasets obtained through two programs 
\#2015.1.01286.S (PI: F. Costagliola) and \#2016.1.01613.S (PI: T. Izumi), as summarized in Table~\ref{tab:alm_dat_lis}. 
In the former program, two observations with short on-source exposures ($\sim$ a few minutes) were 
executed. They covered five molecular lines: \molline{CO}{3}{2} at the rest-frame frequency of $\nu_{\rm rest}$
= 345.796 GHz, \molline{HCO$^{+}$}{3}{2} at $\nu_{\rm rest}$ = 267.558 GHz, \molline{HCO$^{+}$}{4}{3}
at $\nu_{\rm rest}$ = 356.734 GHz, \molline{HCN}{3}{2} at $\nu_{\rm rest}$ = 265.886 GHz, and \molline{HCN}{4}{3}
at $\nu_{\rm rest}$ = 354.505 GHz. We used this program to estimate their line ratios, and finally 
to discuss properties of the molecular gas (Sections~\ref{sec:radex}, \ref{sec:mol_dia}, and \ref{sec:xdr}).
This allowed us to ignore the systematic uncertainty in the absolute flux as much as possible that
can be caused between different observations. 
On the other hand, the latter program yielded much better-quality data for \molline{CO}{3}{2}
and HCO$^{+}$($J$=4--3) using a long exposure ($\sim$ 2 h). In this paper, we analyze only the 
\molline{HCO$^{+}$}{4}{3} data to reveal the dense molecular gas distribution in great detail 
by exploiting the high signal to noise ratio (SNR). An analysis of the \molline{CO}{3}{2} line
data was already reported by \citet{Izu18}. 

HCN($J$=4--3), HCO$^+$($J$=4--3), and CO($J$=3--2) lines were observed simultaneously in 
Band 7 on 2015 December 31, when 38 antennas were operated.
  The maximum recoverable scale (MRS), the largest angular scale structure that
  can be recovered from observations, was 7 arcsec. This is large enough 
  to discuss the central $\approx$100 pc, or 5 arcsec, scale structure.
Out of four spectral windows, three were used to observe the lines. Each window had a total bandwidth
of 1.875 GHz and 240 channels. The raw spectral resolution was 7.8 MHz (6.8 km s$^{-1}$), but we 
binned three spectral elements to achieve a resolution of 23 MHz (20 km s$^{-1}$), as
explained in Section~\ref{sec:alm_spec_ana}. The total on-source time was $\approx$ 0.05 h (3 min).
Standard flux, bandpass, and phase calibrations 
were conducted based on the observations of Titan, J1427-4206, and J1424-6807, respectively.

Band 6 simultaneous observations of HCN($J$=3--2) and HCO$^+$($J$=3--2) lines 
were made on 2015 December 31 using 36 antennas. A larger MRS of 9 arcsec was achieved because of the lower frequencies.
Two of three spectral windows were used for the line detection in the same total bandwidth and number of channels as above.
The objects for the calibrations were also the same. 
The total on-source time was $\approx$ 0.09 h (5 min). As in the case above, a
resolution of 16 MHz (18 km s$^{-1}$) was adopted by binning two spectral elements at the 
raw resolution of 7.8 MHz (8.9 km s$^{-1}$).

We mapped the HCO$^+$($J$=4--3) line using Band 7 data acquired on 2016 November 24 and 26,
and 2017 May 05 using 42, 42, and 47 antennas, respectively.
  The MRS was the same as that for the above HCO$^+$($J$=4--3) line data (i.e., 7 arcsec).
Four spectral windows were employed in 
the observations, each with a total bandwidth of 1.875 GHz and 480 channels. Although two of the spectral
windows can be used to detect the HCO$^+$($J$=4--3) line, we used only one that covered a broader frequency 
range around the HCO$^+$($J$=4--3) line. The total on-source time was $\approx$ 2.1 h (125 min).
The calibration sources for flux, bandpass, and phase were J1427-4206 or J1617-5848, J1427-4206, and
J1424-6807, respectively. Because it is possible to detect broad but faint components using a high SNR, we adopted
a finer resolution of 12 MHz (9.8 km s$^{-1}$), which was achieved by binning three spectral elements at the
raw resolution of 3.9 MHz (3.3 km s$^{-1}$).


%

\section{\textit{CHANDRA} DATA ANALYSES}\label{sec:cxo_ana} 

\begin{table*}
  \caption{Positional Subregion Information \label{tab:pos_inf}}
\begin{center}
\begin{tabular}{ccccccccccc}
  \hline \hline
  Region & R.A. (J2000) & Decl. (J2000) & Sep. angle & Dist. \\
  &  &  & (arcsec) & (pc) \\ 
  (1) &  (2) & (3)  & (4) & (5) \\ \hline 
  Nucleus & 14h13m09.953s & $-$65d20m21.047s &... &... \\
  \rega & 14h13m10.402s & $-$65d20m20.205s & 2.9 & 62 \\  
  \regb & 14h13m10.342s & $-$65d20m22.461s & 2.8 & 60 \\
  \hline
  \multicolumn{1}{@{}l@{}}{\hbox to 0pt{\parbox{98mm}
{\footnotesize
  \textbf{Notes.}\\
 (1) Name of the subregions defined in Figure~\ref{fig:x_ima}
 (2) Right ascension. 
 (3) Declination. 
 (4) Separation angle to the nucleus, or the SMBH position. 
 (5) Deprojected physical distance (see detail in Section~\ref{sec:xray_map}). 
}
\hss}}
\end{tabular}
\end{center}
\end{table*}

\begin{figure}
 \hspace{-.5cm}
 \includegraphics[angle=-90,scale=0.32]{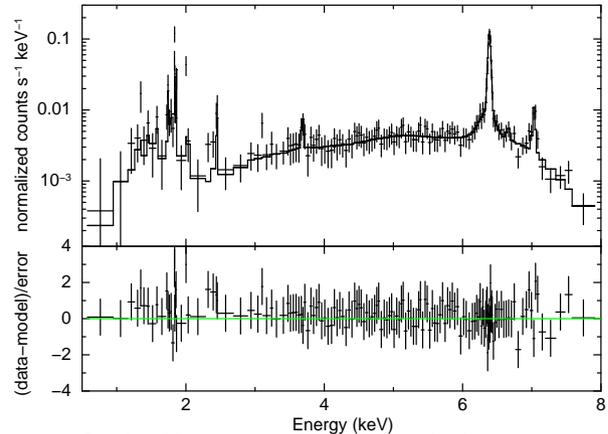} \hspace{-.5cm}
 \caption{\small{
       Grating X-ray spectrum (crosses in the upper panel), folded by the
       response function. The best-fit model is shown by a solid line.
       For clarity, all three spectra used for the fit were combined. The lower panel plots the 
       residuals.
     }
 }\label{fig:nuc_spe}
\end{figure}

\begin{figure*}
  \hspace{-.5cm}
  \includegraphics[scale=0.22]{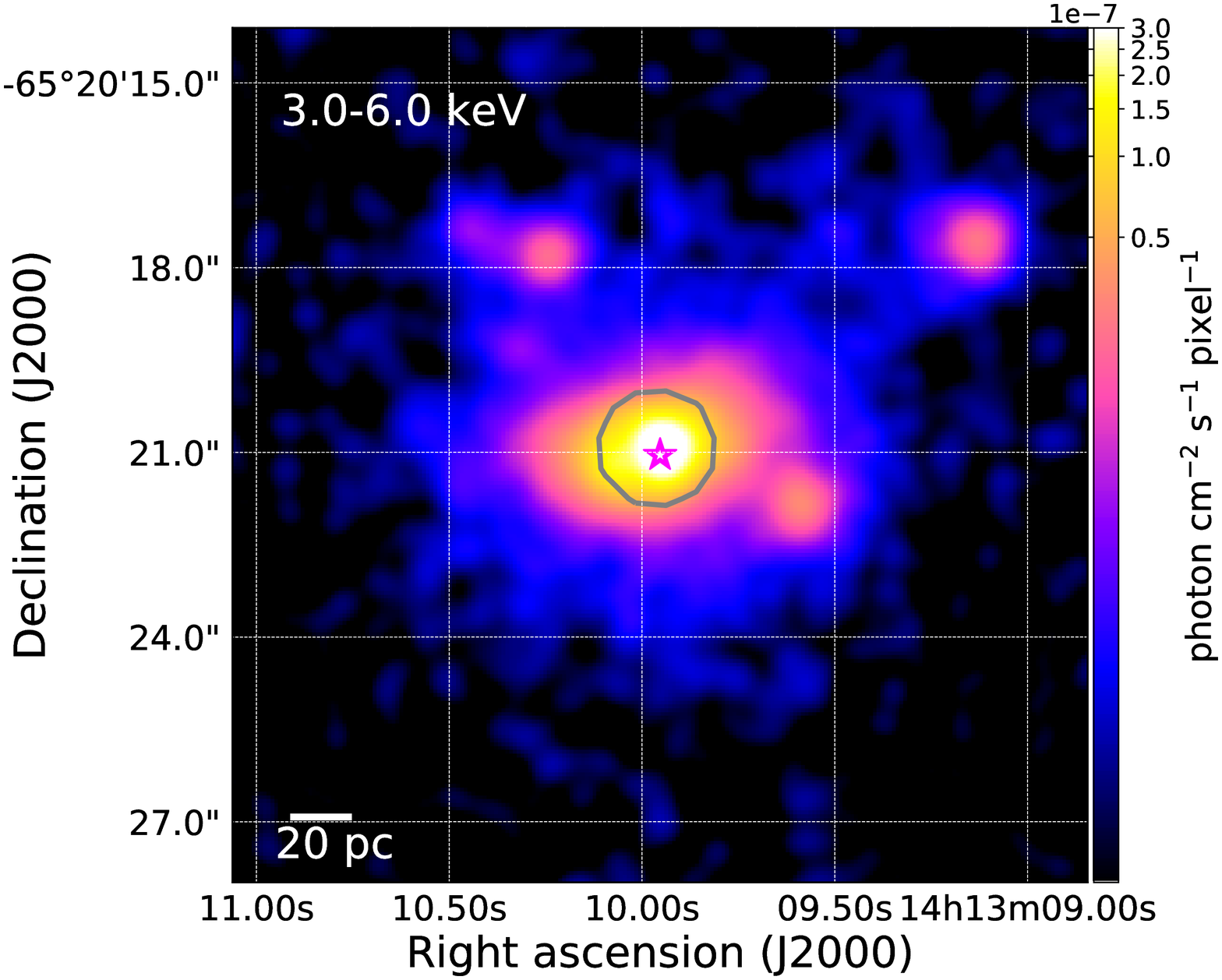}
  \includegraphics[scale=0.22]{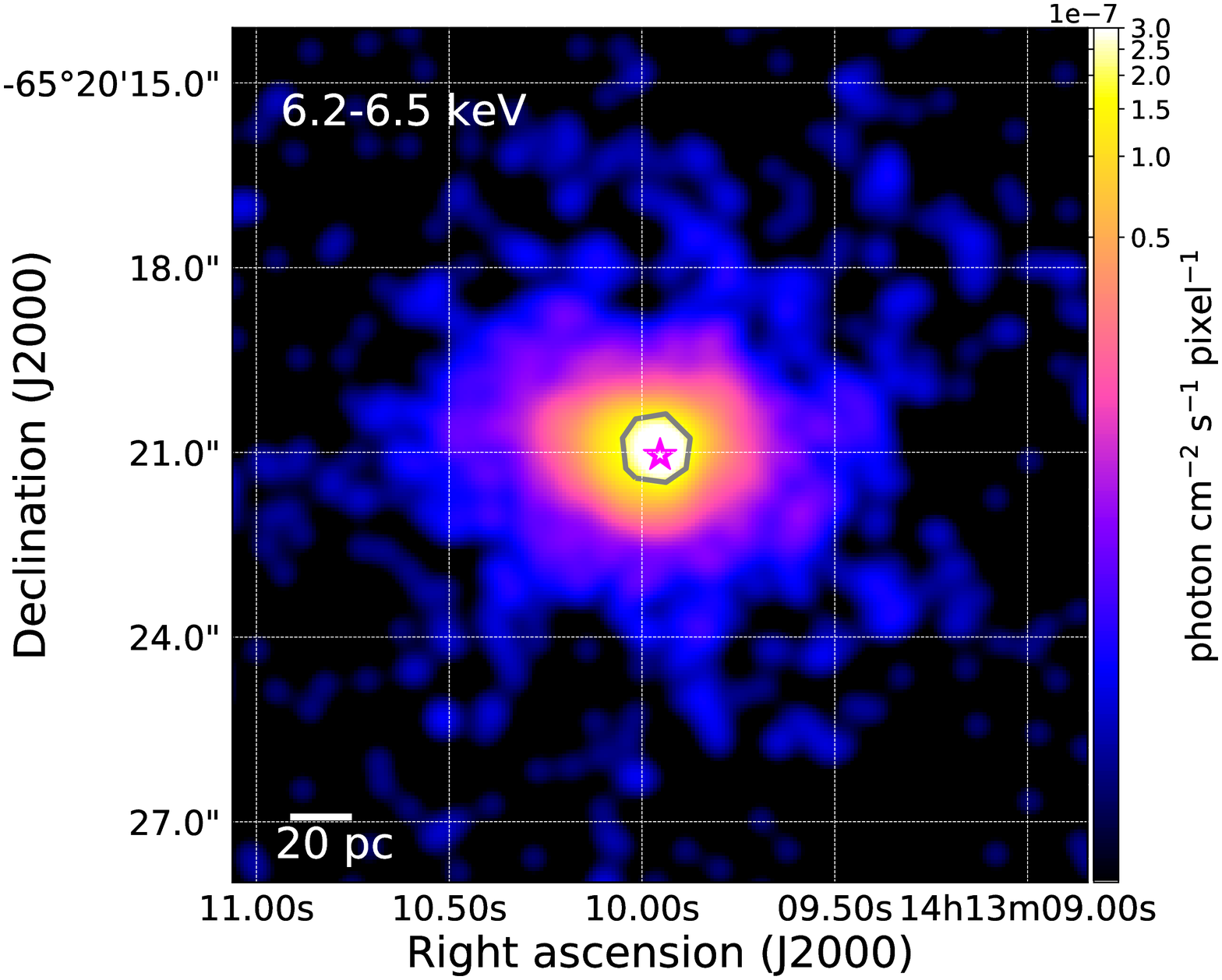} 
  \includegraphics[scale=0.22]{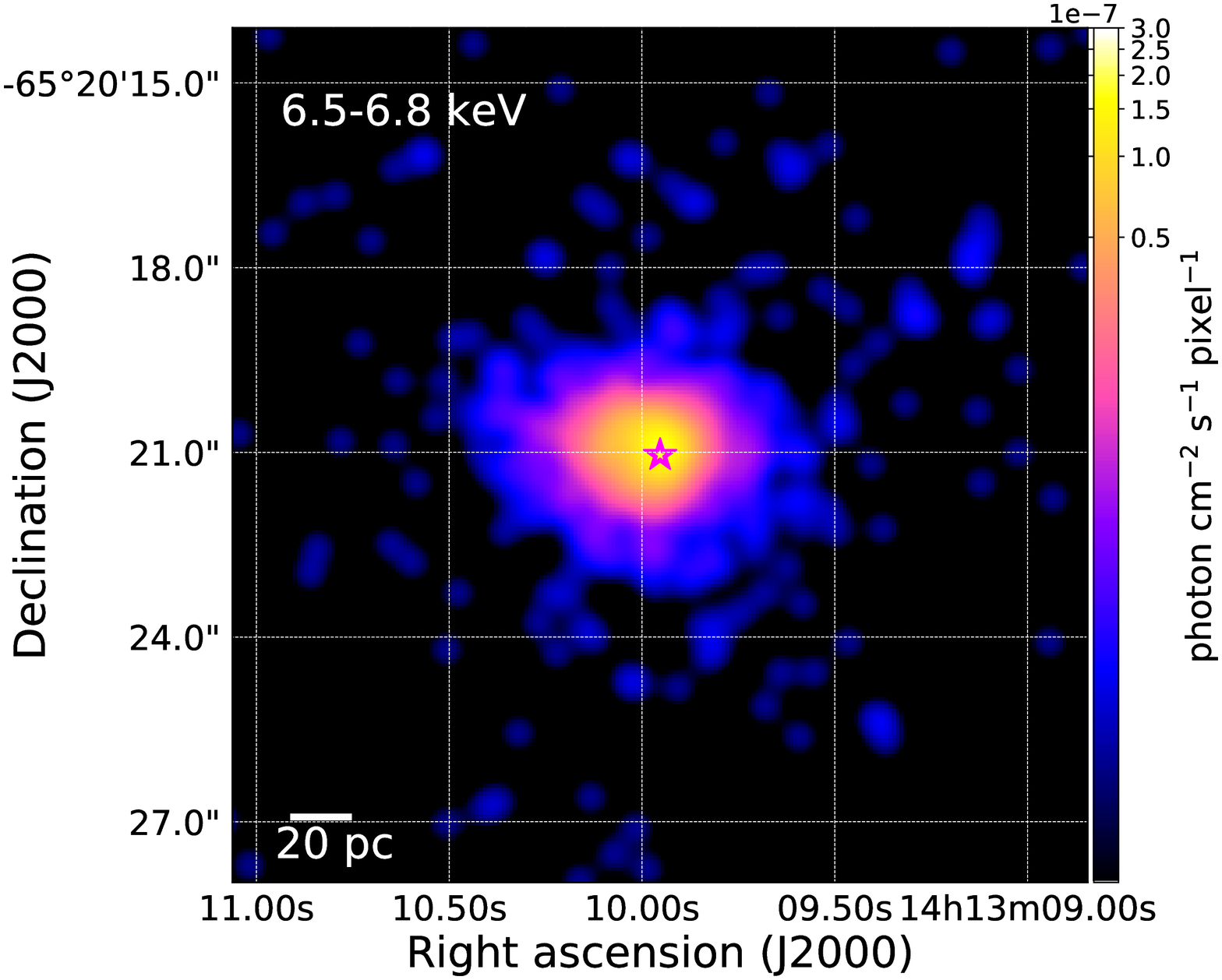}
  \\ 
  \hspace{-.55cm}  
  \begin{center}
  \includegraphics[scale=0.29]{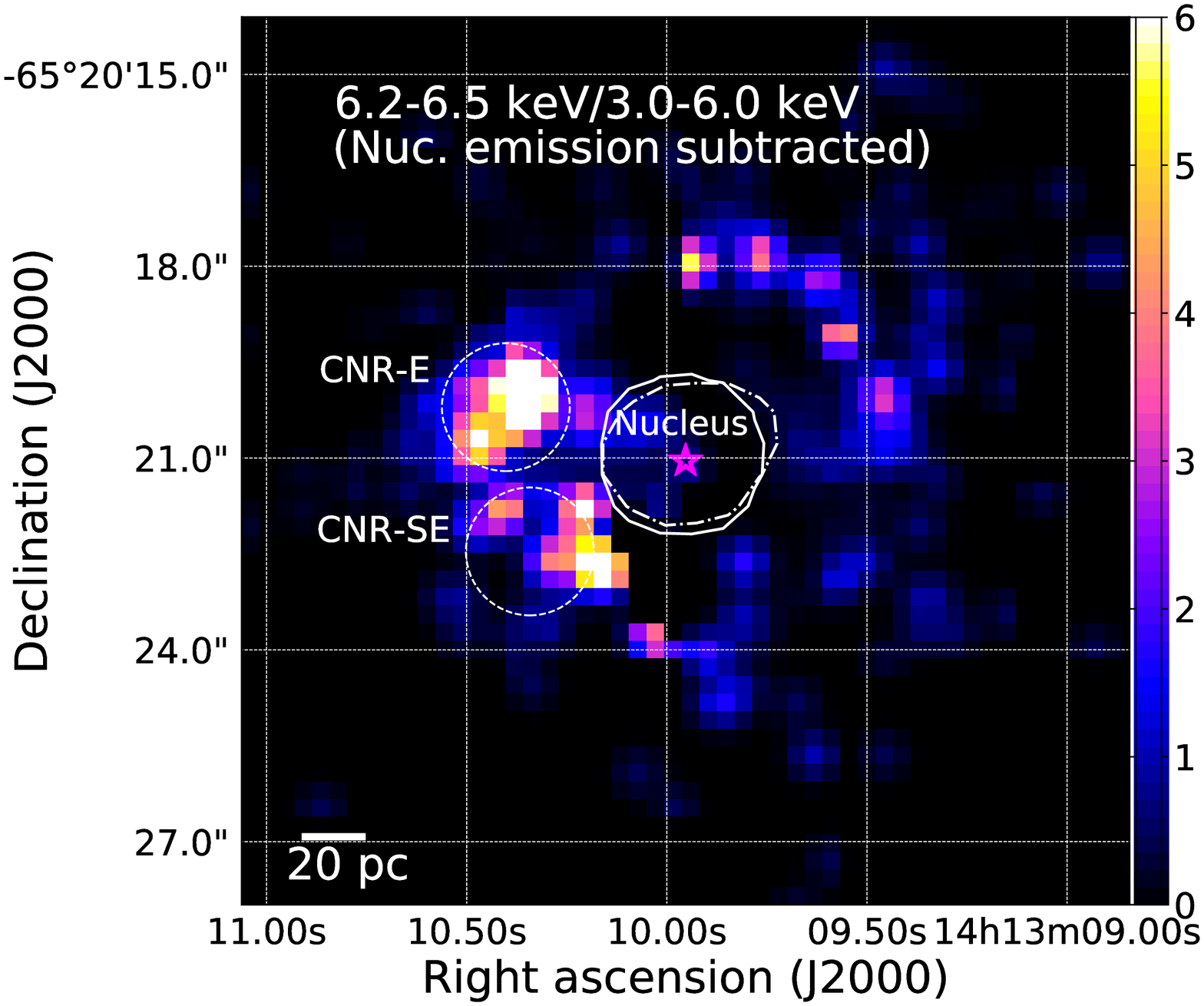}   
  \includegraphics[scale=0.29]{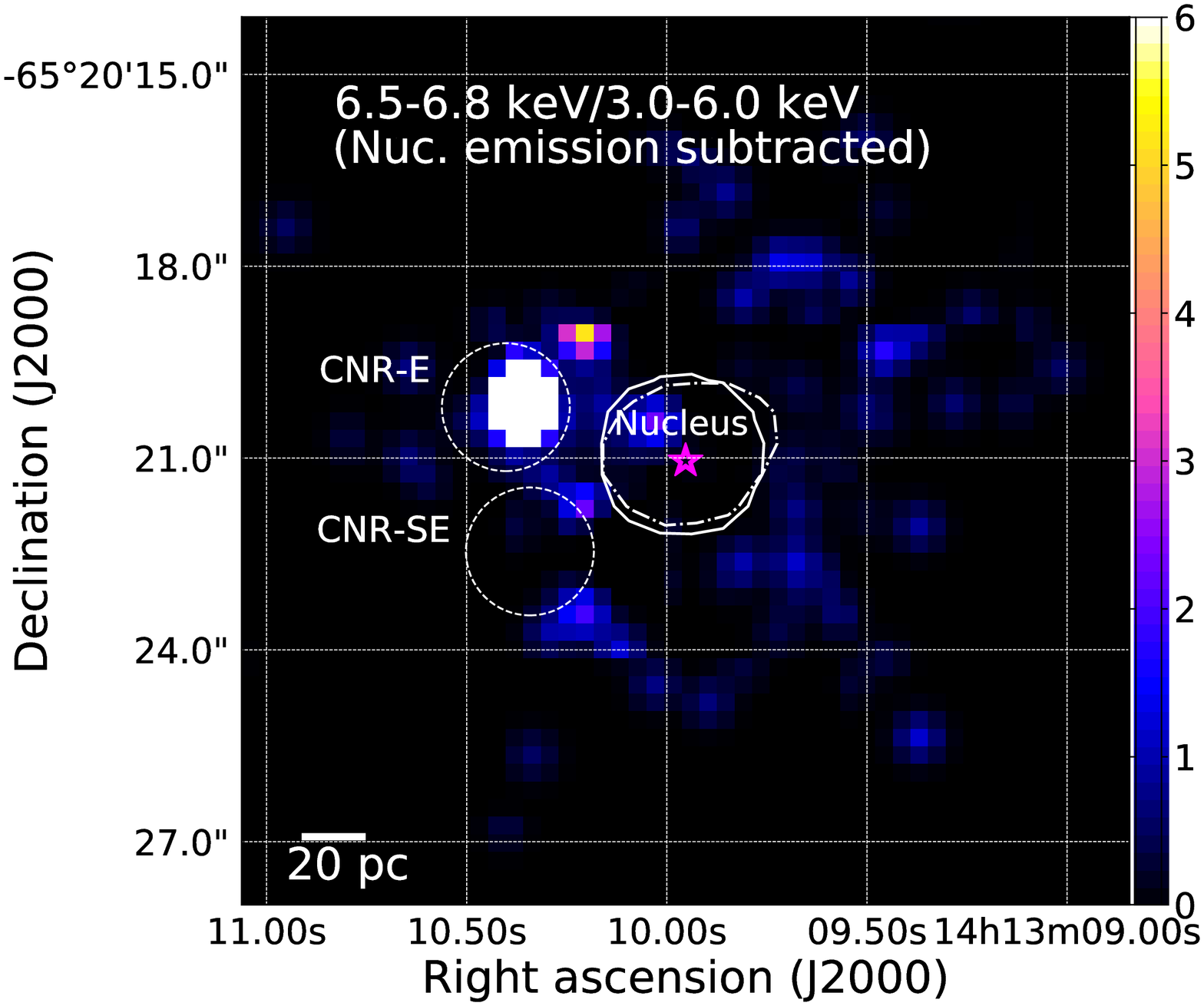}   
  \end{center}
  \vspace{1.cm}
  \caption{\small{
      Top panels show integrated ACIS-S images in different energy bands 
     (from left to right: 3.0--6.0 keV, 6.2--6.5 keV, and 6.5--6.8 keV)
     within the central $\approx$14$\times$14 arcsec$^2$ region. The size
     of each pixel is 0.0625$\times$0.0625 arcsec$^2$, and smoothing was
     performed using a Gaussian kernel with FWHM of 0.495 arcsec. The gray solid contours enclose the area with pile-up
     fractions larger than 5\% for each energy band. Bottom panels show images 
     constructed by dividing two of the three images (from left to right: 
     6.2--6.5 keV/3.0--6.0 keV and 6.5--6.8 keV/3.0--6.0 keV) from which the
     nuclear point source emission was subtracted. We calculated the ratios with 
     images whose pixel size was 0.25$\times$0.25 arcsec$^2$. Smoothing
     was performed with the same Gaussian kernel as above. Contours corresponding
     to the pile-up fraction of 5\% determined from the 0.5--8.0 keV simulated
     data and the observed data via the \texttt{pileup\_map} command are
     shown by solid and dot-dashed lines, respectively. Two dotted circles
     with a 1-arcsec radius indicate the subregions of \rega and \regb.
     In all of the figures, the magenta star corresponds to the SMBH
     position, labeled as the Nucleus, the third subregion.
 }
 }\label{fig:x_ima} 
\end{figure*}

\subsection{Reduction and Reprocessing of \chandra Data}\label{sec:red_cxo}

The \chandra data were reduced using the \chandra Interactive Analyses of Observations (CIAO)
4.9 and the \chandra Calibration Data Base 4.7.6, provided by the \chandra team. Following 
standard procedures, we reprocessed all raw data using the \texttt{chandra\_repro} script. 
Furthermore, we examined the data with \texttt{deflare} to remove background flaring periods
when the rates deviated from the mean by more than 3$\sigma$. 

We checked the absolute astrometry of the \chandra image based on the SMBH position that was 
determined by the H$_2$O maser observation. We representatively examined ObsID 12823 because of 
the longest exposure. The centroid of the nuclear X-ray emission was likely coincident with the
central AGN engine, or the SMBH position. Hence, they were compared to each other. The derived
spatial offset was $\approx$0.1 arcsec, sufficiently small to be ignored when we discuss 
larger-scale structures ($\gtrsim$0.5 arcsec). Thus we used it as a basis for the X-ray image. 

We corrected the remaining imaging data (ObsID = 12824) for a systematic offset by spatially
matching point sources distributing in both images. We started by making exposure-corrected
0.5--7.0 keV images with 
\texttt{fluximage} where exposure maps were created with an effective energy of 2.3 keV. Also, 
we looked up the point spread function (PSF) size of each pixel in all images using the \texttt{mkpsfmap}
script. Combined with these results, \texttt{wavdetect} was run to create X-ray source lists, 
containing information on the source angular size and SNR. At last, the adjustment was made by
adopting only point-like, bright sources detected with PSF size $<$ 1.5 arcsec, source angular
size $<$ 1.0 arcsec, and SNR $>$ 7. We found that the offset was 0.2 arcsec, much smaller
than structures of interest. This suggests that even if we choose the ObsID = 12824 as the basis, 
it does not affect our discussion.

\subsection{X-ray Emission from Nuclear Point Source}\label{sec:sim_xray}

To properly constrain spatial and spectral properties of extended emission 
around the bright AGN, we need to take account of contamination by the AGN, 
possible due to the finite PSF. Here, we simply 
define the AGN as a point source spatially unresolved with \chandra. For that
purpose, we prepared a simulated AGN data.

\subsubsection{Grating Spectra}\label{sec:gra_spe}

We first determined an X-ray spectrum of the AGN. Because the \chandra imaging
suffered from the pile-up effect around the nucleus (Section~\ref{sec:xray_map}), 
we used the grating data, or the HEG data, reprocessed in Section~\ref{sec:red_cxo}. 
We extracted spectra using a 4.8 arcsec full-width in the cross-dispersion direction
centered on the AGN. The plus and minus first order spectra were combined to increase 
the SNR. The spectra were binned so that each energy bin had at least one count. 
Accordingly adopting the $C$-statistic, appropriate even for low photon counts, 
we determined the best-fit on XSPEC Version 12.9.1m \citep{Arn96}. 

We simultaneously fitted the three spectra in the 0.5--8.0 keV band with two power-law components
and emission lines identified by \citet{Sam01}. They conducted a spectral analysis for the ObsID
= 62877 HETG data. Because residuals were left in the fit, we further 
added several Gaussian components and obtained a good fit with $C$-statistic/degrees
of freedom (d.o.f) = 2670/3130. Figure~\ref{fig:nuc_spe} shows the fitting result. 
The precise measurement of the spectral shape is the main aim of this analysis, and
the physical interpretation is beyond the scope of this work.


\subsubsection{MARX Simulation}\label{sec:marx}

Next, we simulated imaging observations of a point source having the 0.5--8.0 keV 
spectral shape determined in Section~\ref{sec:gra_spe}. We utilized the MARX Version 
5.3.3 \citep{Dav12}, which performs ray-tracing simulations that take account of the
mirror and detector responses. Simulation parameters, such as nominal position,
start time, and exposure, were set to those of each of the two observations (ObsID = 12823 and 12824). The 
source position was set to the X-ray peak found in the 3--6 keV band image. Even
if we adopt the SMBH position instead, our conclusion does not change. To reduce
statistical fluctuations of the simulated models, we performed 100 MARX simulations
for each observation, and took their average. Because the pile-up was ignored in
the simulation as default, we also created pile-up affected data according to a
standard procedure\footnote{http://cxc.harvard.edu/ciao/threads/marx\_sim/}. Then,
we confirmed good agreement between the simulated and observed data by comparing
their central 1-arcsec spectra. The data without the pileup is used to confirm regions
affected by the pile-up at different energy bands by taking advantage of the record of the intrinsic count
rates (Section~\ref{sec:xray_map}). The other data is used to subtract the nuclear emission
from observed images to reveal the extended emission (Section~\ref{sec:xray_map}), 
and also is taken into consideration in analyzing spectra at the subregions of
interest (Section~\ref{sec:x_ana}).


\begin{figure*}
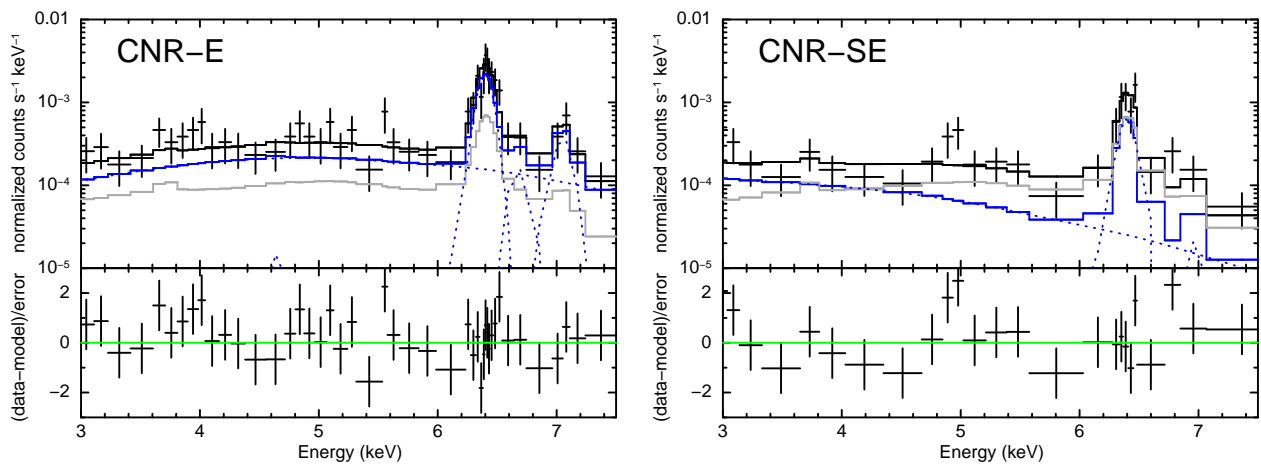

\begin{center}
  \hspace{-.5cm}
 \includegraphics[angle=-90,scale=0.33]{limb_e_spec_pow.ps} \hspace{-.2cm}  
 \includegraphics[angle=-90,scale=0.33]{cone_spec_pow.ps} 
 \end{center} 
 \caption{\small{
       X-ray spectra (black crosses) extracted from the \rega and \regb regions.
       Each figure represents only a single spectrum taken from the ObsID = 12823 data 
       and rebinned for clarity. 
       The spectra were reproduced by the contaminating X-ray from the nuclear point source 
       (gray) and the extended emission of interest (blue). Lower data represent the residuals.
     }
 }\label{fig:x_fit_pow}
\end{figure*}

\subsection{Mapping of Iron Emission Lines}\label{sec:xray_map}

We probe the X-ray-irradiated dense gas mainly through an Fe K$\alpha$ emission line map. 
Other emission from ionized irons is useful for the discussion as well. Thus, it is mapped 
supplementarily. First, we create X-ray images based on Energy Dependent Sub-pixel Event
Repositioning (e.g., \cite{Tsu01}; \cite{Mor01}; \cite{Li03}, \yearcite{Li04}) \footnote{More 
  details are available in \\ \hspace{1cm} http://cxc.harvard.edu/ciao/why/acissubpix.html \\
 \hspace{1cm} http://cxc.harvard.edu/ciao/guides/acis\_subpixel.html}, which makes it possible to 
achieve angular resolutions greater than the CCD pixel size of 0.498 arcsec. The algorithm redistributes 
photons based on the shape of the charge cloud recorded in a 3$\times$3 pixel island, while considering the
energy dependence. At last, all images are merged into one to increase the SNR using \texttt{merge\_obs}. 

The upper panels of Figure~\ref{fig:x_ima} show the 3.0--6.0 keV, 6.2--6.5 keV, and 6.5--6.8 keV band images,
whose exposures maps are created by representatively adopting their intermediate energies of 4.50 keV, 6.35
keV, and 6.65 keV, respectively. A pixel size of 0.0625$\times$0.0625 arcsec$^2$ is adopted, and the images
are smoothed via a Gaussian kernel with full width half maximum (FWHM) = 0.495 arcsec. The energy bands are selected to trace the
continuum component, the neutral Fe K$\alpha$ emission, and the emission from ionized He-like irons (Fe XXV
He$\alpha$), respectively. Although emission from highly ionized irons (Fe XXVI Ly$\alpha$) can be seen at 6.97
keV, we ignore it because of the difficulty to isolate the line emission from the neutral Fe K$\beta$ emission
at 7.06 keV. The lower panels of Figure~\ref{fig:x_ima} show images created by dividing two of the three energy
band images. To increase the SNR, pre-divided X-ray images are made by adopting a larger pixel size of 
0.25$\times$0.25 arcsec$^2$. Also, the simulated AGN emission is subtracted to reveal the extended emission. 
Smoothing is conducted using the Gaussian kernel with FWHM = 0.495 arcsec. The ratio between the 6.2--6.5
keV and 3.0--6.0 keV images and that between the 6.6--7.0 keV and 3.0--6.0 keV images 
correspond to proxies of the Fe K$\alpha$ line and ionized iron line equivalent widths, respectively.

Here, we check regions affected by the pile-up. The fraction of piled events to the total detections
($f_{\rm p}$) can be calculated as $f_{\rm p} = 1 - \alpha/(\exp(\alpha \Lambda) - 1)$ 
\footnote{see also ``The Chandra ABC Guide to Pileup" in http://cxc.harvard.edu/ciao/download/doc/pileup\_abc.pdf}. 
The $\alpha$ parameter is a probability that for each photon event beyond the first, the piled event 
is identified as a real event, and $\Lambda$ represents the intrinsic counts per detector region
per frame time. With $\Lambda$ from the simulated data (Section~\ref{sec:marx}) and an assumption 
of $\alpha = 1$, corresponding to the most conservative case, the fraction can be estimated 
for a given energy band. Figure~\ref{fig:x_ima} shows contours surrounding areas with pile-up fractions
$>$ 5\%. Regarding the images of the ratios, we representatively show contours determined using the 
0.5--8.0 keV events. Note that a similar result can be obtained using the CIAO \texttt{pileup\_map} 
task (Figure~\ref{fig:x_ima}), which approximately calculates the pile-up fraction based on the observed
events. 

We find that the Fe K$\alpha$ emission seems to extend up to $\approx$3 arcsec ($\sim$60 pc) to the
east and southeast (the lower left panel of Figure~\ref{fig:x_ima}), consistent with \citet{Mar13}. 
This configuration looks like a conical structure oriented towards the southeast direction, and may
be a counterpart of the optical ionization cone found in the northwest direction \citep{Mar94}. As
denoted in the lower left panel of Figure~\ref{fig:x_ima}, we define two circum-nuclear regions 
with moderate Fe K$\alpha$ emission as \rega and \regb, which are our main interest in this paper.
Also, the central region is labeled as Nucleus. Table~\ref{tab:pos_inf} lists detailed information on the three 
subregions. The deprojected distances of the \rega and \regb regions are estimated to be $\approx$60
pc from the SMBH. This estimate assumes that the subregions associate with a counterpart of the
northwest H$\alpha$ ionization cone \citep{Elm98}, inclined by 70 degrees with respect to our sightline.

\subsection{Spatially Resolved X-ray Spectral Analyses}\label{sec:x_ana}

\begin{table*}
  \caption{Best-fit Parameters \label{tab:x_fit_pow}}
  \begin{tiny}
    \begin{tabular}{cccccccccccccccccccc}
      \hline \hline
      Region & $\Gamma$ & Norm &  $I_{{\rm Fe~I~K}\alpha}$ & $I_{{\rm Fe~I~K}\beta}$ & $I_{{\rm Fe~XXV~He}\alpha}$
      & $I_{{\rm Fe~XXVI~Ly}\alpha}$ & EW$_{{\rm Fe~I~K}\alpha}$ & $C$-stat./d.o.f \vspace{3mm} \\ 
      \cline{4-7} \vspace{-1mm} \\ 
      & & \scriptsize{(10$^{-7}$ photons } & \multicolumn{4}{c}{\scriptsize{(10$^{-7}$ photons cm$^{-2}$ s$^{-1}$)}} & (keV) & \\
      & & \scriptsize{\hspace{7mm}keV$^{-1}$ cm$^{-2}$ s$^{-1}$)} & & & & & & \\ 
      (1) & (2) & (3)   & (4)  & (5)   & (6)   & (7)    & (8) & (9) \\ \hline
      \rega & $-1.6\pm0.6$        & $0.50^{+0.72}_{-0.38}$ & $21\pm4$ & $6.9^{+3.1}_{-3.0}$ & $1.8^{+1.8}_{-1.5}$ & $0.0~(<1.9)$ & $2.3^{+1.4}_{-0.8}$ 
      & 152/179 \\
      \regb & $1.2^{+1.6}_{-1.3}$ & $12^{+83}_{-10}$       & $6.5^{+2.4}_{-2.2}$            & $0.34~(<2.24)$ & $0.18(<1.40)$ & $0.61~(<2.40)$ & $4.8^{+19.6}_{-3.3}$
      & 97/119 \\
  \hline
  \multicolumn{1}{@{}l@{}}{\hbox to 0pt{\parbox{168mm}
{\footnotesize
  \textbf{Notes.}\\
Columns:
(1) Subregion name. 
(2)-(3) Photon index and normalization of the power-law component. 
(4)-(7) Normalizations of the iron lines. 
(8) Equivalent width of the 6.4 keV Fe K$\alpha$ line. 
(9) $C$-statistic value over degrees of freedom.
}
\hss}}
\end{tabular}
      \end{tiny}  
\end{table*}

Based on X-ray spectral analyses, we confirm that the Fe K$\alpha$ line is the result of X-ray
illumination by the central engine. We extract 3.0--7.5 keV spectra from the two circular regions
(\rega and \regb) with a 1-arcsec radius. We exclude the soft X-ray band ($<$ 3 keV) because
non-AGN emission such as optically thin thermal emission from the SF largely contributes to it
and complicates the analyses. The data above 7.5 keV are filtered because of the low effective 
area, and in order to avoid uncertainty in edge of the AGN model, which is incorporated into spectral 
fits. Note that the pile-up effect is negligible ($<$ 5\%) in the two subregions (Section~\ref{sec:xray_map}). 
Background spectra are estimated from a blank 50-arcsec radius circle located in the same CCD.
Response files are generated using the CIAO tool \texttt{specextract}. We examine time variability 
within each observation based on the Gregory--Loredo algorithm\footnote{http://cxc.harvard.edu/ciao/threads/variable/},
and find it to be insignificant. Thus, we average the individual spectra. We bin the spectra 
so that each energy bin has at least one count. In the same way as in Section~\ref{sec:x_ana},
the best-fits are determined using the $C$-statistic technique on XSPEC. The spectra do not show
significant variability between the two observations, and thus are fitted simultaneously with 
a single model to improve the SNR, without merging them. 

Figure~\ref{fig:x_fit_pow} shows the spectra folded by the response function. We uniformly 
fit a power-law plus four Gaussian functions (\texttt{zgauss} in XSPEC terminology), reproducing
the Fe K$\alpha$ line at 6.40 keV, the Fe K$\beta$ line at 7.06 keV, the Fe XXV He$\alpha$ line
at 6.68 keV, and the Fe XXVI Ly$\alpha$ line at 6.97 keV. The line widths are fixed at $\sigma$
= 20 km s$^{-1}$ or 0.4 eV by assuming that the iron atoms couple with molecules, whose velocities 
are roughly estimated from the ALMA data (Section~\ref{sec:bas_mol_lin}). Note that even
  if the line widths are fixed at 0 eV, our conclusion is not affected. The AGN emission determined
in Section~\ref{sec:gra_spe} is further incorporated through response files produced by assuming
a point source located in the AGN position. This component is fixed for simplicity. In summary,
the free parameters are the normalization and photon index ($\Gamma$) of the power-law component,
and the line normalizations. The fit results are summarized in Table~\ref{tab:x_fit_pow}. We
significantly detect Fe K$\alpha$ lines in the \rega and \regb regions, whose EWs are $2.3^{+1.4}_{-0.8}$
keV and $4.8^{+19.6}_{-3.3}$ keV, respectively. Such high EWs can be reproduced if a direct photoionizing
X-ray source is not seen in the sightline and the reflected X-ray emission dominates the continuum
(e.g., \cite{Ter01}; \cite{Nob10}; \cite{Tan18}).

Motivated by the above results, we further fit the spectra by replacing the power-law with a 
more physically-motivated model of \texttt{pexrav}. The model  calculates a reflected continuum
spectrum from an optically thick neutral slab with a solid angle ($R \equiv \Omega/2\pi$) irradiated
by a cut-off power-law component \citep{Mag95}. We fix the normalization, photon index, and cut-off
energy of the incident power-law component to 0.7 photons at keV$^{-1}$ cm$^{-2}$ s$^{-1}$ at 1 
keV, 2.31, and 160 keV based on the result of \citet{Are14}. The inclination angle to the slab,
which has little impact on results, is representatively fixed at 60$^\circ$. The same 
power-law and inclination angle are adopted also in below models if required. Thus, the \texttt{pexrav} 
model has the only free parameter of the solid angle ($R$). The best-fit models of \rega 
($C$-statistics/d.o.f = 160/180) and \regb ($C$-statistics/d.o.f = 98/120) are obtained with 
$R = 5.7^{+0.8}_{-0.7}\times10^{-4}$ and $R = 2.4^{+1.0}_{-0.9}\times10^{-4}$. 
Their EWs are estimated to be $4.1^{+1.5}_{-1.1}$~keV and $2.6^{+2.4}_{-1.5}$~keV, respectively. 
Although the
face values of the solid angles suggest very tiny reflection material, these can be interpreted
as lower limits, given that the normalization of the power-law component may be suppressed by
absorption.
If we alternatively fix the solid angle to 1 and leave the power-law normalization free, 
  we can get $4.0\pm0.5\times10^{-4}$ photons keV$^{-1}$ cm$^{-2}$ s$^{-1}$ for \rega and
  $1.7\pm0.4\times10^{-4}$ photons keV$^{-1}$ cm$^{-2}$ s$^{-1}$ for \regb. 
  We can also reproduce the spectra well using \texttt{pexmon} \citep{Nan07} instead
  of \texttt{pexrav}. The model gives the sum of the continuum from the \texttt{pexrav} model
  \citep{Mag95}, the fluorescence lines of Fe-K$\alpha$, Fe-K$\beta$, and Ni-K$\alpha$, and the Compton 
  shoulder of the Fe-K$\alpha$ line, in a self-consistent way. The fitting returns almost the same 
  solid angles ($R = 7.6^{+0.8}_{-0.7}\times10^{-4}$ with $C$-statistics/d.o.f = 200/182 
  and $R = 2.8\pm0.6\times10^{-4}$ with $C$-statistics/d.o.f = 101/122 for \rega and \regb, respectively).
  In the same manner as in the case of \texttt{pexrav}, we fit the spectra by fixing
    $R$ at 1 and allowing the normalization to vary, and obtain 
    $5.3^{+0.6}_{-0.5}\times10^{-4}$ photons keV$^{-1}$ cm$^{-2}$ s$^{-1}$ for \rega and 
    $1.9\pm0.4\times10^{-4}$ photons keV$^{-1}$ cm$^{-2}$ s$^{-1}$ for \regb.
  Finally, we fit the \texttt{xillver} model \citep{Gar13}, a more complicated reflection spectral
  model than the \texttt{pexmon} model. That allows us to change the ionization state. In the fits,
  the normalization (an alternative parameter of the solid angle) and the ionization parameter
  ($\xi_{\rm X}$) for X-ray (1--1000 Ry energy) emission (see Equation~(10) in \cite{Gar13} for its
  definition) are left as free parameters. In either region of interest, we obtain the lowest ionization 
  parameter accepted by the model ($\xi_{\rm X} = 1$). This favors $\xi_{\rm X} \sim$ 0.2, 
  estimated by assuming gas at 60 pc distance from the nucleus being irradiated by X-ray emission with
  $\approx 3\times10^{42}$ \ergs through an medium with $N_{\rm H} = 10^{23.9}$ cm$^{-1}$ (i.e.,
  optical depth = 1 for X-rays at the K-edge energy of neutral iron). Thus, we suggest that the gas in the
\rega and \regb regions is irradiated by X-ray emission from the central AGN.

\begin{figure*}
 \includegraphics[scale=0.22]{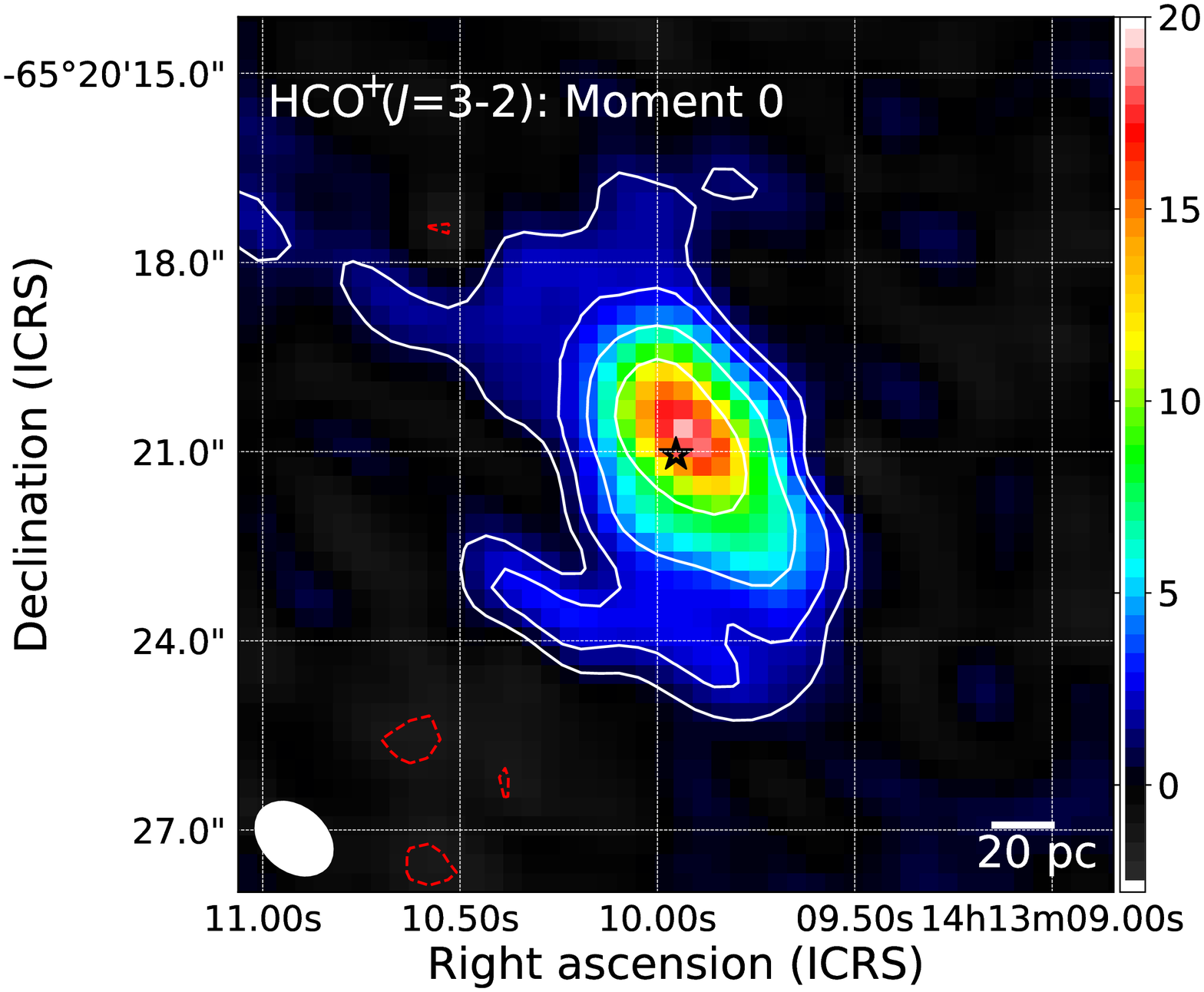}
 \includegraphics[scale=0.22]{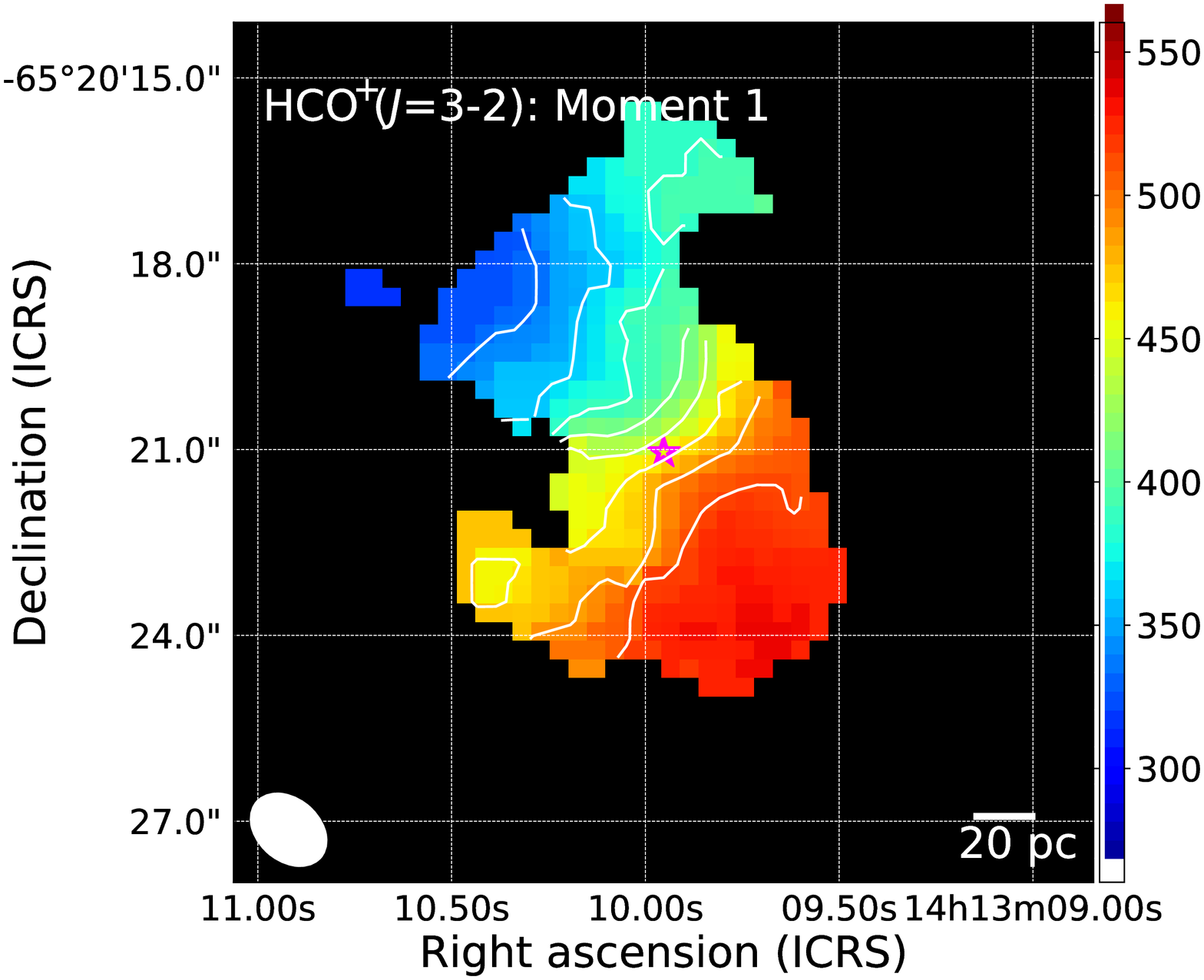} 
 \includegraphics[scale=0.22]{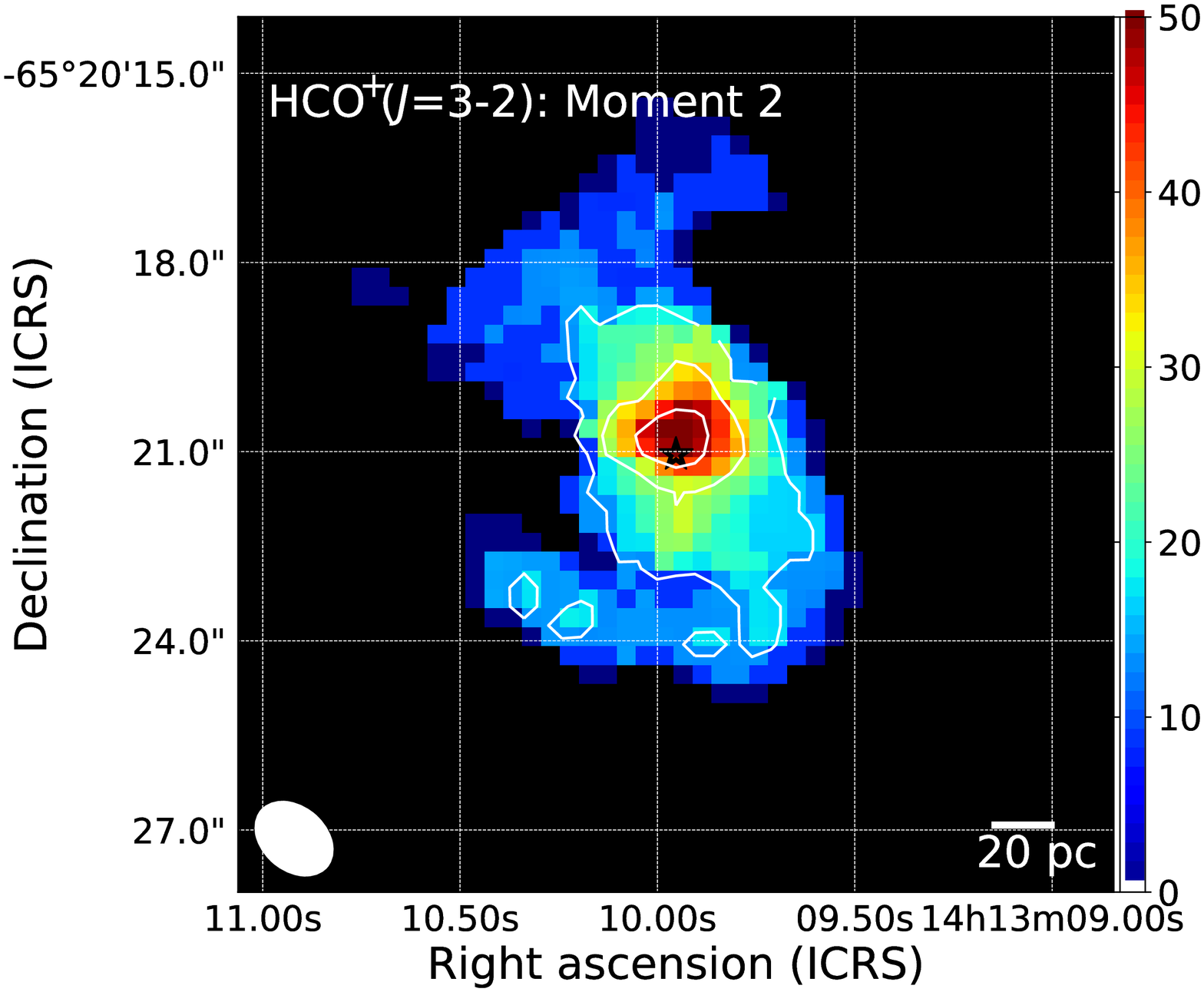} \\ 
 \includegraphics[scale=0.22]{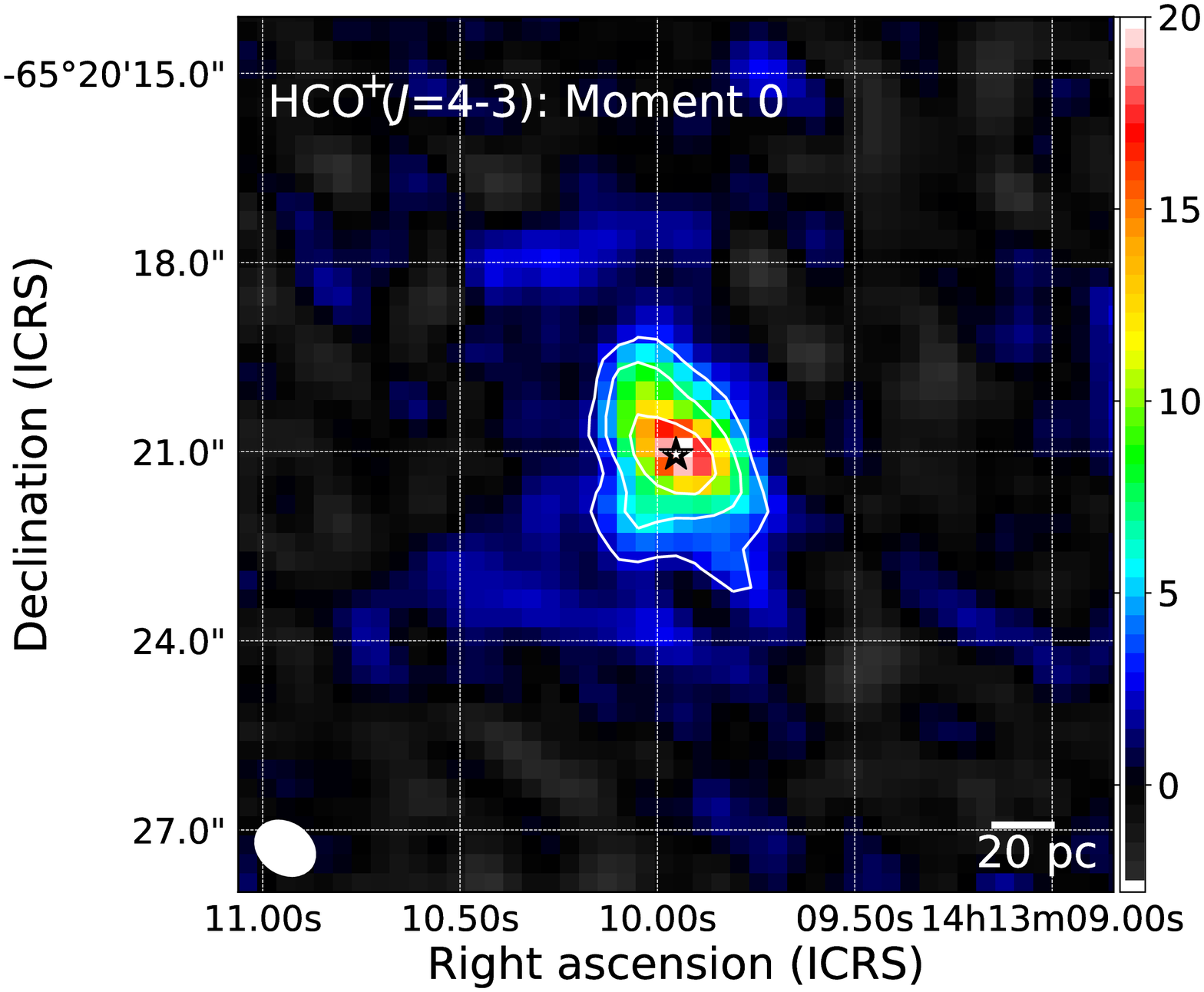}
 \includegraphics[scale=0.22]{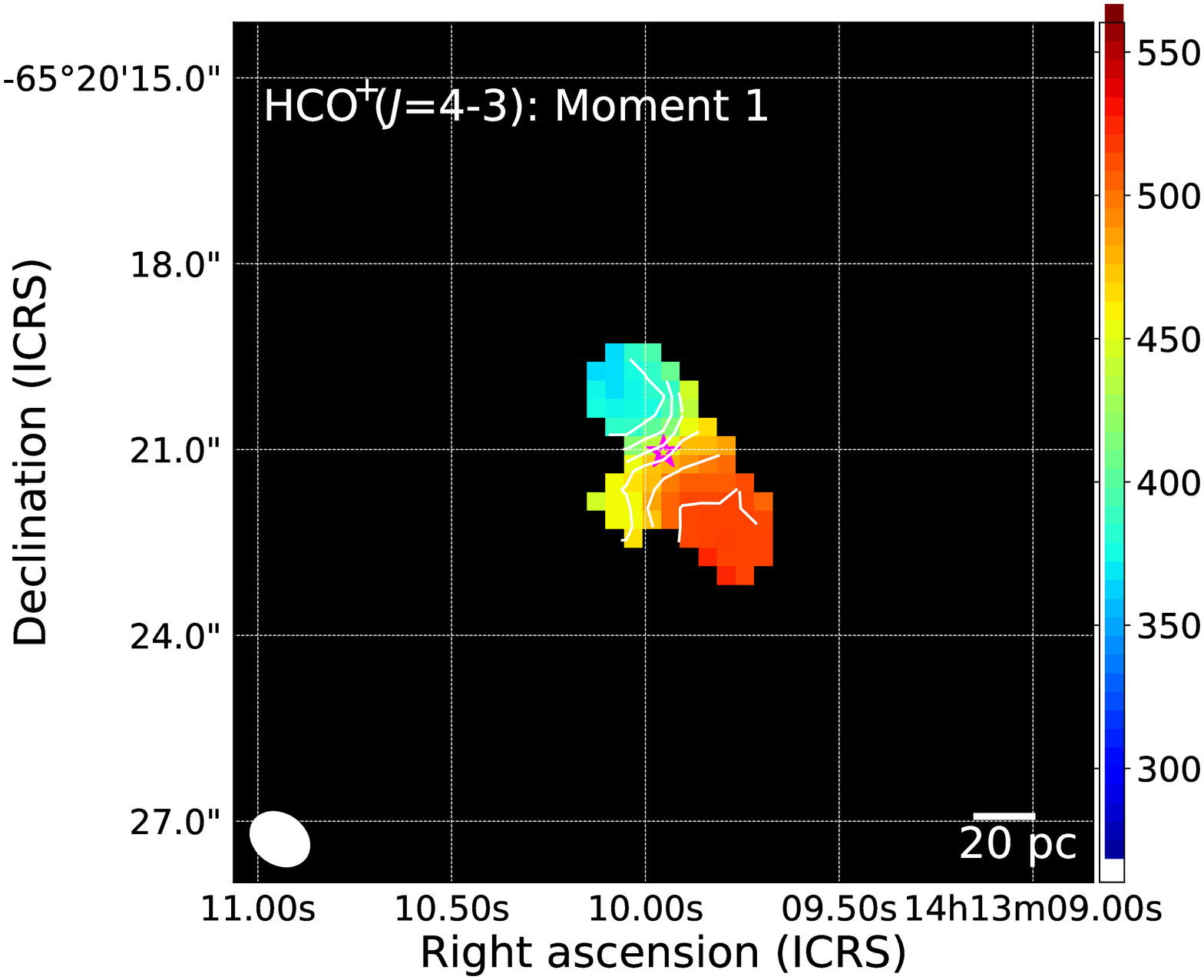} 
 \includegraphics[scale=0.22]{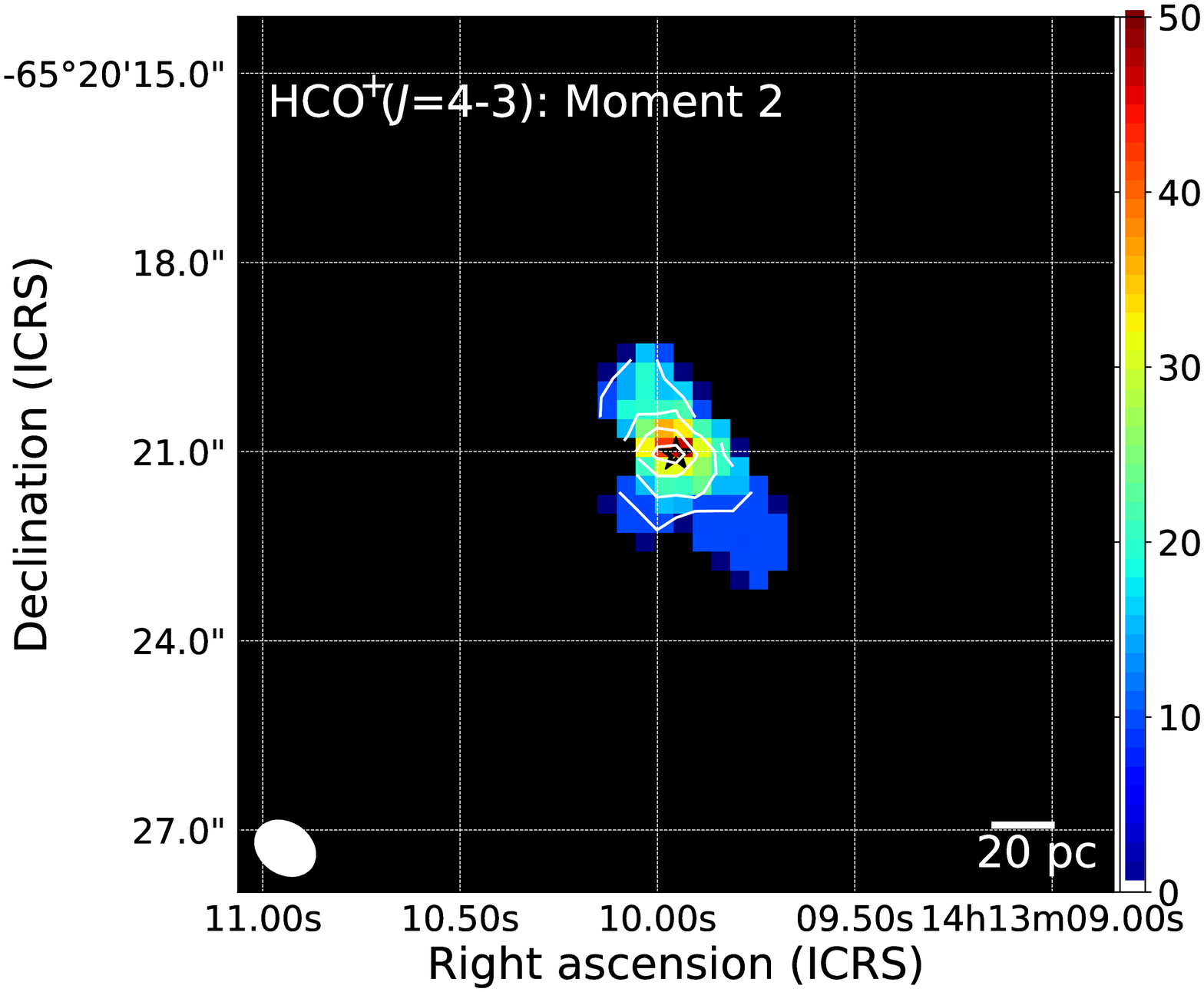} \\ 
 \includegraphics[scale=0.22]{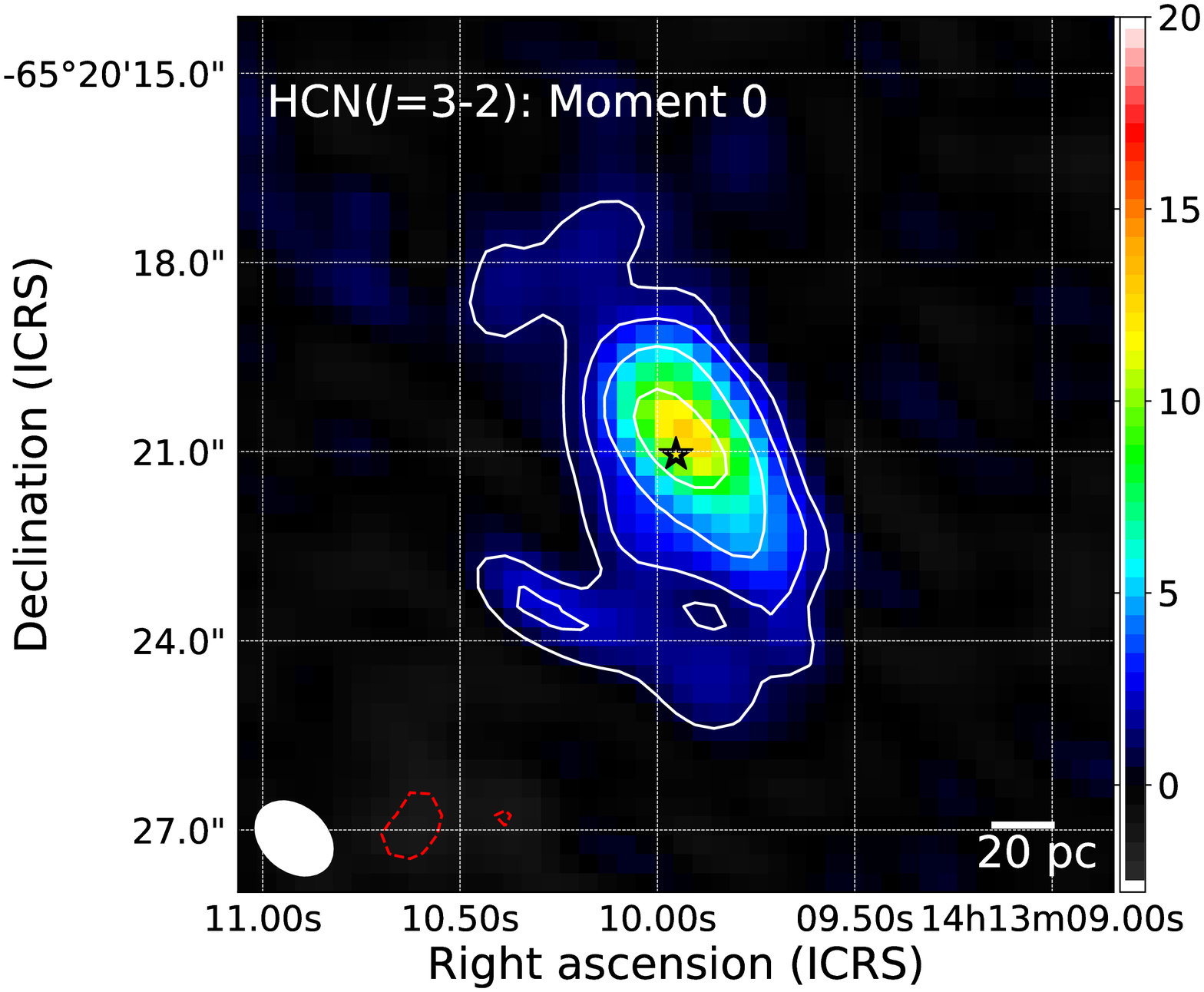} 
 \includegraphics[scale=0.22]{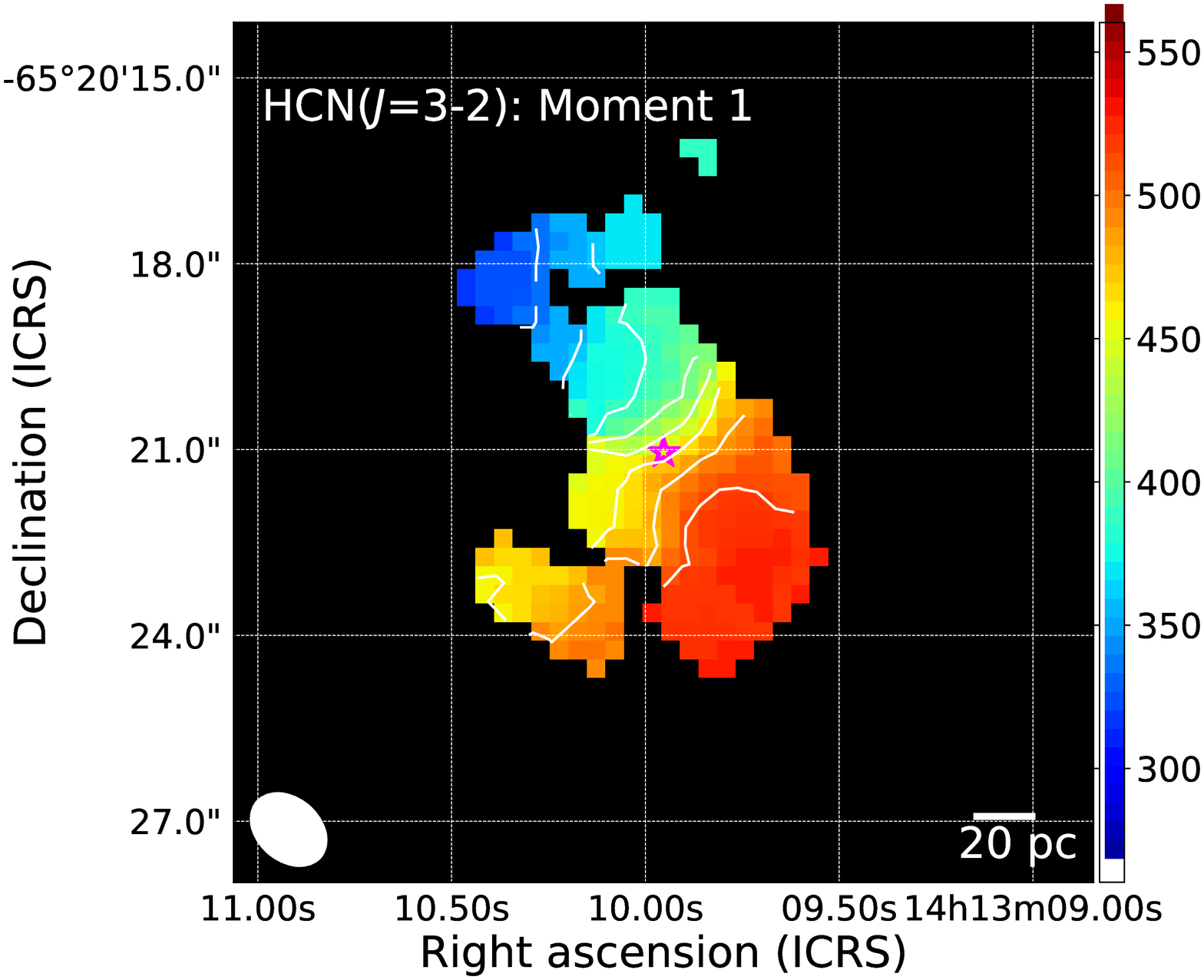} 
 \includegraphics[scale=0.22]{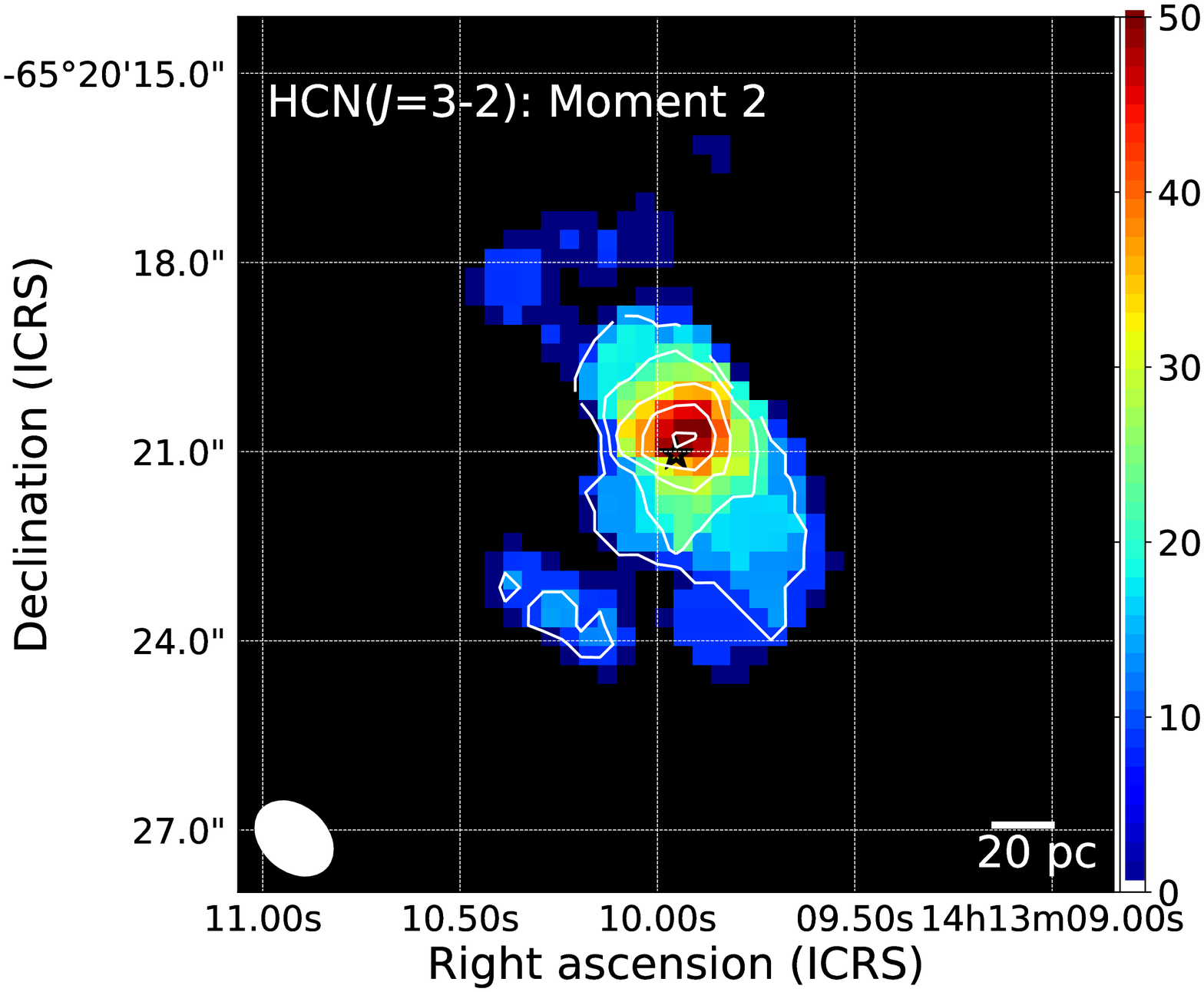} \\ 
 \includegraphics[scale=0.22]{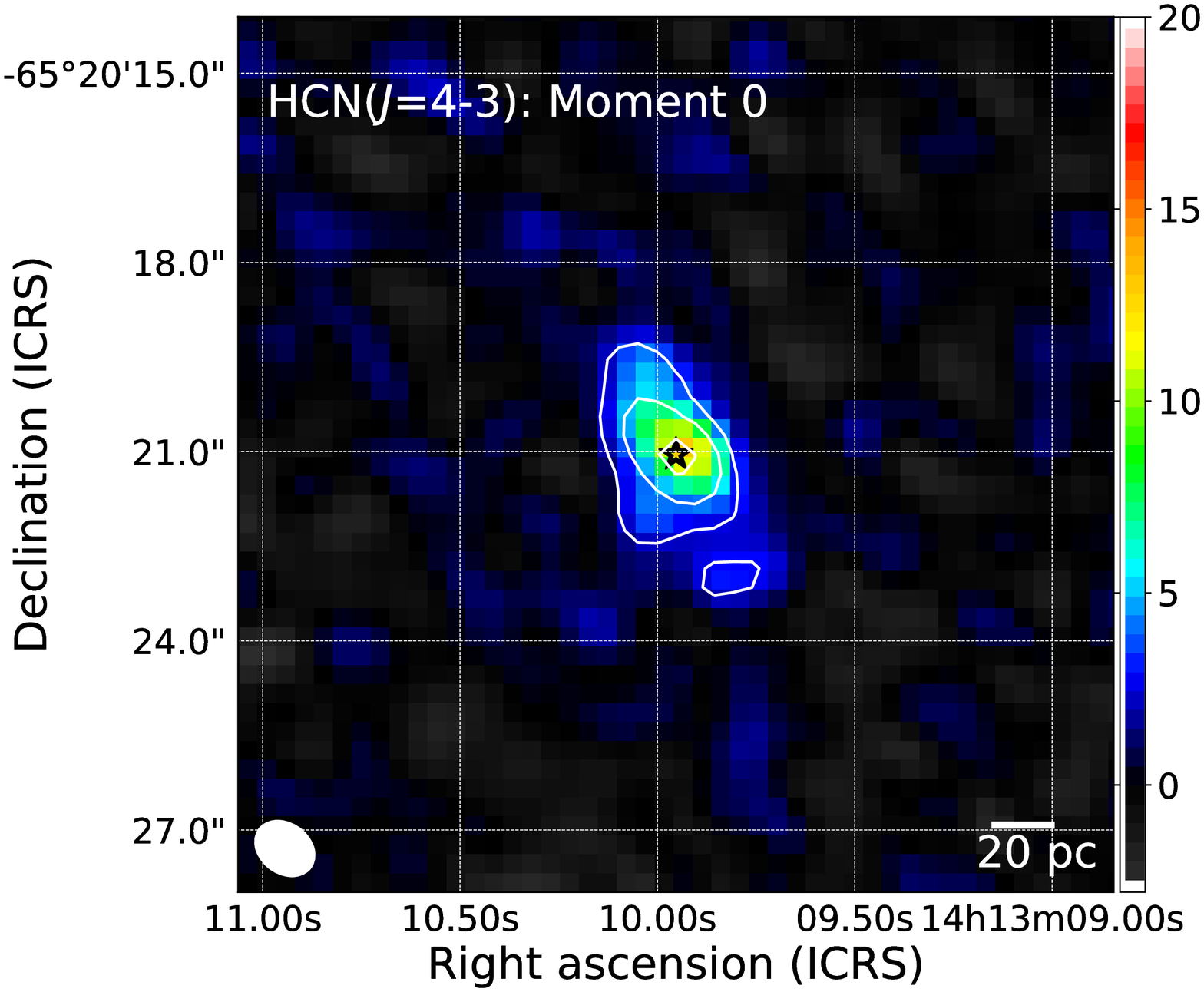}
 \includegraphics[scale=0.22]{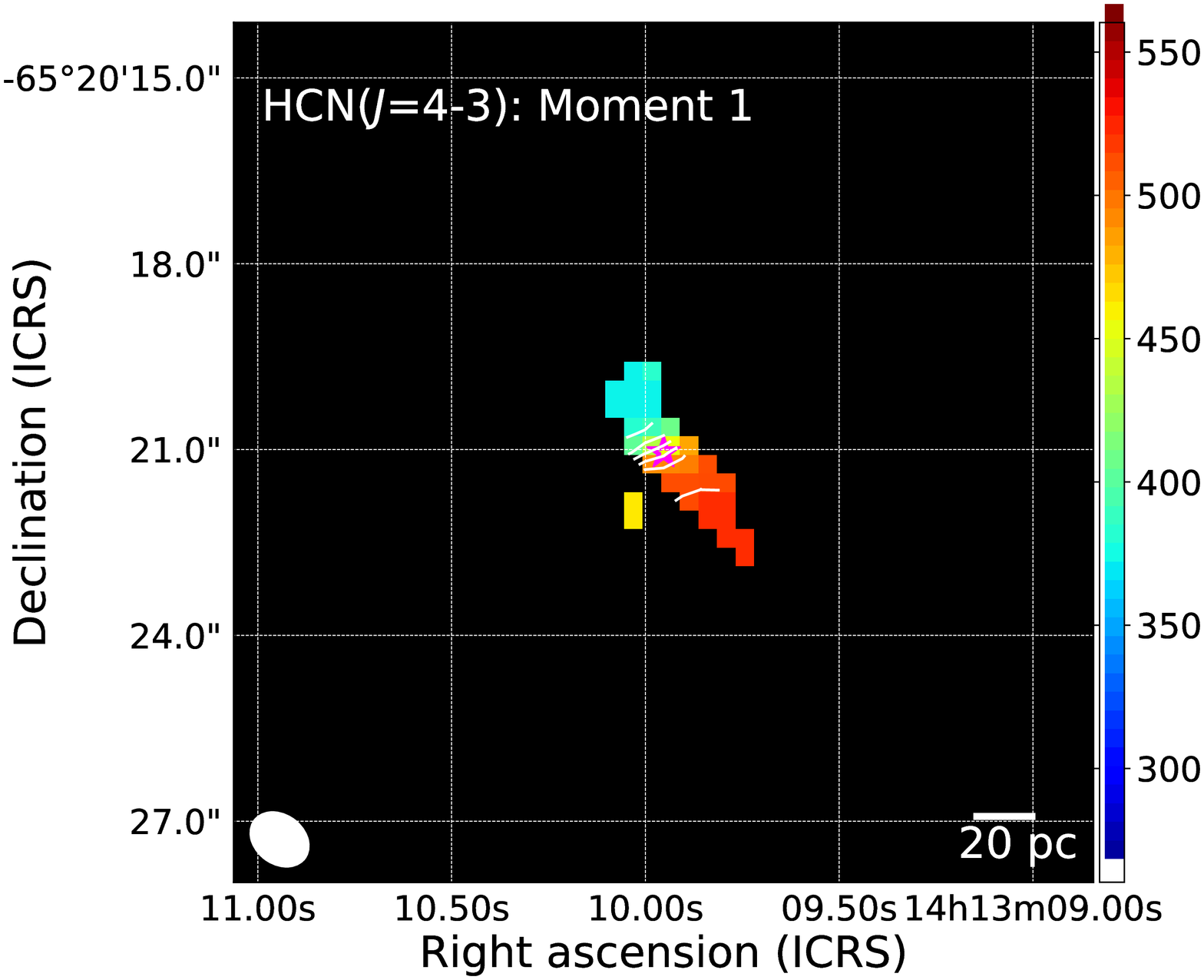} 
 \includegraphics[scale=0.22]{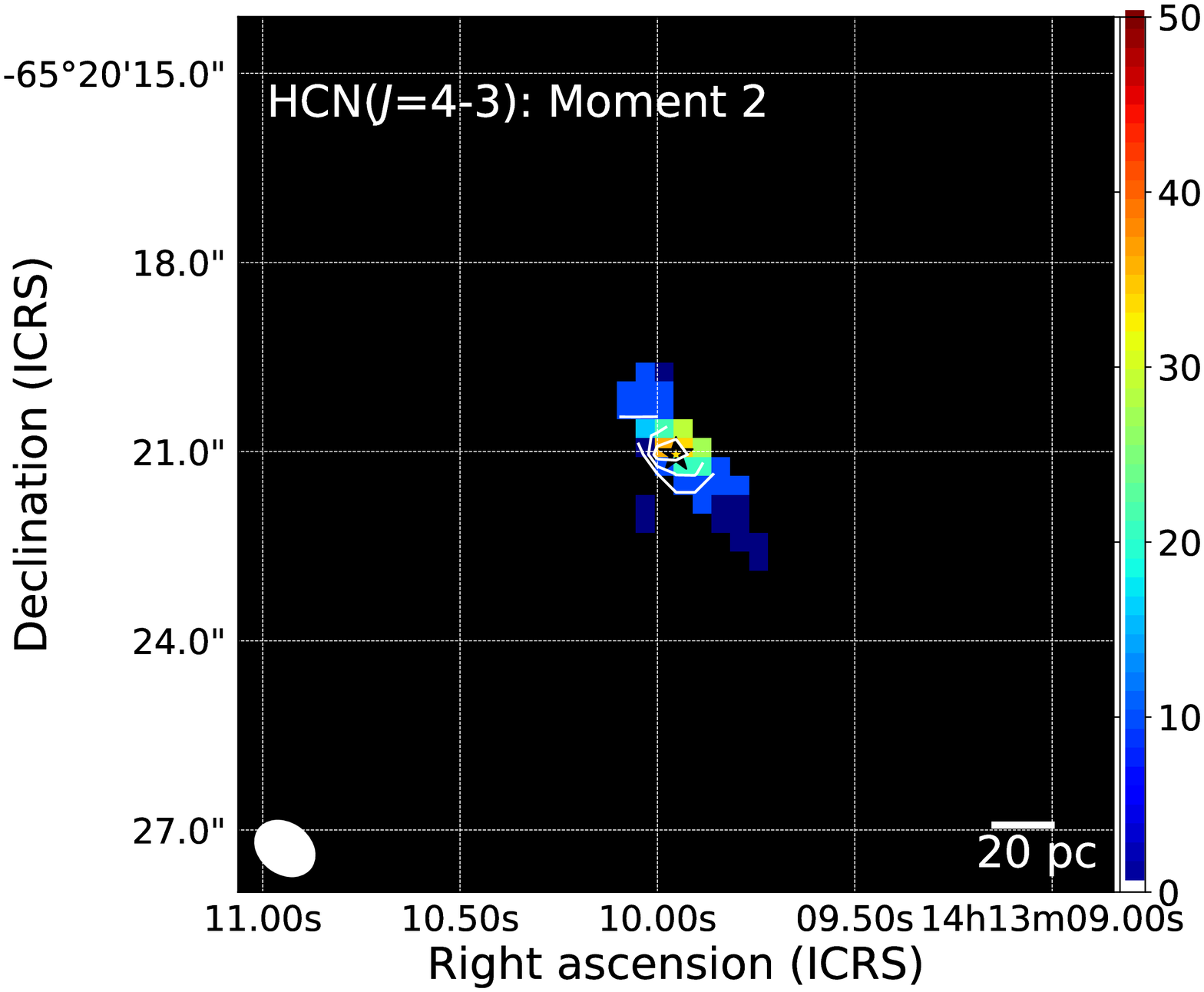}
 \vspace{1.cm}
 \caption{\small{
 (Left) Velocity-integrated intensity maps created using channels from VLSR = 200 to 650 km s$^{-1}$. 
 The contour levels are 5$\sigma$, 10$\sigma$, 20$\sigma$, 40$\sigma$, and 80$\sigma$,
 where $\sigma$ is
 0.24 Jy beam$^{-1}$ km s$^{-1}$,
 0.61 Jy beam$^{-1}$ km s$^{-1}$,
 0.23 Jy beam$^{-1}$ km s$^{-1}$, and
 0.55 Jy beam$^{-1}$ km s$^{-1}$ for
 HCO$^+$($J$=3--2),
 HCO$^+$($J$=4--3),
 HCN($J$=3--2), and
 HCN($J$=4--3), respectively.
   Negative signals at $-5\sigma$ are also shown in red contours, if present.
 (Middle) Intensity-weighted mean velocity maps. The contours represent the VLSR with
 steps of 25 km s$^{-1}$. 
 (Right) Intensity-weighted velocity dispersion maps with the 
 contours separated by 10 km s$^{-1}$.
 The bottom-left filled ellipses represent beam sizes of 
 1.37$\times$1.00 arcsec$^2$ with PA = 49.2 degrees, 
 1.02$\times$0.80 arcsec$^2$ with PA = 54.9 degrees,
 1.37$\times$1.01 arcsec$^2$ with PA = 48.8 degrees, and 
 1.02$\times$0.79 arcsec$^2$ with PA = 52.8 degrees for
 HCO$^+$($J$=3--2),
 HCO$^+$($J$=4--3),
 HCN($J$=3--2), and
 HCN($J$=4--3), respectively.
    The SMBH position is denoted with the black (Moment 0 and 2) or magenta (Moment 1) star.
 }
 }\label{fig:mol_lin_map_all}
\end{figure*}
\addtocounter{figure}{-1}
\begin{figure*}
 \includegraphics[scale=0.22]{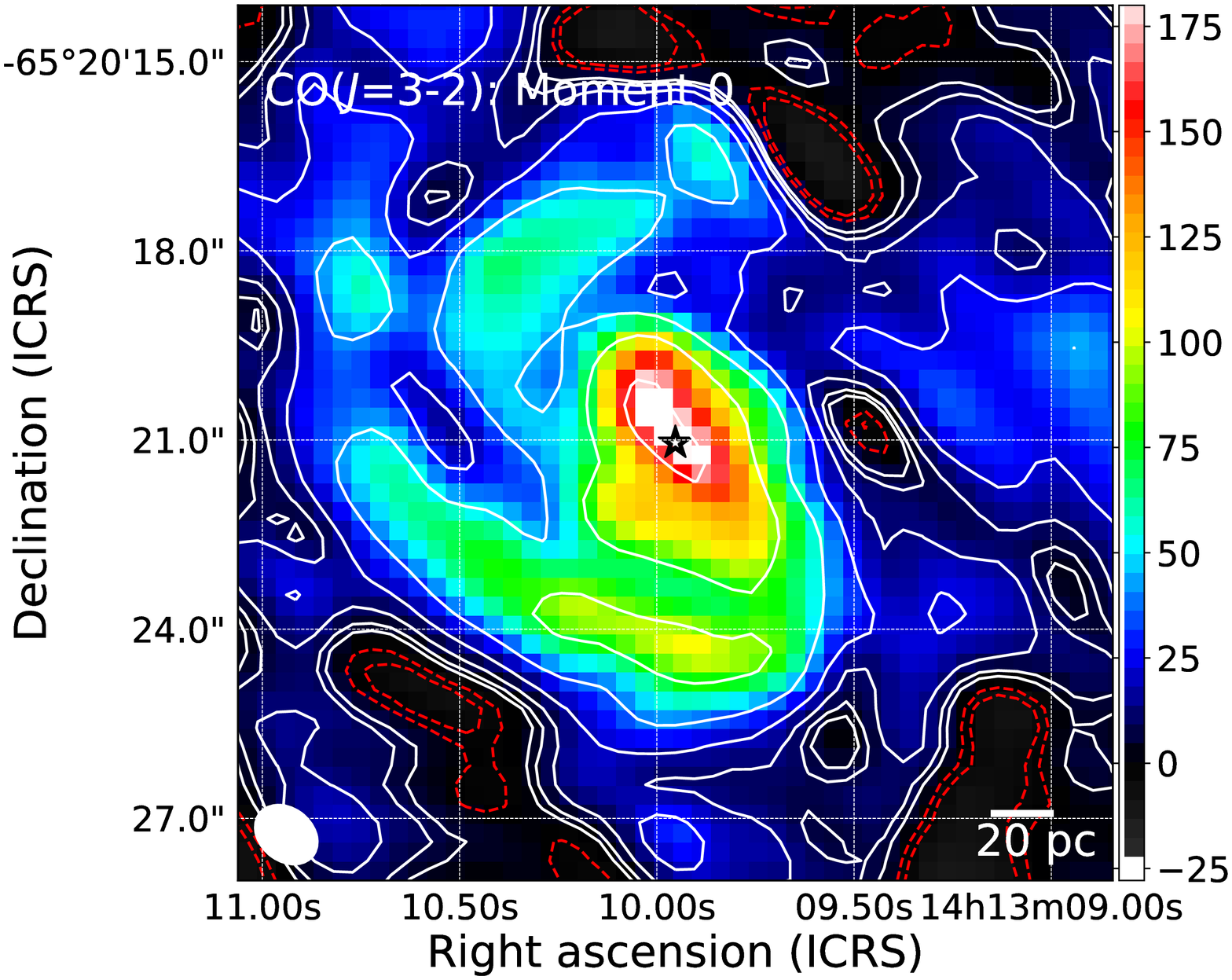}
 \includegraphics[scale=0.22]{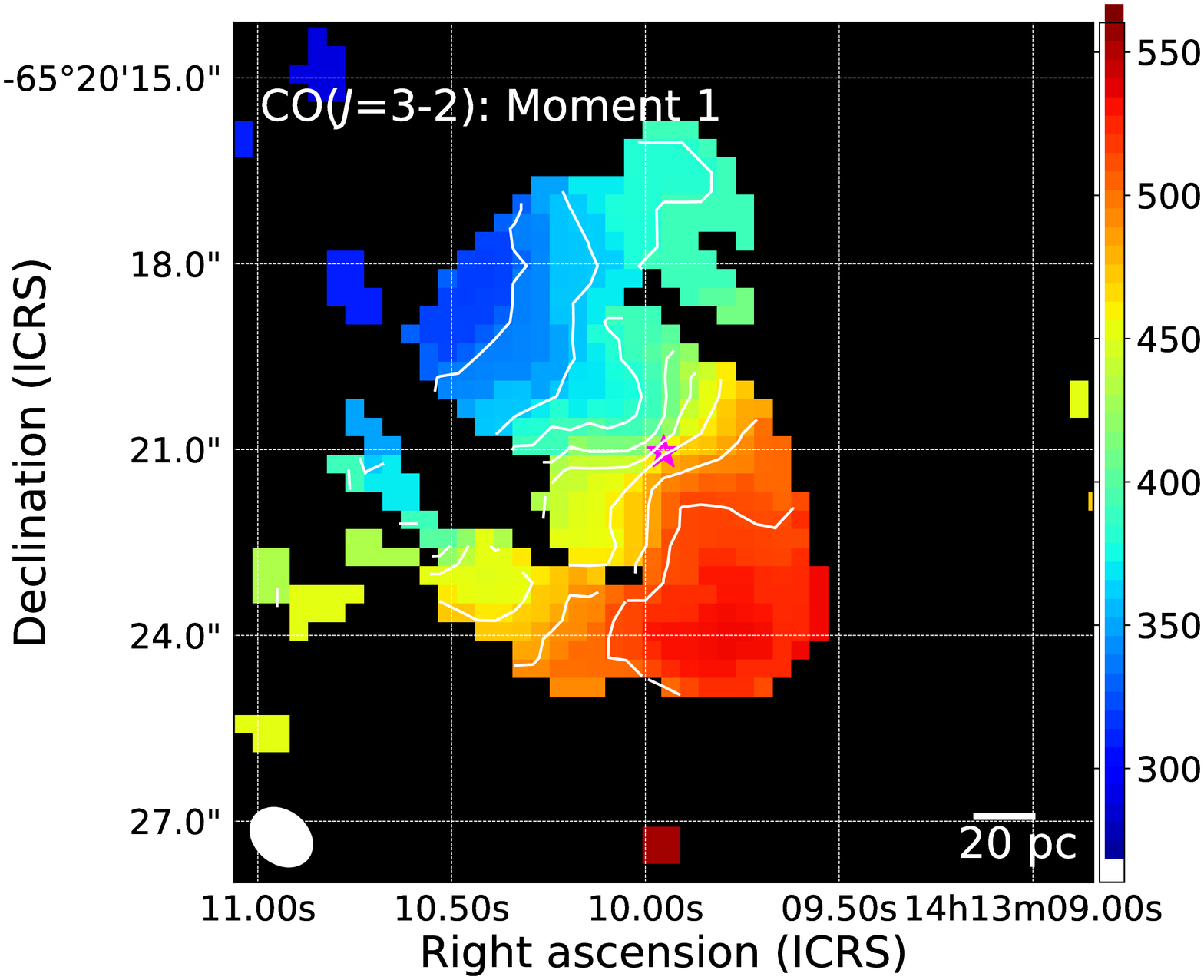} 
 \includegraphics[scale=0.22]{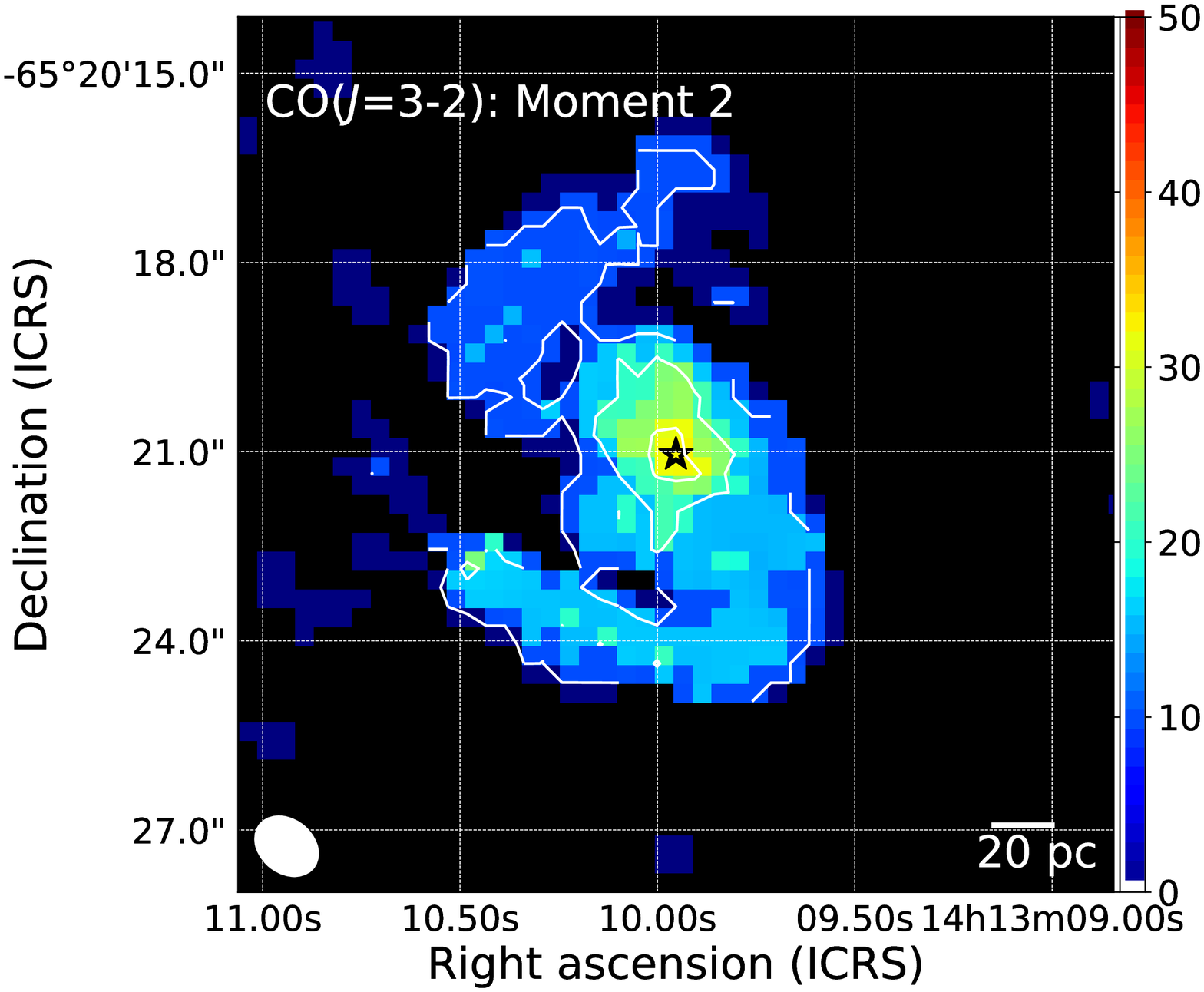}
 \vspace{1.cm}
 \caption{\small{
     (Left) Velocity-integrated intensity map of CO($J$=3--2) created using channels from VLSR = 200 to 650
     km s$^{-1}$. The white contour levels are 5$\sigma$, 10$\sigma$, 20$\sigma$, 40$\sigma$, 80$\sigma$,
     160$\sigma$, and 320$\sigma$, where $\sigma$ is 0.521 Jy beam$^{-1}$ km s$^{-1}$. 
     Negative signals at $-10\sigma$ and $-5\sigma$ are also shown in red contours. 
     (Middle) Intensity-weighted mean velocity map. 
     The contours represent the VLSR with steps of 25 km s$^{-1}$. (Right) Intensity-weighted 
     velocity dispersion map with the contours separated by 10 km s$^{-1}$. 
     The bottom-left filled ellipses represent a beam size of 1.09$\times$0.84 arcsec$^2$ with PA = 50.8 degrees.
     The SMBH position is denoted with the black (Moment 0 and 2) or magenta (Moment 1) star.
 }
 }\label{fig:mol_lin_map_all}
\end{figure*}

\begin{figure*}
 \includegraphics[scale=0.22]{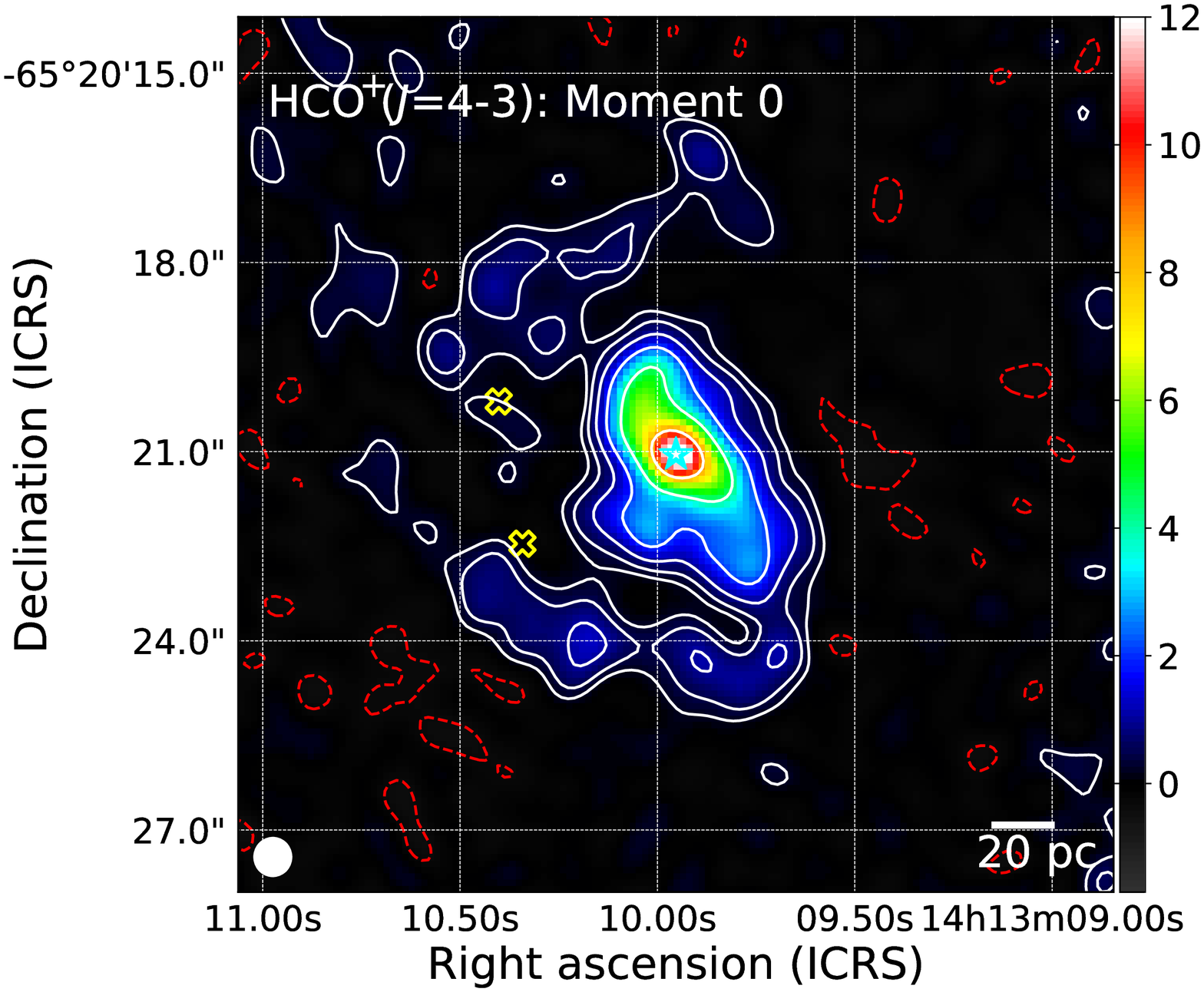}
 \includegraphics[scale=0.22]{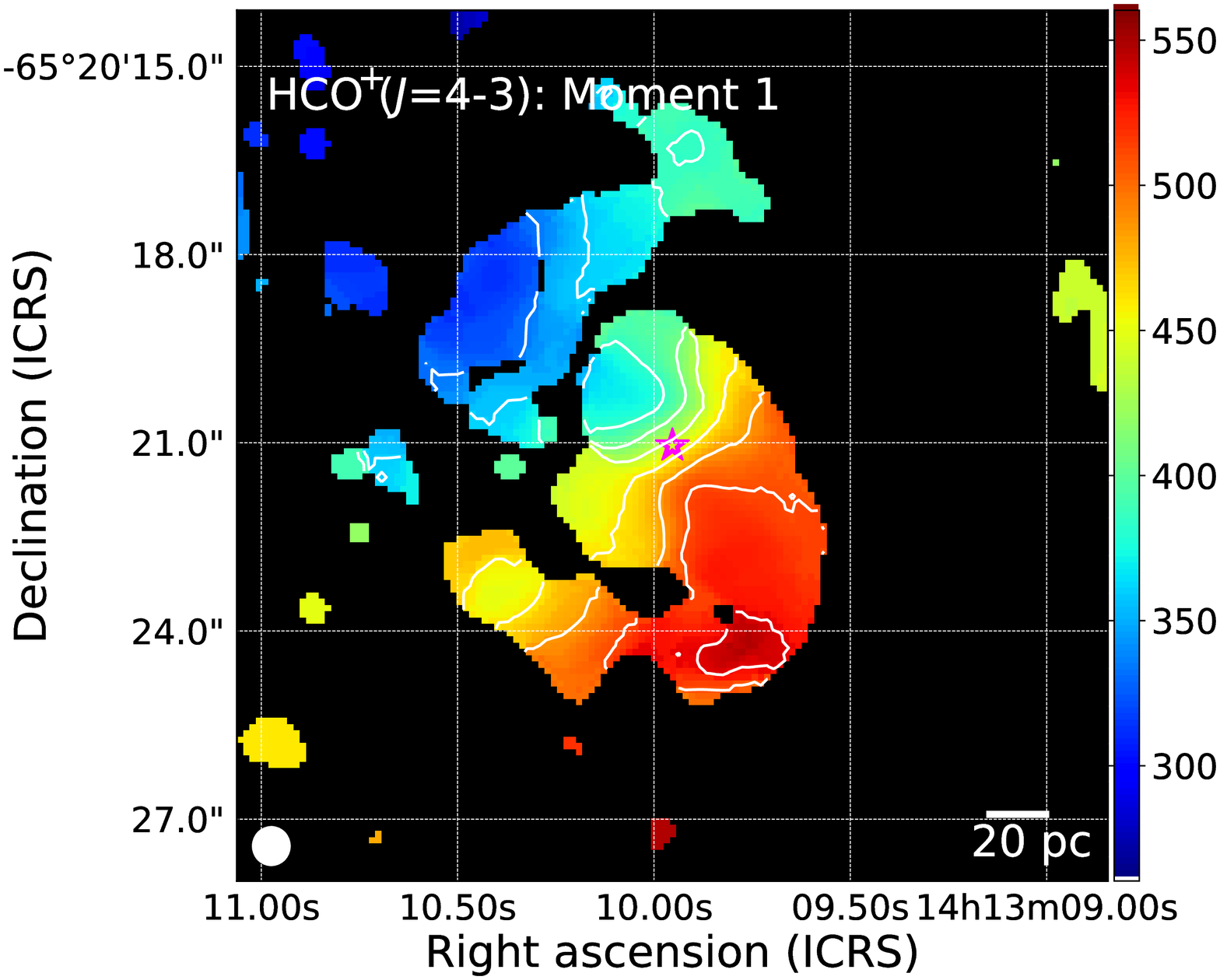} 
 \includegraphics[scale=0.22]{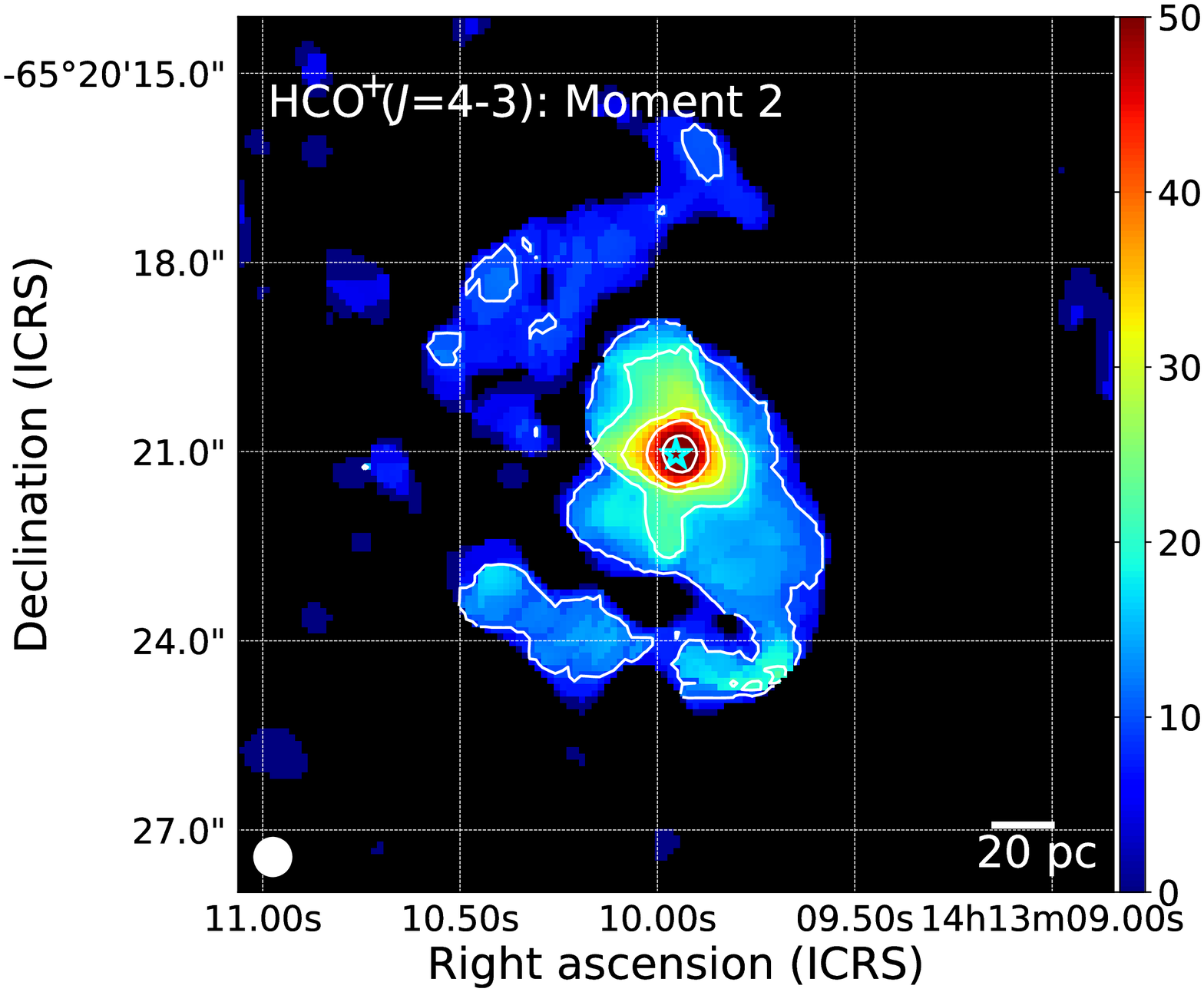} 
 \vspace{1.cm}
 \caption{\small{
 (Left) Velocity-integrated intensity map of HCO$^+$($J$=4--3) created using channels from VLSR = 150 to 700
 km s$^{-1}$. The contour levels are 5$\sigma$, 10$\sigma$, 20$\sigma$, 40$\sigma$, 80$\sigma$, and 160$\sigma$, 
 where $\sigma$ is 0.054 Jy beam$^{-1}$ km s$^{-1}$.
 Negative signals at $-5\sigma$ are also shown in red contours. 
 The spectra in Figure~\ref{fig:mol_spec} were extracted 
 using the synthesized beams centered at the yellow crosses, except for the Nucleus, the spectra of which are taken from the 
 center (the cyan star). 
 (Middle) Intensity-weighted mean velocity map.
 The contours represent the VLSR with steps of 25 km s$^{-1}$. (Right)
 Intensity-weighted velocity dispersion map with the contours separated by 
 10 km s$^{-1}$. The bottom-left filled ellipses represent a beam size of 0.61$\times$0.59
 arcsec$^2$ with PA = $-$1.25 degrees.
   The SMBH position is denoted with the cyan (Moment 0 and 2) or magenta (Moment 1) star.}
 }\label{fig:hcop43_img}
\end{figure*}

\begin{figure*}
  \begin{center}
  \includegraphics[scale=0.23]{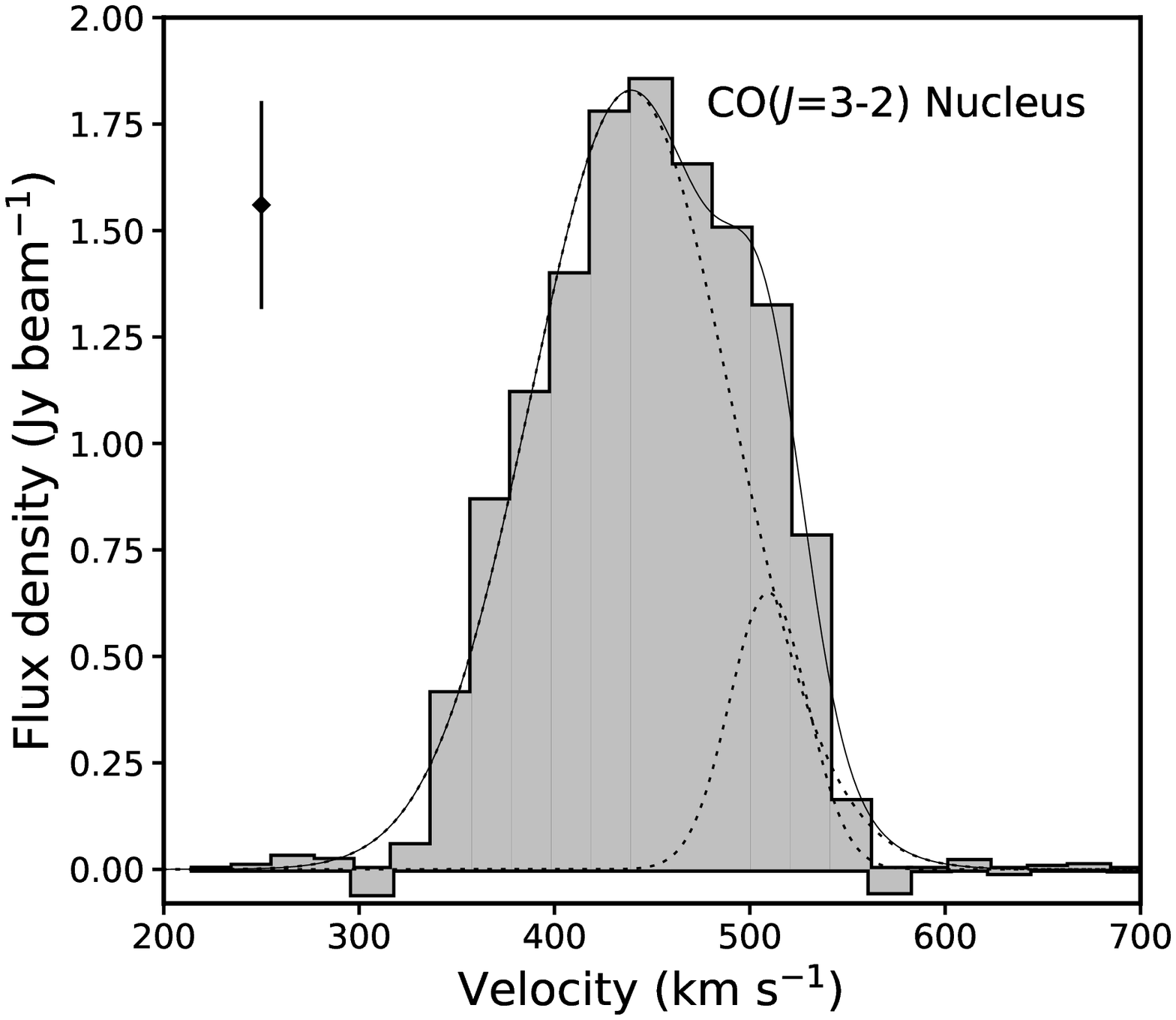}
 \includegraphics[scale=0.23]{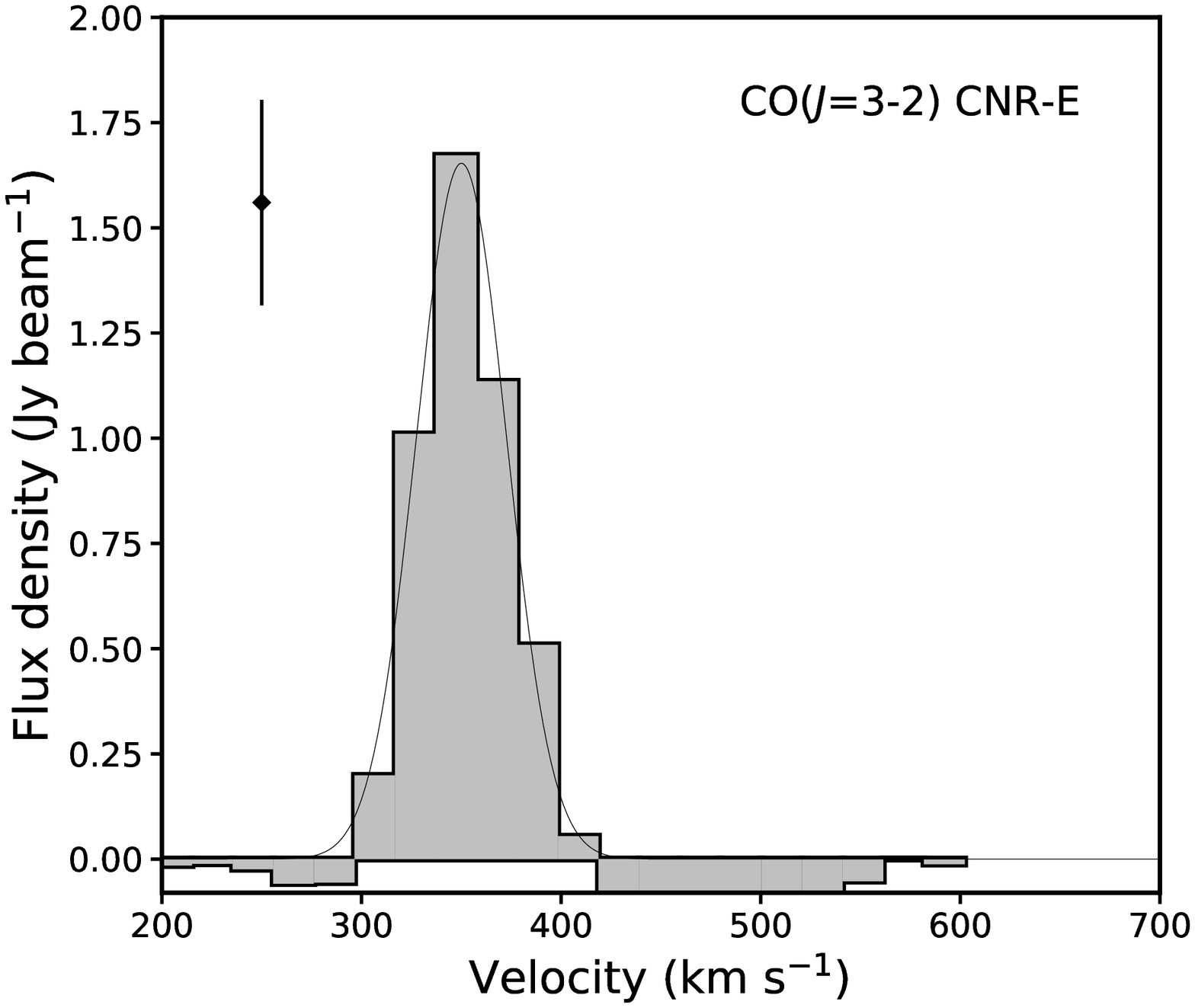}  
 \includegraphics[scale=0.23]{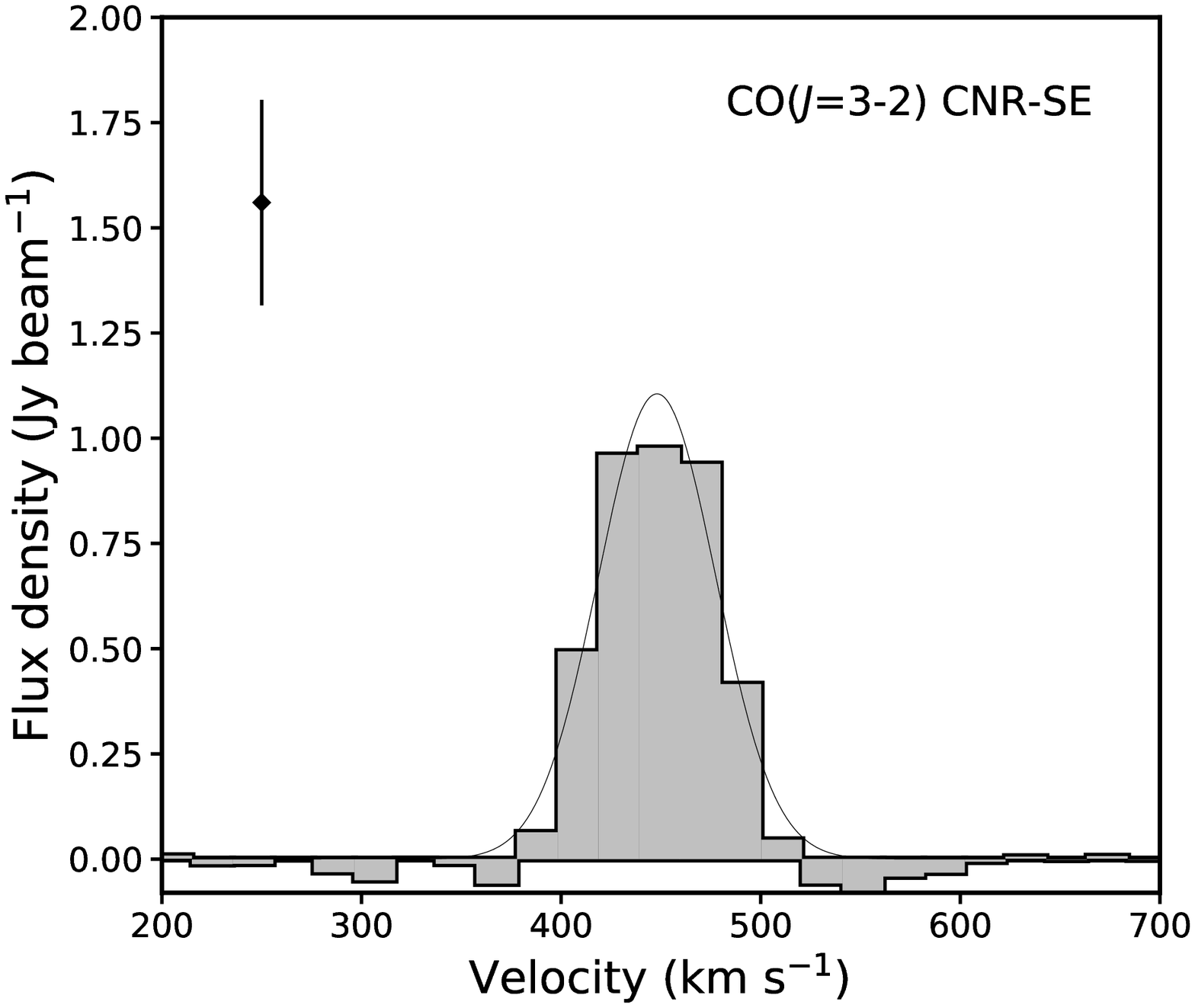} \\ \vspace{-0.5cm}   
 \includegraphics[scale=0.23]{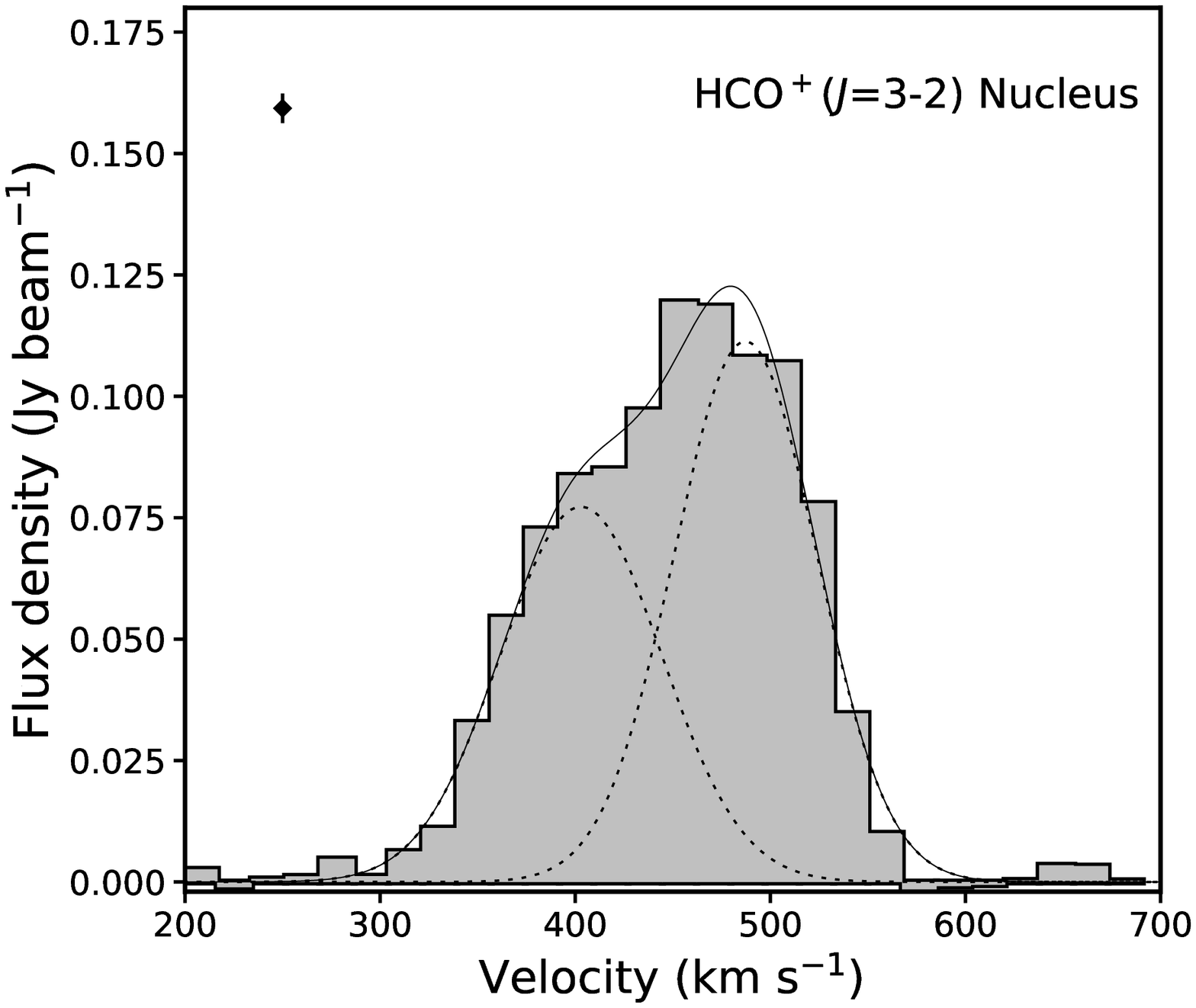} 
 \includegraphics[scale=0.23]{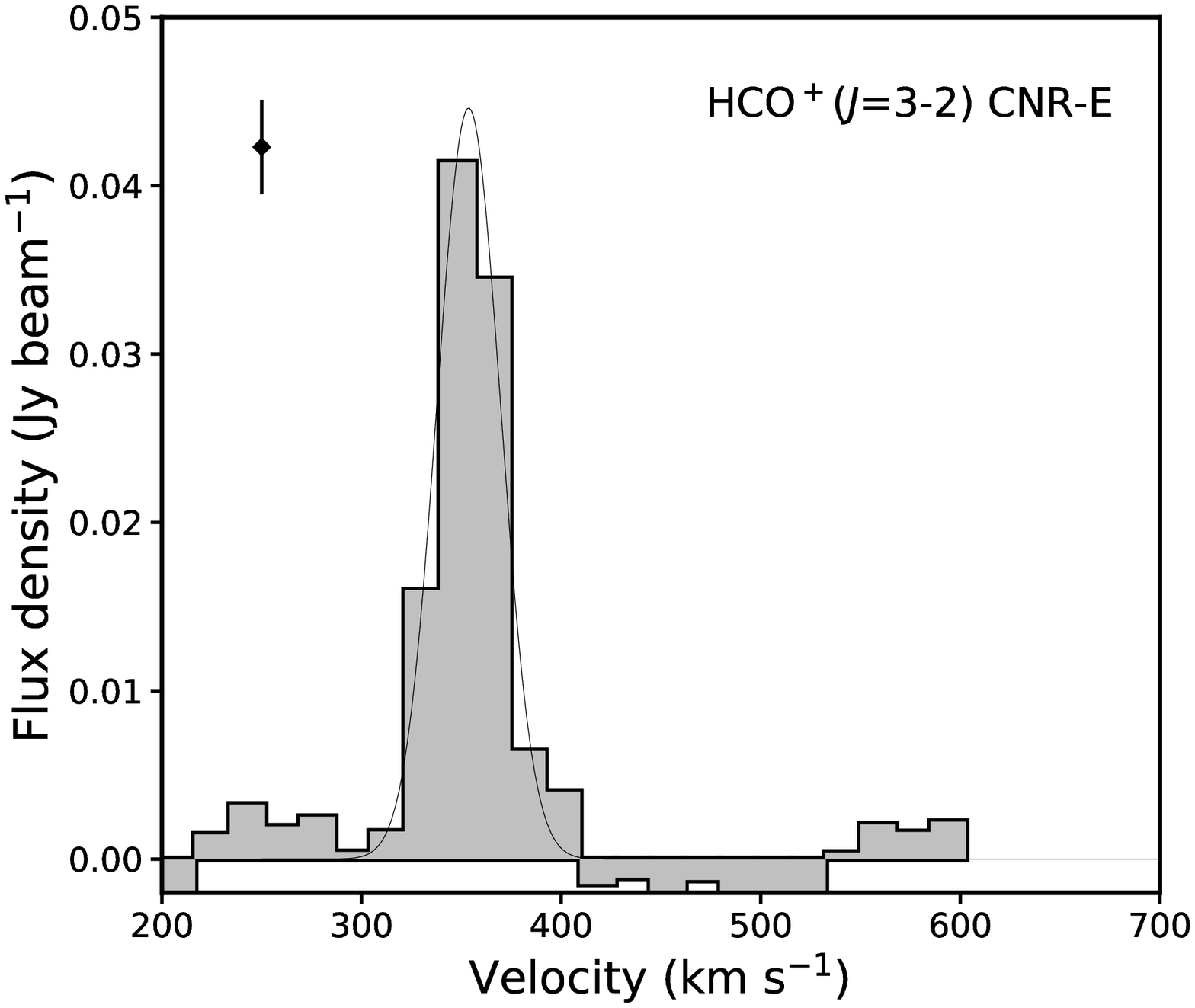}
 \includegraphics[scale=0.23]{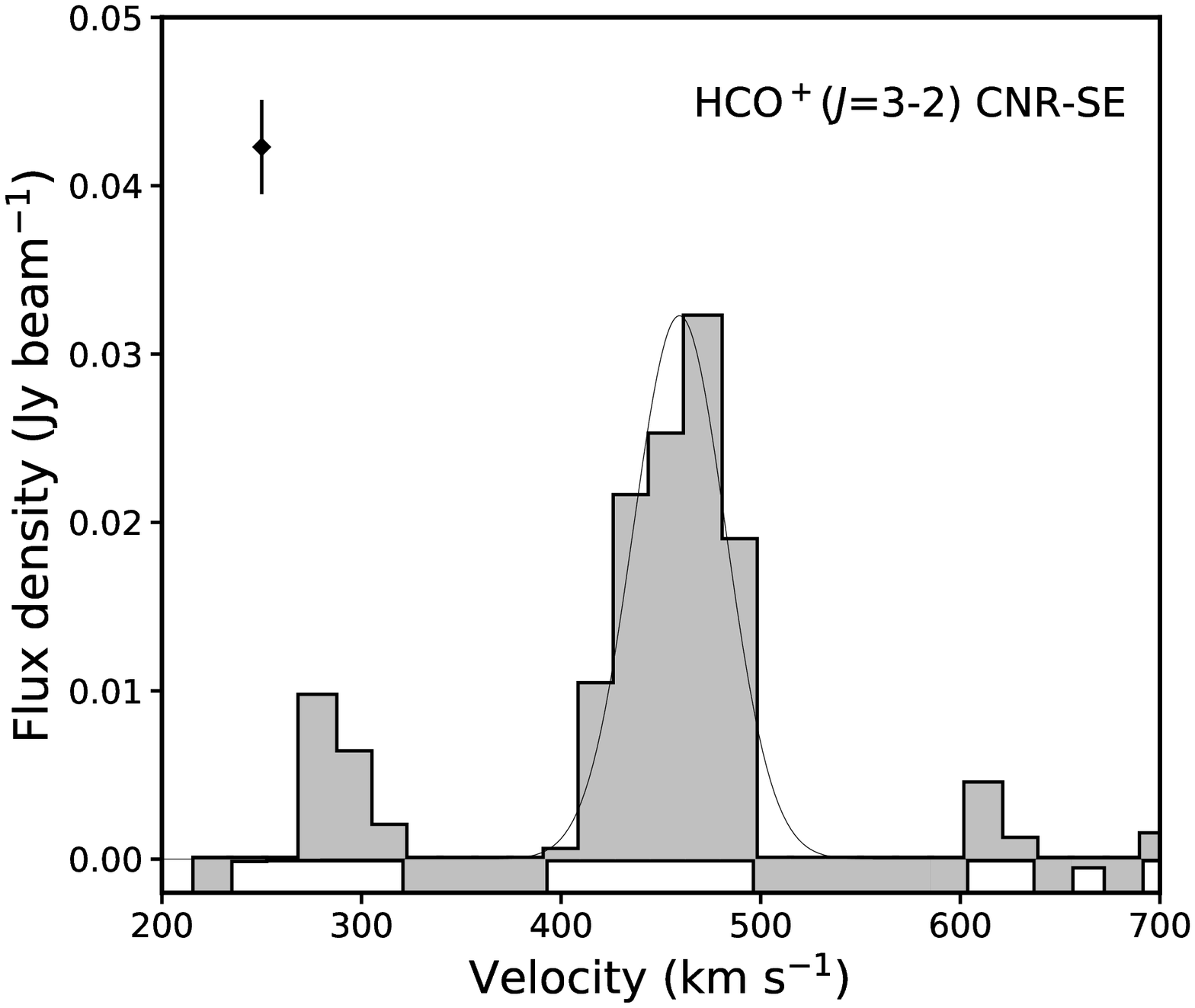} \\ \vspace{-0.5cm}  
 \includegraphics[scale=0.23]{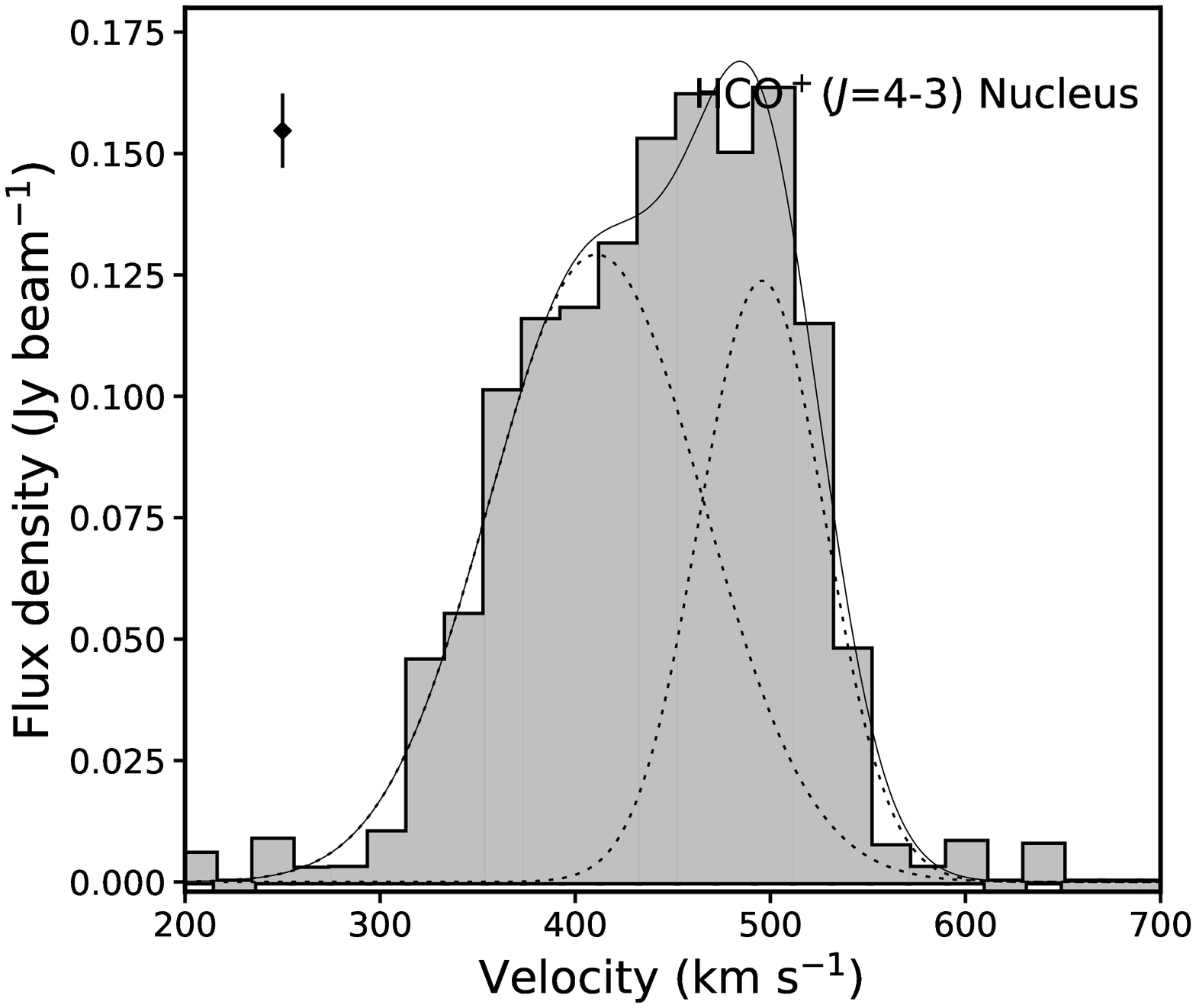} 
 \includegraphics[scale=0.23]{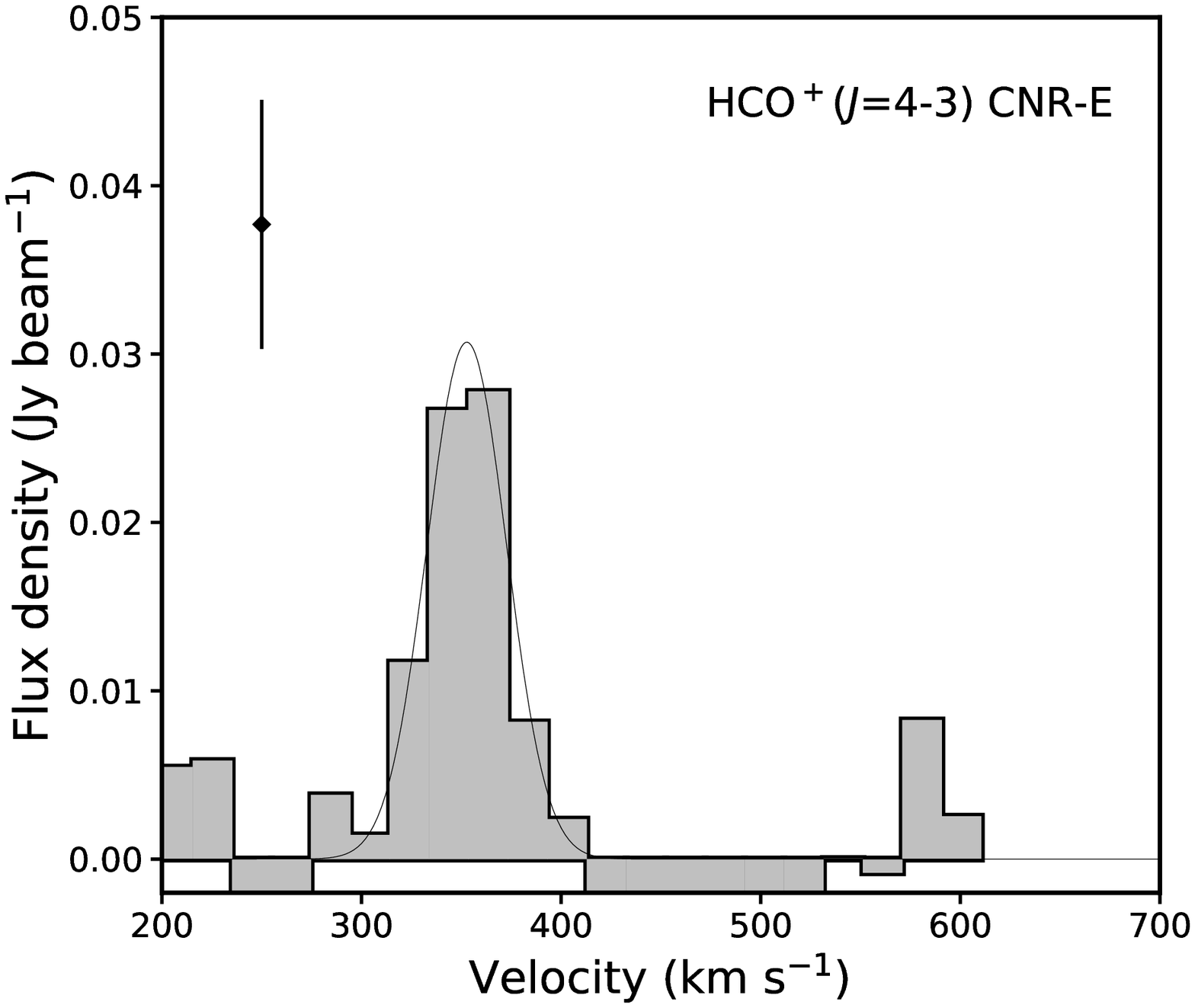}
 \includegraphics[scale=0.23]{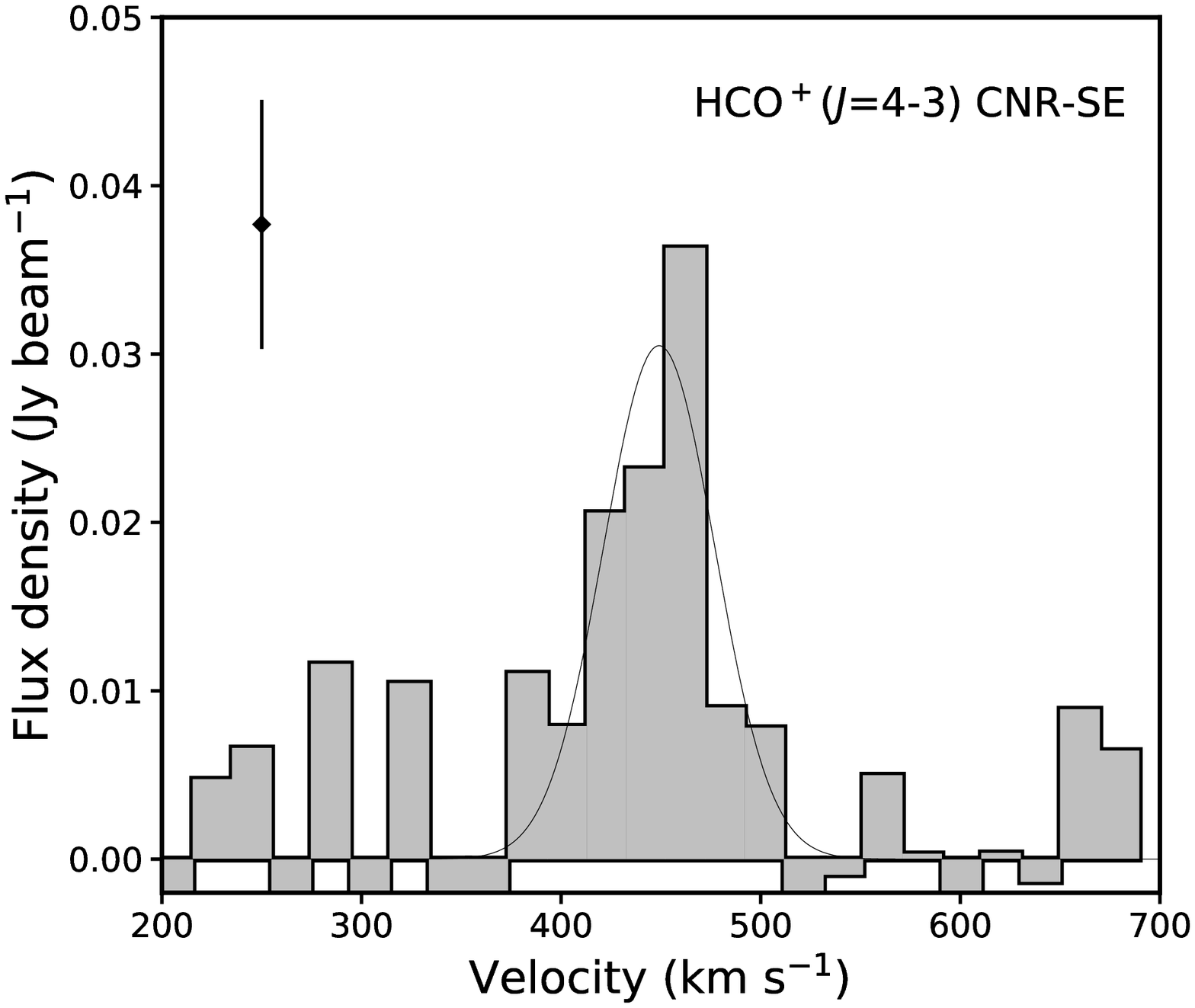} \\ \vspace{-0.5cm}  
 \includegraphics[scale=0.23]{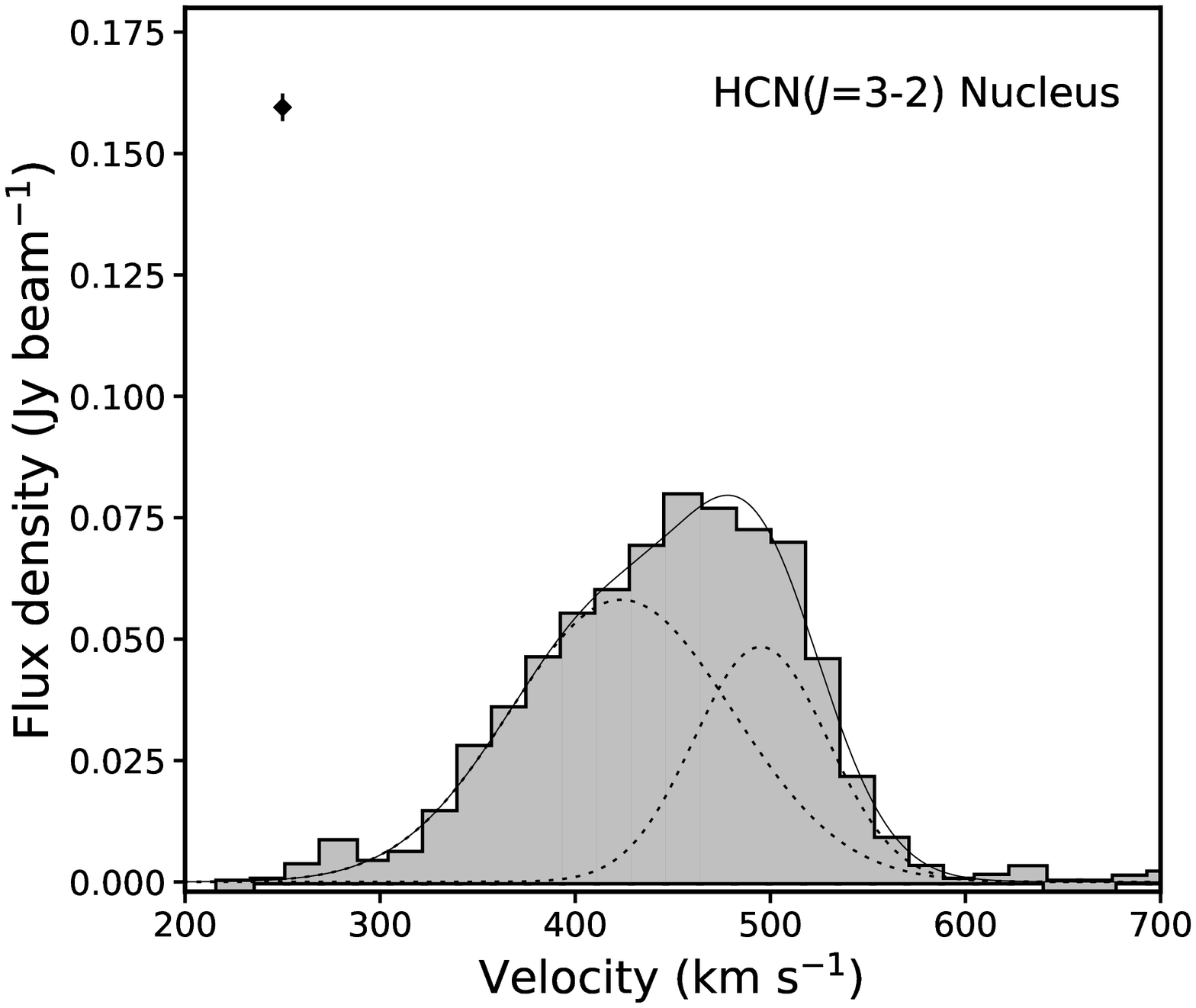} 
 \includegraphics[scale=0.23]{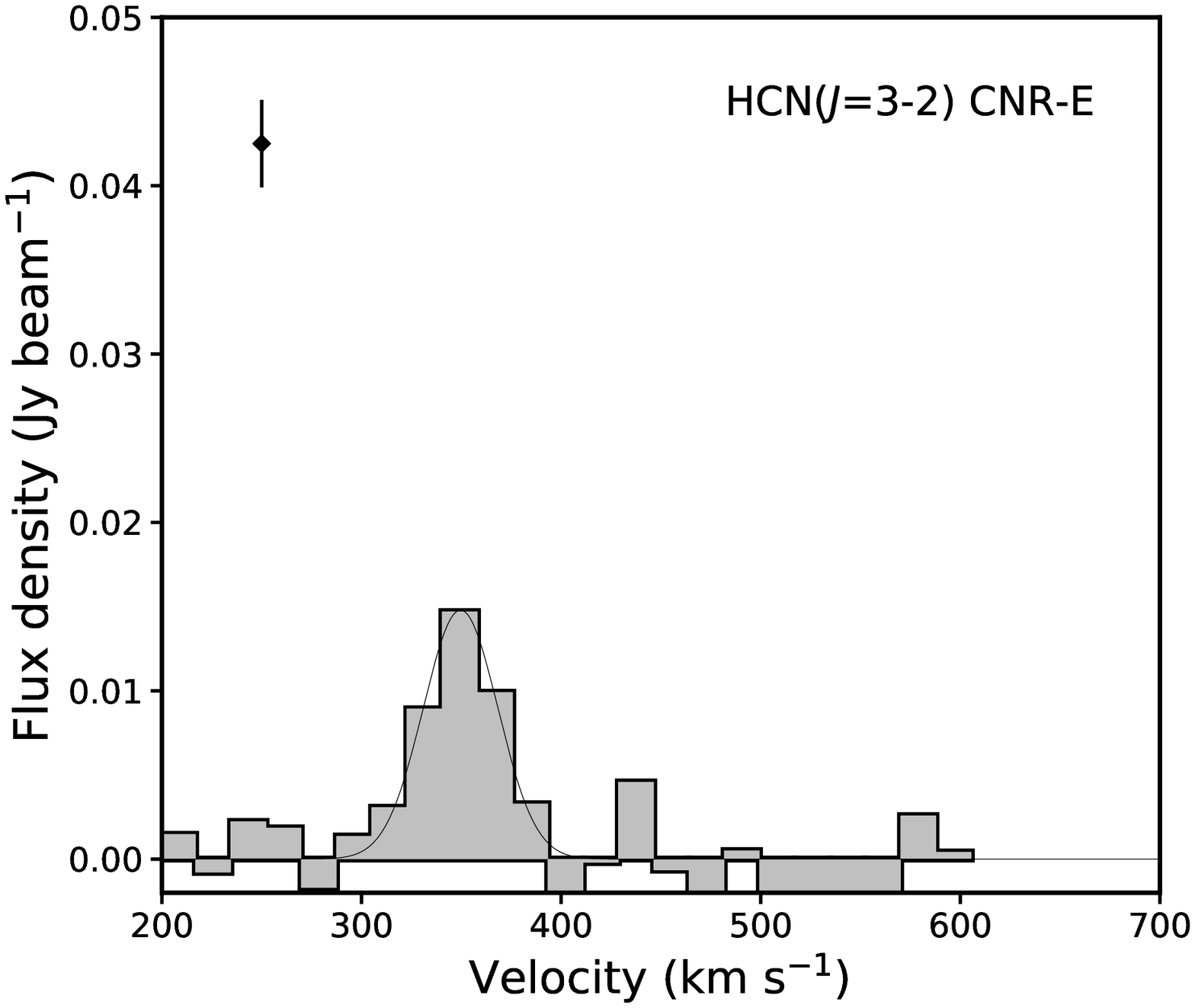}
 \includegraphics[scale=0.23]{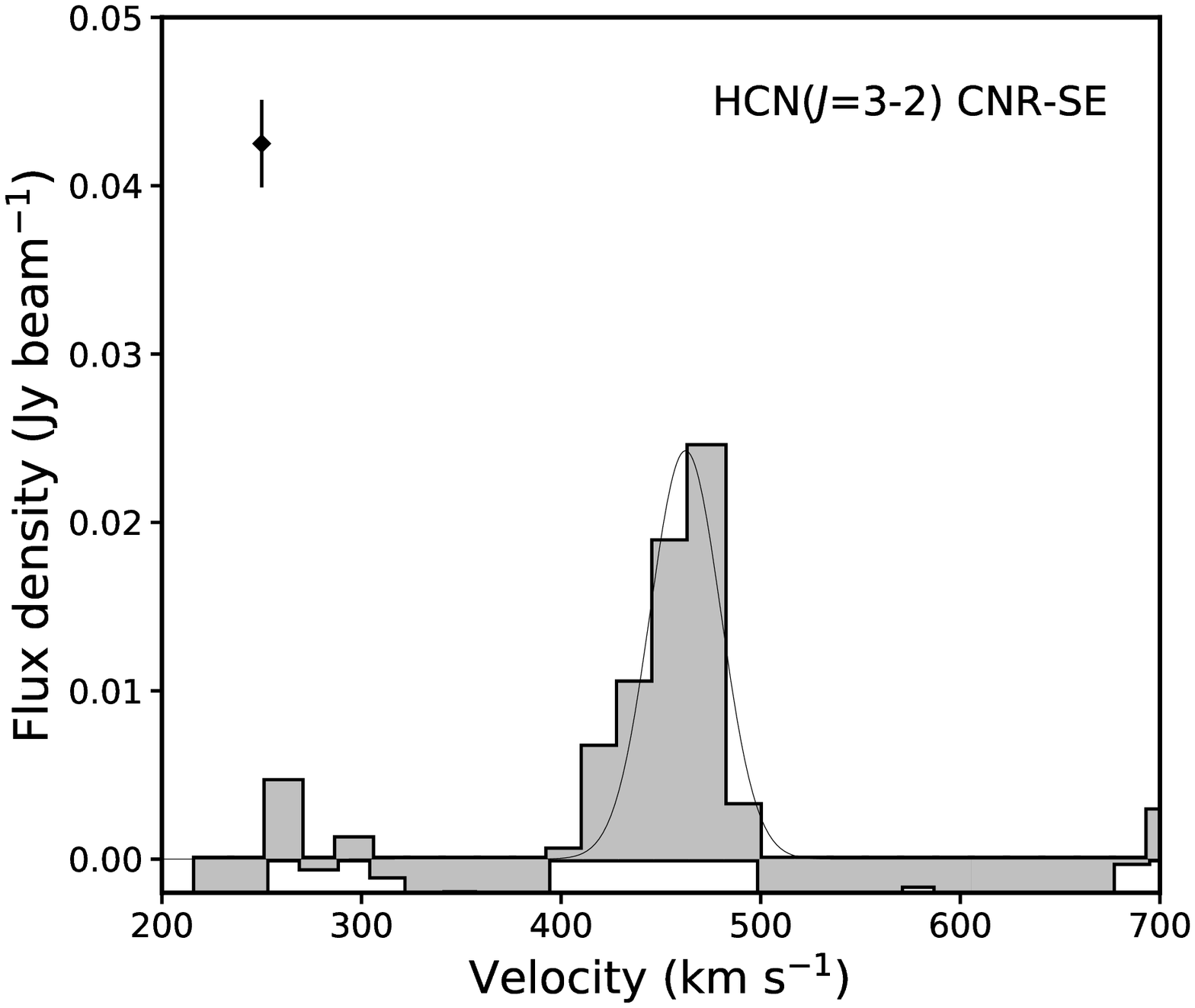} \\ \vspace{-.5cm}  
 \includegraphics[scale=0.23]{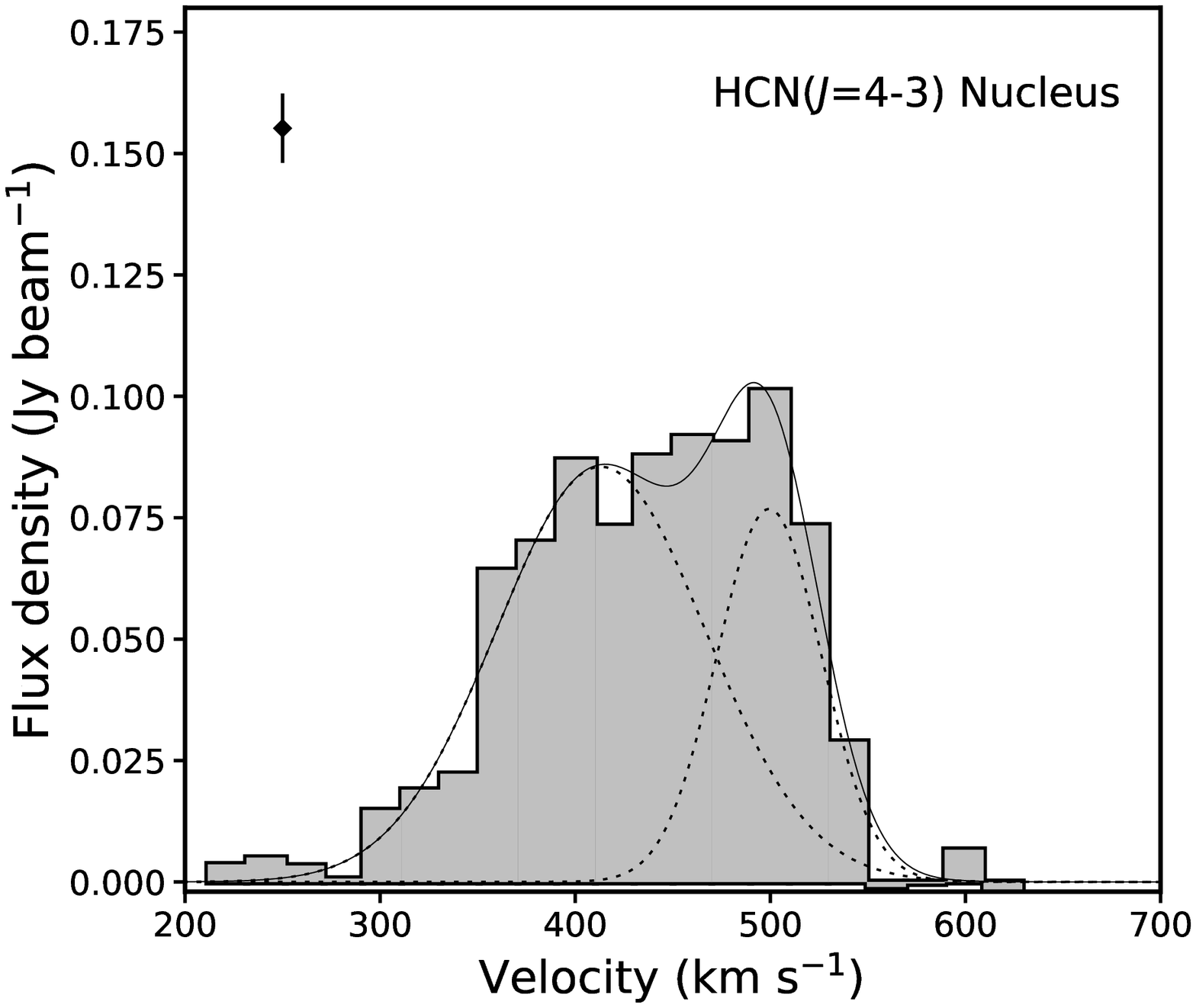}
\includegraphics[scale=0.23]{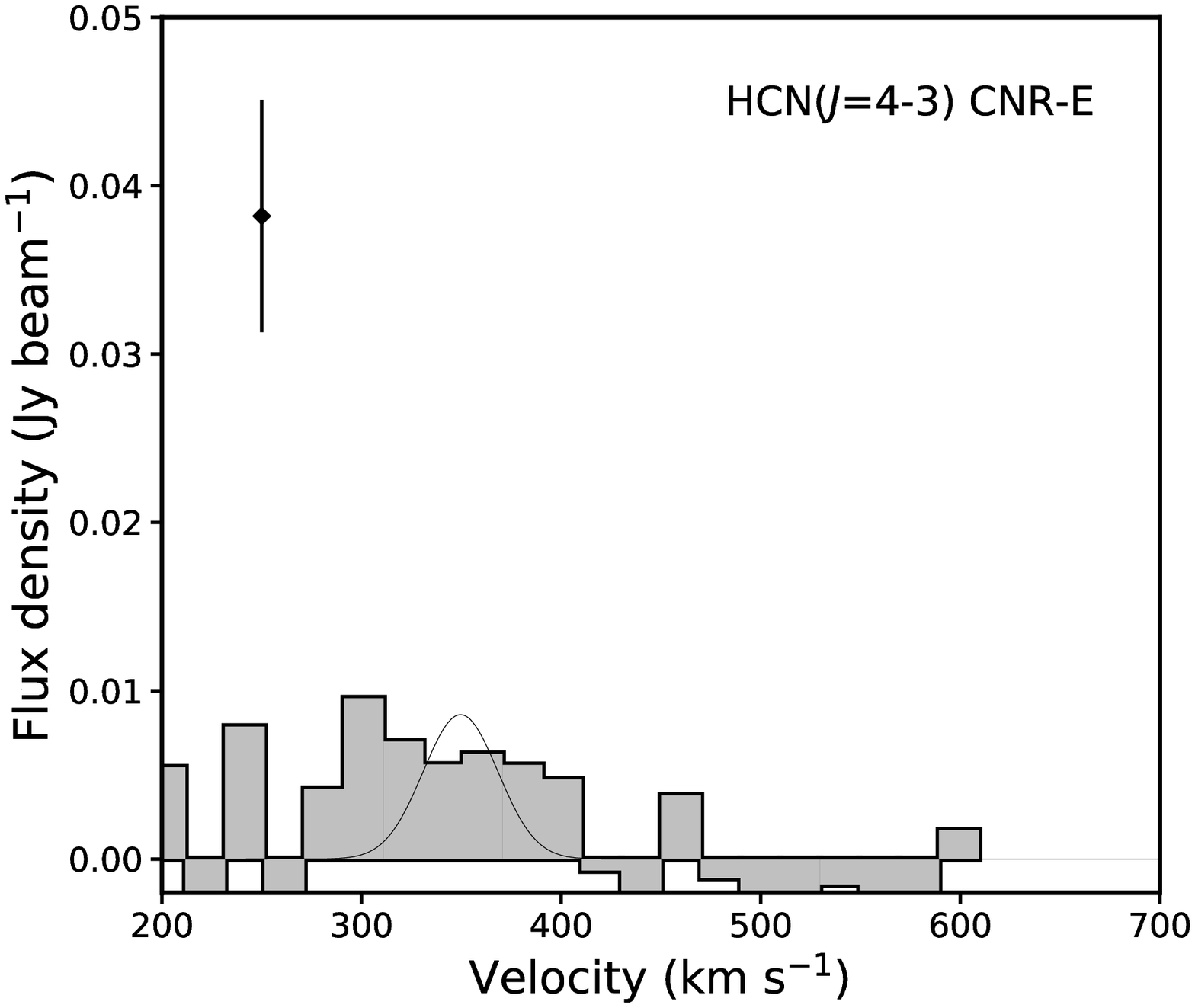} 
 \includegraphics[scale=0.23]{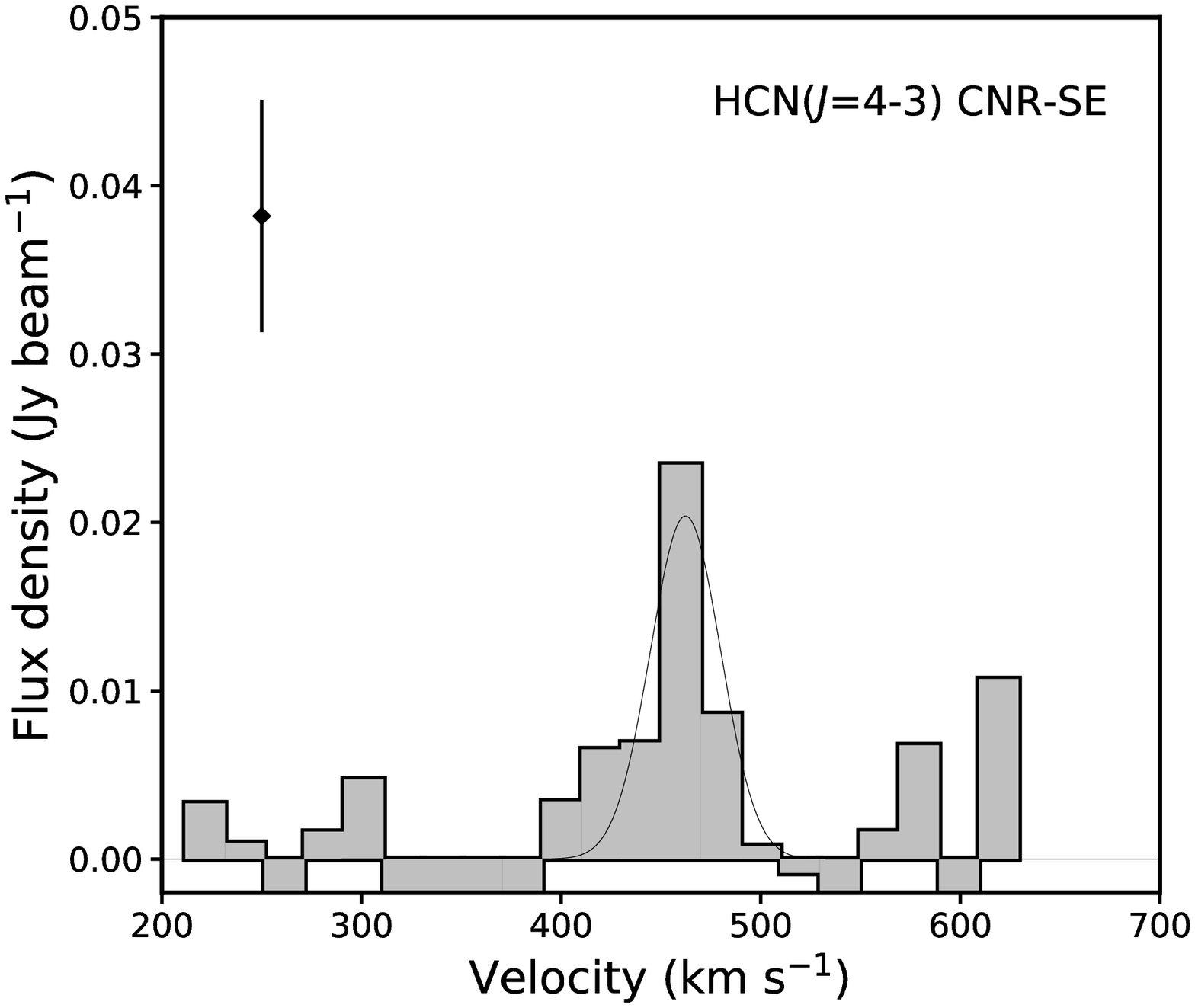} \vspace{-.5cm}  
 \vspace{1cm}
 \end{center}  
 \caption{\small{
 Molecular emission lines of CO($J$=3--2), HCO$^+$($J$=3--2), HCO$^+$($J$=4--3), HCN($J$=3--2), 
 and HCN($J$=4--3). From left to right, the figures present the spectra extracted from the Nucleus,
 \rega, and \regb regions with a beam size of $\approx$1.01$\times$1.37 arcsec$^2$ 
 ($\approx$20$\times$27 pc$^2$). Each scale bar in each top left corner corresponds 
 to 1$\sigma$ error. In the Nucleus spectra, the two Gaussian functions and the total are plotted
 with dotted and solid lines, respectively. In the remaining spectra, each single Gaussian function
 is plotted with a solid line. 
 }
 }\label{fig:mol_spec}

\end{figure*}

\section{ALMA DATA ANALYSES}\label{sec:alm_ana} 

\subsection{Reduction and Reprocessing of ALMA Data} \label{sec:alm_spec_ana}

We reprocessed the ALMA data via the Common Astronomy Software Application (CASA) 
software package \citep{McM07} and adopted the same system versions as those used in the 
Quality Verification by the ALMA Regional Center. To extract signals from the molecular 
emission lines, we subtracted the continuum component from the data cube via the task 
\texttt{uvcontsub}. The continuum level was determined by fitting line-free channels 
with the first-order function. Then, we analyze the ALMA (a) and (b) datasets 
according to the scientific purpose of use. 

We used the (a) dataset mainly to constrain the molecular gas properties through the 
molecular line ratios. The dataset was composed of two observations. We matched their 
spatial and velocity resolutions to extract the spectra under the same conditions as
much as possible. We thus applied the \texttt{clean} task with \texttt{uvcoverage} 
= $13$--$250k\lambda$ to match the $uv$-coverage. The range was covered by the two
observations. The channels were binned such that the velocity resolution was
$\approx$20~km s$^{-1}$: 20.3 km s$^{-1}$ for CO($J$=3--2), 17.8 km s$^{-1}$ for HCN($J$=3--2), 
19.9 km s$^{-1}$ for HCN($J$=4--3), 17.7 km s$^{-1}$ for HCO$^+$($J$=3--2), and 19.8 km s$^{-1}$ 
for HCO$^+$($J$=4--3). We adopted Briggs-weighting with robust = 0.5 and gain = 0.1 
to produce images with high resolution and high sensitivity per beam. The primary beam 
image correction was made via \texttt{impbcor}. Finally, we applied the \texttt{imsmooth} task 
to match the beam sizes to the worst of 1.01$\times$1.37 arcsec$^2$ ($\approx$20$\times$27 pc$^2$) 
with PA = 48.8 degrees, which was achieved for the HCN($J$=3--2) line. Note that the
task was not applied for the HCO$^+$($J$=3--2) line because of almost the same beam size
(i.e., 1.00$\times$1.37 arcsec$^2$). The final continuum-subtracted data had RMS noise
of 5.3 mJy beam$^{-1}$ for CO($J$=3--2), 2.5 mJy beam$^{-1}$ for HCN($J$=3--2), 6.8 mJy
beam$^{-1}$ for HCN($J$=4--3), 2.7 mJy beam$^{-1}$ for HCO$^+$($J$=3--2), and 7.3 mJy beam$^{-1}$ 
for HCO$^+$($J$=4--3). These values were estimated from the emission-free channels. For the
CO($J$=3--2) line, some negative signals were found around the line, likely due to the high
brightness and sparse $uv$-coverage. Thus, to make the correction, we simply added the 
systematic error of 240 mJy beam$^{-1}$ when fitting the spectra. 

Figure~\ref{fig:mol_lin_map_all} shows the velocity-integrated intensity maps 
(Moment 0), intensity-weighted mean velocity maps (Moment 1), and intensity-weighted velocity
dispersion maps (Moment 2) of the five lines. Smoothing was not 
conducted. The Moment 0 maps were made by calculating the 0th moment over the VLSR
range of 200--650 km s$^{-1}$. The other moment maps were produced in 
the same VLSR range, but 10$\sigma$ clipping was further applied.
Note that the Moment maps of the CO($J$=3--2) line only take account of the statistical error. 

In contrast to the (a) dataset, the (b) dataset was used to reveal the molecular gas distributions 
in detail by exploiting the high SNR. We produced the HCO$^+$($J$=4--3) line data cube via the
\texttt{clean} task using the same options adopted above. However, we increased the velocity resolution
to 9.8 km s$^{-1}$ so as not to miss the broad components. Setting a different velocity resolution
does not largely affect our discussion. The primary beam correction was conducted using the
\texttt{impbcor} task. Figure~\ref{fig:hcop43_img} shows the Moment 0, 1, and 2 maps of HCO$^+$($J$=4--3).
Except for the Moment 0 map, 10$\sigma$ clipping was applied. We adopted a wider channel range of
VLSR = 150--700 km s$^{-1}$ to cover broad but faint components. High RMS noise of 0.73 mJy beam$^{-1}$
was achieved, while the beam size was 0.61$\times$0.59 arcsec$^{2}$ with PA = $-$1.25. Note that the
Moment 2 map shows a bit higher values than that created using the (a) dataset. We confirmed that
line fluxes and velocity-integrated brightness temperatures used for discussion were however consistent
between the (a) and (b) datasets.

\subsubsection{Basic Molecular Line Properties}\label{sec:bas_mol_lin}

We constrain the basic quantities of the emission lines in the three subregions 
(Nucleus, \rega, and \regb) through the spectra taken with a 
single synthesized beam ($\approx$1.01$\times$1.37 arcsec$^2$) centered 
at each region (see Figure~\ref{fig:hcop43_img}). The
Gaussian functions are fitted as shown in Figure~\ref{fig:mol_spec}. The best-fits 
are determined using the least chi-square method. We cannot achieve 
acceptable fits to the Nucleus spectra using a single Gaussian function. Thus, two
functions are adopted. The spectra taken from \rega and \regb 
are well reproduced using a single Gaussian function. Because the HCN($J$=4--3) 
lines in the \rega and \regb regions are weak, we fix their velocity widths 
and line centers at those constrained at each HCN($J$=3--2) line. Table~\ref{tab:mol_dat_lis}
summarizes the parameters of the Gaussian functions, and the resultant
velocity-integrated intensities and brightness temperatures. 

\begin{table*}
  \caption{Molecular Line Properties\label{tab:mol_dat_lis}}
  \begin{center}
    \begin{tiny}    
\begin{tabular}{cccccccccccccccccccccc}
  \hline \hline
 Molecular line & $\nu_{\rm rest}$ & Region & $a$  & $\sigma$ & $v_{\rm LSR}$ & Peak flux  & $I$    & $I$  \\ 
   & (GHz)  &  & (Jy beam$^{-1}$ km s$^{-1}$) & (km s$^{-1}$) & (km s$^{-1}$) & (Jy beam$^{-1}$) & (Jy beam$^{-1}$ km s$^{-1}$) & (K km s$^{-1}$) \\ 
 (1)  & (2)  & (3) & (4)  & (5)  & (6)  & (7)  & (8)    & (9) \\ \hline   
\molline{CO}{3}{2} & 345.796 & Nucleus$^{2}$ & 32.6$\pm$13.6 & 20.0$\pm$4.4 & 510$\pm$3 & 0.649$\pm$0.307 & 13.9$\pm$7.2   & 102$\pm$53 \\ 
   &  &   & 234$\pm$16 & 51.1$\pm$3.2 & 439$\pm$4 & 1.83$\pm$0.17 & 99.6$\pm$11.1  & 736$\pm$82 \\
   &  &   &  &   &  &   & 113.4$\pm$13.2$^\ast$ & 838$\pm$98$^\ast$ \\
   &  & \rega$^{1}$ & 93.7$\pm$5.6 & 22.6$\pm$1.6 & 350$\pm$2 & 1.65$\pm$0.15 & 39.8$\pm$4.6   & 294$\pm$34 \\ 
   &  & \regb$^{1}$ & 81.6$\pm$3.3 & 29.5$\pm$1.4 & 448$\pm$1 & 1.11$\pm$0.07 & 34.7$\pm$2.7   & 256$\pm$20 \\ 
\hline 
\molline{HCO$^+$}{3}{2} & 267.558 & Nucleus$^{2}$ & 7.99$\pm$2.00 & 41.3$\pm$6.3 & 403$\pm$11 & 0.077$\pm$0.023 & 3.40$\pm$1.13  & 42.3$\pm$14.0 \\
   &  &   & 10.2$\pm$2.0 & 36.6$\pm$3.5 & 487$\pm$6 & 0.111$\pm$0.024 & 4.34$\pm$1.03  & 54.0$\pm$12.8 \\ 
   &  &   &  &   &  &   & 7.73$\pm$1.52$^\ast$  & 96.4$\pm$19.0$^\ast$ \\
   &  & \rega$^{1}$ & 1.78$\pm$0.13 & 15.9$\pm$1.3 & 354$\pm$1 & 0.045$\pm$0.005 & 0.757$\pm$0.103  & 9.43$\pm$1.28 \\ 
   &  & \regb$^{1}$ & 1.88$\pm$0.26 & 23.2$\pm$3.7 & 459$\pm$4 & 0.032$\pm$0.007 & 0.797$\pm$0.212  & 9.93$\pm$2.64 \\ 
\hline 
\molline{HCO$^+$}{4}{3} & 356.734 & Nucleus$^{2}$ & 17.8$\pm$4.1 & 54.9$\pm$9.3 & 411$\pm$14 & 0.129$\pm$0.037 & 7.56$\pm$2.52  & 52.5$\pm$17.5 \\
   &  &   & 9.98$\pm$3.95 & 32.2$\pm$4.8 & 496$\pm$5 & 0.124$\pm$0.052 & 4.24$\pm$1.91  & 29.4$\pm$13.2 \\
   &  &   &   &  &  &   & 11.8$\pm$3.2$^\ast$   & 81.9$\pm$21.9$^\ast$ \\
   &  & \rega$^{1}$ & 1.54$\pm$0.36 & 19.9$\pm$5.4 & 353$\pm$5 & 0.031$\pm$0.011 & 0.652$\pm$0.291  & 4.53$\pm$2.02 \\
   &  & \regb$^{1}$ & 2.14$\pm$0.38 & 28.0$\pm$5.7 & 449$\pm$6 & 0.030$\pm$0.008 & 0.908$\pm$0.309  & 6.30$\pm$2.15 \\
\hline 
\molline{HCN}{3}{2} & 265.886 & Nucleus$^{2}$ & 8.31$\pm$2.77 & 57.0$\pm$9.9 & 424$\pm$20 & 0.058$\pm$0.022 & 3.53$\pm$1.46  & 44.1$\pm$18.3 \\
   &  &   & 4.16$\pm$2.68 & 34.3$\pm$6.3 & 495$\pm$5 & 0.048$\pm$0.032 & 1.77$\pm$1.23  & 22.1$\pm$15.4 \\
   &  &   &   &  &  &   & 5.30$\pm$1.91$^\ast$  & 66.2$\pm$23.9$^\ast$ \\
   &  & \rega$^{1}$ & 0.694$\pm$0.126 & 18.7$\pm$3.9 & 350$\pm$4 & 0.015$\pm$0.004 & 0.295$\pm$0.102  & 3.69$\pm$1.27 \\
   &  & \regb$^{1}$ & 1.07$\pm$0.14 & 17.6$\pm$2.6 & 462$\pm$3 & 0.024$\pm$0.005 & 0.456$\pm$0.113  & 5.70$\pm$1.41 \\
\hline 
\molline{HCN}{4}{3} & 354.505 & Nucleus$^{2}$ & 11.5$\pm$2.0 & 53.5$\pm$8.5 & 413$\pm$11 & 0.086$\pm$0.020 & 4.87$\pm$1.39  & 34.2$\pm$9.8 \\
   &  &   & 5.17$\pm$1.85 & 26.9$\pm$4.7 & 499$\pm$4 & 0.077$\pm$0.031 & 2.20$\pm$0.96  & 15.5$\pm$6.7 \\ 
   &  &   &   &  &  &   & 7.07$\pm$1.69$^\ast$  & 49.7$\pm$11.8$^\ast$ \\
   &  & \rega$^{1}$ & 0.403$\pm$0.217 & 18.7$^\dagger$ & 350$^\dagger$ & 0.009$\pm$0.005 & 0.171$\pm$0.092  & 1.20$\pm$0.65 \\
   &  & \regb$^{1}$ & 0.902$\pm$0.174 & 17.6$^\dagger$ & 462$^\dagger$ & 0.020$\pm$0.004 & 0.383$\pm$0.074  & 2.69$\pm$0.52 \\
  \hline
  \multicolumn{1}{@{}l@{}}{\hbox to 0pt{\parbox{165mm}
{\footnotesize
  \textbf{Notes.}\\
  Columns:
   (1) Molecular line name.
 (2) Rest frequency. 
 (3) Subregion name. The exact positions are listed in Table~\ref{tab:pos_inf} and are marked in Figure~\ref{fig:hcop43_img}.
 Here, the superscripts represent the number of the Gaussian function(s) used to fit each spectrum. 
 (4)--(6)
 Parameters defined in the Gaussian function of $a/\sqrt{2\pi\sigma^2}\exp\{-(v-v_{\rm LSR})^2/2\sigma^2\}$.
 (7)--(8) Peak flux density and velocity-integrated flux density. 
 (9) Velocity-integrated brightness temperature, derived from (8) by adopting a beam size of 1.01$\times$1.37 
 arcsec$^2$, except for the HCO$^+$($J$=3--2) line, for which the 1.00$\times$1.37 arcsec$^2$ beam is adopted. 
 Note that all estimates are based on the spectra binned at $\approx$ 20 km s$^{-1}$ resolutions and 
   corrected for  the primary beam attenuation.
 ~$^\ast$ These values take account of the two Gaussian functions.
 ~$^\dagger$ These values are fixed at those constrained in each HCN($J$=3--2) line. 
}
\hss}}
\end{tabular}
  \end{tiny}
  \end{center}
\end{table*}

\subsubsection{RADEX Non-LTE Modeling}\label{sec:radex}

\begin{table*}
  \caption{Molecular Line Ratios\label{tab:mol_rat_lis}}
\begin{center}
\begin{tabular}{cccccccccccccccccccccc}
  \hline \hline
   Region
 & $\frac{{\rm HCO}^+(J=3\mathchar`-\mathchar`-2)}{{\rm CO}(J=3\mathchar`-\mathchar`-2)}$
 & $\frac{{\rm HCO}^+(J=4\mathchar`-\mathchar`-3)}{{\rm CO}(J=3\mathchar`-\mathchar`-2)}$
 & $\frac{{\rm HCN}(J=3\mathchar`-\mathchar`-2)}{{\rm CO}(J=3\mathchar`-\mathchar`-2)}$
 & $\frac{{\rm HCN}(J=4\mathchar`-\mathchar`-3)}{{\rm CO}(J=3\mathchar`-\mathchar`-2)}$
 & $\frac{{\rm HCO}^+(J=4\mathchar`-\mathchar`-3)}{{\rm HCO}^+(J=3\mathchar`-\mathchar`-2)}$
 \\
 (1) & (2)    & (3)    & (4)
 & (5)    & (6)  \\ 
 & $\frac{{\rm HCN}(J=3\mathchar`-\mathchar`-2)}{{\rm HCO}^+(J=3\mathchar`-\mathchar`-2)}$
 & $\frac{{\rm HCN}(J=4\mathchar`-\mathchar`-3)}{{\rm HCO}^+(J=3\mathchar`-\mathchar`-2)}$
 & $\frac{{\rm HCO}^+(J=4\mathchar`-\mathchar`-3)}{{\rm HCN}(J=3\mathchar`-\mathchar`-2)}$
 & $\frac{{\rm HCN}(J=4\mathchar`-\mathchar`-3)}{{\rm HCN}(J=3\mathchar`-\mathchar`-2)}$
 & $\frac{{\rm HCN}(J=4\mathchar`-\mathchar`-3)}{{\rm HCO}^+(J=4\mathchar`-\mathchar`-3)}$
 \\
 & (7)  & (8)  & (9)  & (10)  & (11) \\ \hline 
 Nucleus & 0.0527$\pm$0.0142 & 0.107$\pm$0.031 & 0.0359$\pm$0.0145 & 0.0639$\pm$0.0170 & 2.03$\pm$0.74 \\ 
 & 0.681$\pm$0.280 & 1.21$\pm$0.41 & 2.99$\pm$1.47 & 1.78$\pm$0.81 & 0.596$\pm$0.214 \\ \hline
 \rega & 0.0147$\pm$0.0026 & 0.0169$\pm$0.0078 & 0.0057$\pm$0.0021 & 0.0044$\pm$0.0024 & 1.15$\pm$0.54 \\
 & 0.387$\pm$0.144 & 0.299$\pm$0.166 & 2.97$\pm$1.67 & 0.773$\pm$0.494 & 0.261$\pm$0.182 \\ \hline 
 \regb & 0.0178$\pm$0.0055 & 0.0270$\pm$0.0094 & 0.0101$\pm$0.0030 & 0.0113$\pm$0.0024 & 1.52$\pm$0.69 \\ 
 & 0.569$\pm$0.207 & 0.637$\pm$0.228 & 2.67$\pm$1.19 & 1.12$\pm$0.39 & 0.419$\pm$0.164 \\ 
  \hline
  \multicolumn{1}{@{}l@{}}{\hbox to 0pt{\parbox{135mm}
{\footnotesize
  \textbf{Notes.}\\
  Columns:
   (1) Subregion name. 
 (2)--(11) Line ratios derived from fluxes in units of erg cm$^{-2}$ s$^{-1}$. The absolute flux uncertainty (10\%) 
 was taken into account if required. 
}
\hss}}
\end{tabular}
\end{center}
\end{table*}

The physical and chemical properties of the molecular gas are constrained by comparing the observed
molecular line ratios to those predicted by the non-local thermodynamic equilibrium (non-LTE) code of
RADEX \citep{van07}. We simulate 10 combinations of line ratios taken from the five lines 
for various kinetic temperatures ($T_{\rm kin}$), hydrogen densities ($n_{\rm H2}$), and hydrogen
column densities ($N_{\rm mol}$). A spherically, homogeneous geometry is assumed in the code. The
molecular line transition rates are compiled from the Leiden Atomic and Molecular Database (LAMDA;
\cite{Sch05}). We cover the gas kinetic temperature within $T_{\rm kin}$ = 10--600 K for steps of
$\Delta T_{\rm kin}$ = 10 K, the gas density within $\log n_{\rm H2}/{\rm cm}^{-3}$ = 2--7 for
steps of $\Delta \log n_{\rm H2}/{\rm cm}^{-3}$ = 0.5, and the column density within $\log
N_{\rm H2}/{\rm cm}^{-2}$ = 20--25 for steps of $\Delta \log N_{\rm H2}/{\rm cm}^{-2}$ = 0.5. In
addition, the HCN to HCO$^+$ abundance ratio ([HCN]/[HCO$^+$]) is varied from 1 to 10 for steps of
1 by considering previous studies that indicated the enhanced HCN abundance in AGN host galaxies
(e.g., \cite{Koh05}; \cite{Izu13}, \yearcite{Izu16}). For the other abundance ratios, we adopt canonical values 
of [CO]/[H$_{\rm 2}$] = $6\times10^{-5}$ and [HCO$^+$]/[H$_{\rm 2}$] = $8\times10^{-9}$, equivalent
to those observed in galactic molecular clouds \citep{Bla87}. We confirm that even if we vary the
abundances to [CO]/[H$_{\rm 2}$] = $5\times10^{-5}$ and [HCO$^+$]/[H$_{\rm 2}$] = $2\times10^{-9}$, 
estimated in other galactic molecular clouds \citep{Bla87}, our conclusion in Section~\ref{sec:xdr} 
is not strongly affected. The line velocities are fixed at the average 3$\sigma$ width of the
five lines: 330 km s$^{-1}$, 120 km s$^{-1}$, and 140 km s$^{-1}$ for the Nucleus, \rega and \regb
regions, respectively. Our calculation also takes account of the blackbody radiation with $T_{\rm bg}$
= 2.73 K as the uniform background emission. In summary, the free parameters
are $T_{\rm kin}$, $n_{\rm H2}$, $N_{\rm mol}$, and [HCN]/[HCO$^+$]. These parameters are constrained
by the chi-square method. The line ratios are calculated using fluxes in units of erg cm$^{-2}$ s$^{-1}$.
Those observed are listed in Table~\ref{tab:mol_rat_lis}. The fitting results are summarized in
Table~\ref{tab:radex_fit}.

\begin{table*}
  \caption{Physical Parameters of Molecular Gas\label{tab:radex_fit}}
\begin{center}
\begin{tabular}{cccccccccccccccccccccc}
  \hline \hline
  Region & $\log N_{\rm H_2}/{\rm cm}^{-2}$ & $\log n({\rm H}_2)/{\rm cm}^{-3}$
  & $T_{\rm k}$ & [HCN]/[HCO$^{+}$] & $\chi^2$ \\ 
  &  &  & (K)  &   &  \\ 
 (1) & (2)  & (3) & (4)  & (5)  & (6) \\ \hline 
 Nucleus & 24.5$^{+0.5}_{-0.0}$  & 5.0$^{+0.0}_{-0.5}$ 
 & 290$^{+110}_{-100}$  & 3$\pm0$   & 0.45 \\
 \rega & 24.5$^{+0.5}_{-4.5}$  & 3.5$^{+1.5}_{-0.5}$ 
  & 200$^{+130}_{-150}$  & 2$\pm1$ & 0.71 \\ 
 \regb & 23.5$^{+1.5}_{-0.0}$  & 4.5$^{+0.0}_{-1.5}$ 
  & 130$^{+270}_{-50}$  & 4$^{+1}_{-2}$  & 0.42 \\ 
  \hline
  \multicolumn{1}{@{}l@{}}{\hbox to 0pt{\parbox{100mm}
{\footnotesize
  \textbf{Notes.}\\
 Columns: 
 (1) Subregion name. 
 (2) Molecular hydrogen gas column density. 
 (3) Molecular hydrogen gas density. 
 (4) Kinetic temperature. 
 (5) Abundance ratio between the HCN and HCO$^+$ molecules. 
 (6) Chi-square value. 
}
\hss}}
\end{tabular}
\end{center}
\end{table*}

\section{DISCUSSION}\label{sec:dis}

\subsection{Spatial Comparison between Fe K$\alpha$ Line and Molecular Emission Line}\label{sec:ima_com}

\begin{figure*}
 \hspace{.5cm} 
 \includegraphics[scale=0.33]{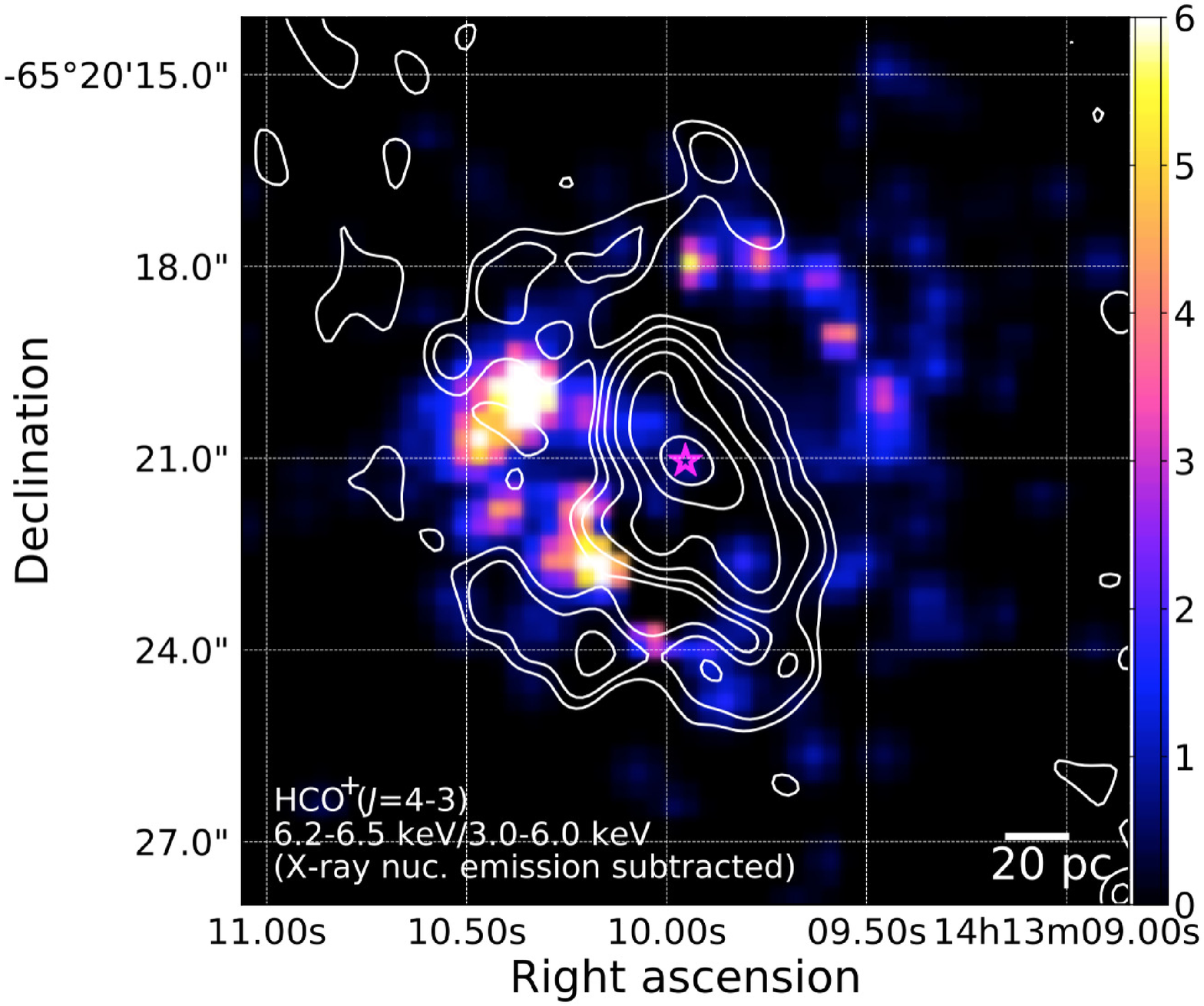} 
 \includegraphics[scale=0.33]{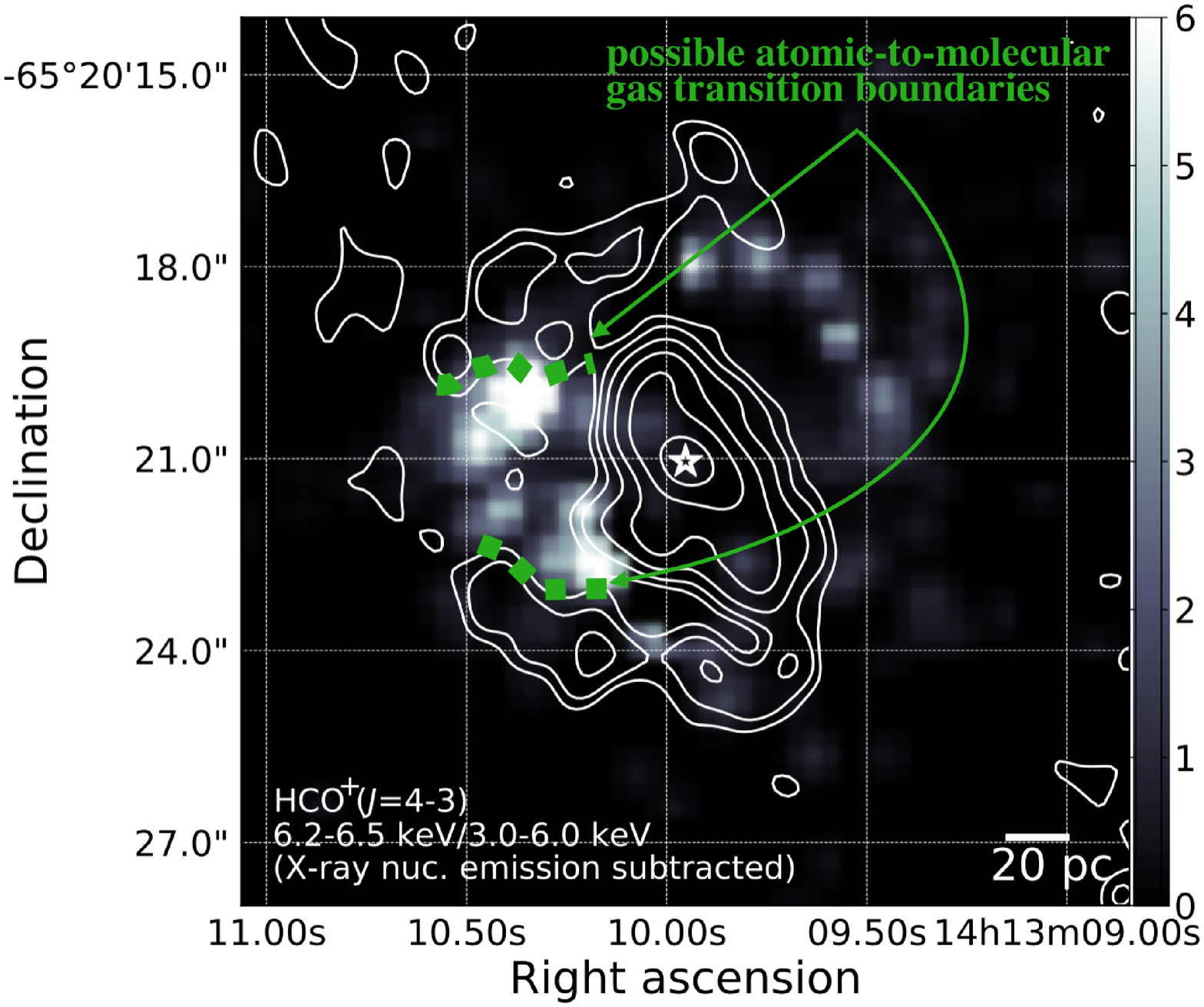}  
 \caption{\small{
 (Left) Color-coded ratio between the 6.2--6.5 keV and 3.0--6.0 keV images, 
 corresponding to a proxy of the Fe K$\alpha$ line equivalent width, and
 the HCO$^+$($J$=4--3) velocity-integrated intensity map (white contour). 
 (Right) The same figure as in the left panel, except for the use of color and 
 the illustration that represents possible 
 atomic to molecular gas transition boundaries (green dashed lines). 
 The magenta and white stars are located at the SMBH position.
 }
 }\label{fig:iron2hcop43}
\end{figure*}

Figure~\ref{fig:iron2hcop43} shows the spatial distributions of the 6.4 keV Fe K$\alpha$ 
and HCO$^{+}$($J$=4--3) lines at $\sim$0.5 arcsec ($\approx$10 pc) resolution. The most 
interesting feature is the spatial anti-correlation between the two emission lines. 
The iron line traces the gas irradiated by the hard X-rays from the AGN (see Section~\ref{sec:x_ana}), 
and it is irrelevant whether the gas is in the molecular or atomic phase. In addition, the 
line is effectively emitted when X-rays propagate into the dense gas given the low cross-section. 
The HCO$^+$($J$=4--3) line may also become luminous in such dense gas regions if abundant
molecular gas is available as well, but is faint in regions with bright Fe K$\alpha$ line
emission. The molecular emission seems to be brighter rather in the outer side. Thus, this
anti-correlation indicates that molecules are dissociated into atoms in the proximity of
the AGN due to the harsh X-ray radiation. Then, outer surfaces of the \rega and \regb 
regions may correspond to the atomic-to-molecular gas transition boundary (right panel 
of Figure~\ref{fig:iron2hcop43}) within which photoelectrons produced 
by X-rays can greatly dissociate molecules. Although the Nucleus region is apparently closest 
to the central AGN, an amount of molecular gas is present. This is likely because we observe  
the molecular gas that is located far enough away from the central engine not to be effectively
dissociated into atomic gas, like those found outside the other subregions. High inclination
angles ($\sim 70^\circ$) of the host galaxy as well as the molecular gas around the center, 
kinematically modeled by \citet{Izu18}, support this idea.

\subsection{Diagram of Molecular Gas Line Ratios}\label{sec:mol_dia} 

\begin{figure}
 \includegraphics[angle=-90,scale=0.33]{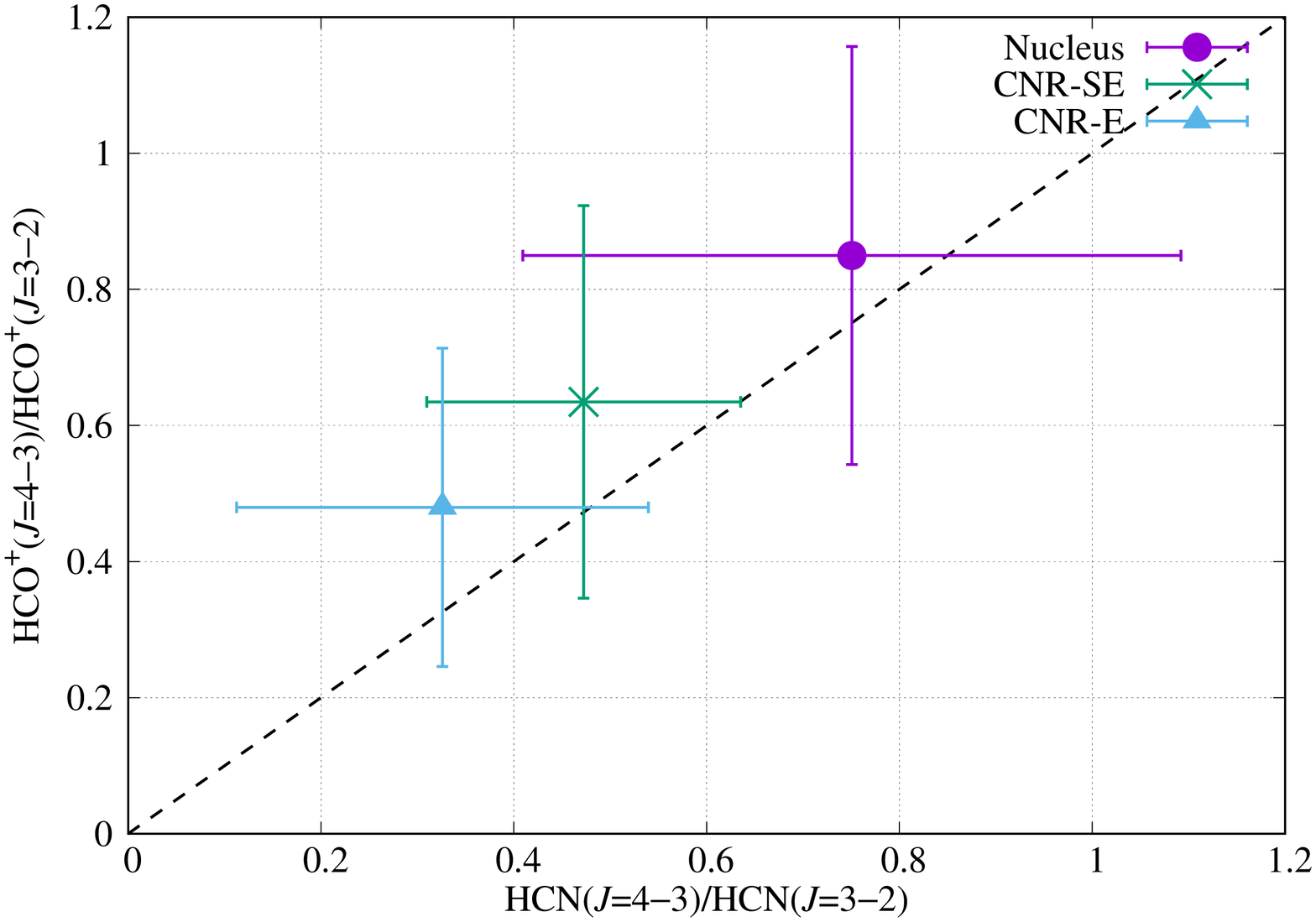}\vspace{2mm}
 \caption{\small{
     Ratio of HCN($J$=4-–3) to HCN($J$=3–-2) versus that of HCO$^+$($J$=4-–3) to HCO$^+$($J$=3-–2)
     on the velocity-integrated brightness temperature scale.
     The dashed line represents the one-to-one line. 
   }
 }\label{fig:mol_dia_1} 
\end{figure}

\begin{figure}
 \includegraphics[angle=-90,scale=0.33]{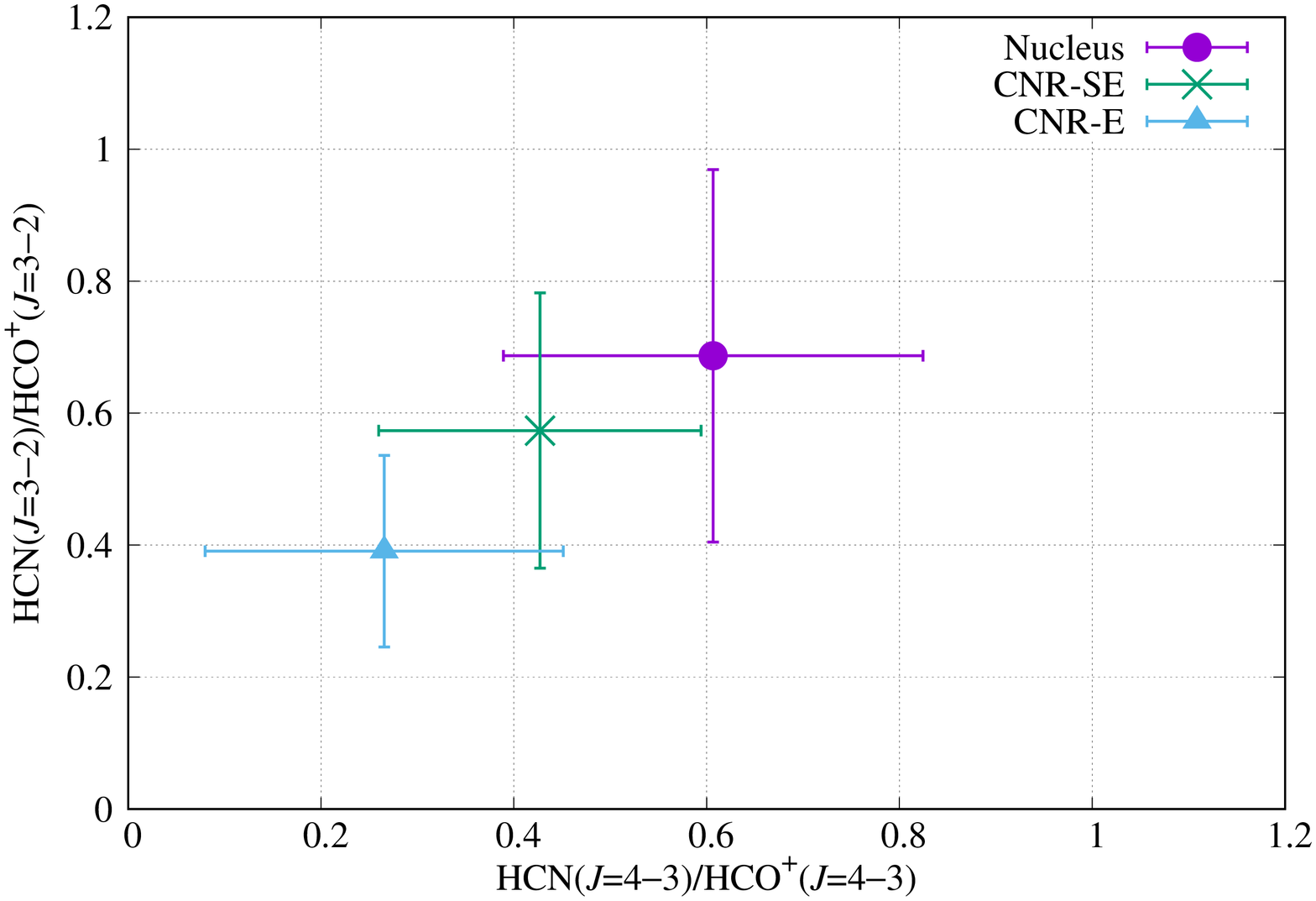}\vspace{0mm}
 \caption{\small{
     Ratio of HCN($J$=4-–3) to HCO$^+$($J$=4-–3) versus that of HCN($J$=3-–2) to HCO$^+$($J$=3-–2) 
     on the velocity-integrated brightness temperature scale.     
 }
 }\label{fig:mol_dia_2} 
\end{figure}

Based on the velocity-integrated temperature ratios, we briefly discuss
molecular gas properties in the subregions. We can focus on the AGN effect
rather than that of the SF, given that our spectra were taken in the proximity
of the nucleus at the 20 pc scale and the SF is not influential due to
the low SFR of 0.1 $M_\odot$ yr$^{-1}$ in the central 75 pc \citep{Esq14}. 

Figure~\ref{fig:mol_dia_1} shows a diagram of the HCN($J$=4--3)/HCN($J$=3--2) and HCO$^+$($J$=4--3)/HCO$^+$($J$=3--2)
ratios. Note that the ratios take into account an absolute flux uncertainty of 10\%. The Nucleus region shows the
highest values in both ratios. The same diagram was made by \citet{Ima18}, who compiled AGN
host and starburst galaxies. As judged from their Figure~26, the ratios of Circinus are low and
comparable to those of starburst galaxies. NGC~7469, an unobscured AGN, also shows similar,
low line ratios. If an AGN is able to heat the surrounding gas more effectively via X-ray
emission rather than SF via UV emission, elevated line ratios would be expected for the AGN host galaxy.
However, this was not observed. A possible cause is molecular gas dissociation, as discussed in Section~\ref{sec:ima_com}. Indeed,
if the molecular gas density is lower than the critical densities ($\sim$ 10$^{6-7}$ cm$^{-3}$),
the excitation level can become not so  sensitive to the kinetic temperature or heating. 
At last, we note that the slightly higher HCO$^+$ line ratios than the HCN ones probably
reflect that HCO$^+$ can be more easily excited due to a lower critical density than that
of HCN (e.g., \cite{Gre09}). 

Figure~\ref{fig:mol_dia_2} shows a scatter plot between the HCN and HCO$^+$ line ratios at
$J$=4--3 and $J$=3--2. These ratios have been studied extensively (e.g., \cite{Koh05}; \cite{Ima10}; \cite{Ima14};
\cite{Izu13}; \cite{Izu16}; \cite{Gar14}), 
and several studies have reported enhanced HCN emission in AGN host galaxies. Thus, questions remain as to 
whether AGN can be probed from the HCN line. Using data with high spatial resolution,
\citet{Izu16} showed that AGN host galaxies may have HCN($J$=4--3)/HCO$^+$($J$=4--3)~$\gtrsim$
1 (see also \cite{Vit14}; \cite{Gar14}). However, Circinus has significantly lower values, 
implying that it may be difficult to identify all AGN based on HCN emission. 
As argued by \citet{Izu16}, the HCN($J$=4--3)/HCO$^+$($J$=4--3) ratio does not correlate with 
AGN X-ray luminosity. Thus, an alternative mechanism, such as mechanical energy inputs by 
AGN jets/outflows or cosmic-rays, would be more influential regarding the HCN enhancement. No report 
on such a signature for Circinus is consistent with this idea. 

\subsection{XDR Model}\label{sec:xdr} 

We quantitatively discuss the gas in the three subregions along with the XDR model, developed
by \citet{Mal96}. The gas-phase abundance of C and O with respect to H is assumed to be
$3.0\times10^{-4}$ and $5.0\times10^{-4}$, respectively. The X-ray spectrum incident on the
medium is made up of single power-law emission. A key parameter for the discussion is the
effective ionization parameter, which determines the fractional abundances of atomic and molecular
gas species (see their Figure~3). The parameter is calculated as 
\begin{eqnarray}\label{eqn:ion}
  \xi_{\rm eff} 
  \simeq 0.1 \frac{L_{{\rm X},44}}{n_{{\rm H}_2,4} R^2_{100} N^\alpha_{{\rm H},22}}, 
\end{eqnarray} 
where $L_{{\rm X},44}$, $n_{{\rm H}_2,4}$, $R_{100}$, $N_{\rm H,22}$, and $\alpha$ represents 
the 1--100 keV luminosity in units of $10^{44}$ erg s$^{-1}$, the hydrogen molecular gas density
in units of $10^{4}$ cm$^{-3}$, the distance from the X-ray source in units of 100 pc, the hydrogen
column density of the gas that attenuates the incident X-ray flux in units of $10^{22}$ cm$^{-2}$,
and an X-ray photon index-dependent value, respectively. The last factor ($\alpha$) is specifically
expressed as $\alpha =$ ($\Gamma$+2/3)/(8/3), and takes into account that softer spectra are more
heavily absorbed. As canonical X-ray parameters for discussion, we adopt the 1--100 keV luminosity
  ($L_{\rm 1-100~keV}$) of 1.3$\times$10$^{43}$ erg s$^{-1}$ and the photon index of 2.31, derived 
  from the \nustar data \citep{Are14}. The photon index corresponds to $\alpha \approx$1.1.
  As detailed below, we can draw the same conclusion based on estimates from other hard X-ray 
  observatories of \textit{BeppoSAX}, \textit{INTEGRAL}, and \textit{Suzaku} (\cite{Sol05}; \cite{Yan09}).

The molecular gas density has been derived from the non-LTE analyses (Section~\ref{sec:radex}). 
The distances from the center to the \rega and \regb regions were set to 62 pc and 60 pc, respectively. 
The distance appropriate for the Nucleus region is difficult to determine due to the sightline. 
We here representatively consider $R >$ 60 pc. This is based on the discussion in Section~\ref{sec:ima_com} 
that the Nucleus molecule we observe is equivalent to those seen outside the other subregions ($\gtrsim$ 60 pc).
Here, it may be stressed that robust constraint on the plausible atomic to molecular transition boundary is
essential to calculate $\xi_{\rm eff}$, and is achieved by exploiting the high spatial resolution ($\approx$
0.5 arcsec) data taken by ALMA and \chandra.
Finally, we assume $\log N_{\rm H}/{\rm cm}^{-2} = 23.9$, where 
the medium becomes optically thick for the 7.1 keV photon, the K-edge of neutral iron. This assumption is
reasonable given that such photons are used up in the subregions, except for the Nucleus, and the fluorescent 
line is not bright beyond them (Figure~\ref{fig:iron2hcop43}).

\begin{table}
  \caption{Effective ionization parameters\label{tab:ion_eff}}
\begin{center}
\begin{tabular}{cccccccccccccccccccccc}
  \hline \hline
 Region & $\log \xi_{\rm eff}$ \\
 (1) & (2) \\ \hline   
 Nucleus & [$< -$4.0] \\
 \rega & [$-$4.6 $\sim$ $-$2.6] \\ 
 \regb & [$-$4.0 $\sim$ $-$2.5] \\ 
  \hline
  \multicolumn{1}{@{}l@{}}{\hbox to 0pt{\parbox{48mm}
{\footnotesize
  \textbf{Notes.}\\
 Columns: 
 (1) Subregion name.
 (2) Effective ionization parameters defined in Equation~\ref{eqn:ion}.  
}
\hss}}
\end{tabular}
\end{center}
\end{table}

As summarized in Table~\ref{tab:ion_eff}, the ionization parameters in the \rega and \regb 
regions are $\log \xi_{\rm eff} = -4.6\sim-2.6$ and $-4.0\sim-2.5$, respectively. These are
consistent with $\log \xi_{\rm eff} \approx-3.0$, above which the XDR model predicts a large 
fraction of the molecular hydrogen to be dissociated into atomic hydrogen. The low ionization
parameter of $\log \xi_{\rm eff} < -4.0$ in the Nucleus is consistent with a moderate amount 
of the molecular gas still remaining therein (Figure~\ref{fig:iron2hcop43}). Seemingly, this
is contrary to the prediction of the XDR model, that is, the increase of the ionization 
parameter with decreasing distance. As already discussed (Section~\ref{sec:ima_com}), this apparent 
discrepancy can be resolved by considering that the molecular gas in the Nucleus is located far 
away from the central source and makes it difficult to confirm the model prediction.
Even if the abundances different from the default values are assumed (see Section~\ref{sec:radex}),
our conclusion does not change within 2$\sigma$. 
  X-ray parameters we adopt are also the factors of uncertainty. \citet{Sol05} derived 
  $L_{\rm 1-100~keV}$ = $1.6\times10^{42}$ with $\Gamma = $1.8 and $L_{\rm 1-100~keV}$ =
  $1.4\times10^{42}$ with $\Gamma = 1.5$ by fitting \textit{INTEGRAL}/ISGRI+SPI and
  \textit{BeppoSAX}/PDS spectra, respectively. Also, \citet{Yan09} reproduced
  \textit{Suzaku}/XIS+HXD broadband X-ray spectra with  $L_{\rm 1-100~keV}$ = $3.8\times10^{42}$
  and $\Gamma = 1.58$. In either case, low ionization parameters of 
  $\log \xi_{\rm eff} \approx -3$ are derived, however.
Thus, generally the spatial anti-correlation between the Fe K$\alpha$ and molecular gas emission
lines can be interpreted as the molecular gas dissociation by the AGN X-ray radiation.

  Our study provides new insights into the nuclear structure of Circinus as
  well as highly obscured Compton-thick objects. As proposed by \citet{Izu18}, Circinus 
  is expected to have a radiation-driven fountain torus within $\sim10$~pc (e.g., \cite{Wad12}
  and \cite{Wad18}). Because the model has low molecular/atomic gas densities in its
  polar regions, emission from the vicinity of the SMBH should preferably escape through
  them. Optical ionization cones in the northwest direction \citep{Mar94} is likely the
  result, while its counterpart is invisible due to dust lanes even if any. This time, 
  the Fe~K$\alpha$ emission is detected in the corresponding region, and supports the
  presence of a kind of the putative narrow line region behind the dust. Another interesting point 
  comes from the fact that the Fe~K$\alpha$ emission is effectively produced in regions
  with high column densities ($N_{\rm H} \sim 10^{24}$~cm$^{-2}$). The Fe~K$\alpha$ 
  line at $\sim$60 pc suggests that a fraction of the X-ray emission that escaped from 
  the torus is blocked by CNR-scale thick structures. That is, such CNR-scale 
  material has potential to strongly contribute to obscuration if it is located in 
  the sightline.


\subsection{SF in X-ray-irradiated Regions} 

We discuss the subsequent impact of X-ray radiation on SF. A suppressed SFR is expected in the
X-ray-irradiated regions, given the positive correlation between the molecular gas and SFR surface
densities (e.g., \cite{Ken07}; \cite{Big08}). \citet{Roc06} reported a Gemini South/T-ReCS long-slit
spectroscopy observation of Circinus at 0.27-arcsec resolution with the slit width of 0.36 arcsec 
\citep{Tel98}. They observed the 11.3 $\mu$m polycyclic aromatic hydrocarbon (PAH) emission, whose
luminosity is a proxy of the SFR. The less dust extincted PAH emission is convenient for tracing
SFs, even in the dusty nuclear region. The T-ReCS slit was placed at two PAs of 100$^\circ$ 
and 10$^\circ$. The former roughly corresponds to the orientation towards the \regb region from
the Nucleus (see Figure~1 of \cite{Roc06} for the slit configuration). They detected significant
PAH emission around the \regb region. However, they concluded that the emission 
originated from the foreground against the dust obscuration, and were not associated with the ionization 
cone behind the dust. Hence, the SF may be inactive in the \regb region, and the X-ray-irradiated region as well. 
This is consistent with expectations involving molecular gas dissociation. In the case of Circinus,
the X-ray effect is limited within $\sim$100 pc, but if the X-ray (1--100 keV) luminosity reaches $10^{45}$
\ergs, like the most luminous AGN, the size could be $\sim$ 1 kpc where $\log n_{\rm H2}/{\rm cm}^{-3} = 3$ 
and $\log \xi_{\rm eff}$ = $-3$ are assumed. Thus, in this case, the X-ray emission may have non-negligible 
impact, even on galaxy-scale SF. 

\section{SUMMARY} \label{sec:sum}

Our main aim was to understand the effects of AGN X-ray emission on gas. Using \chandra 
and ALMA, we investigated the physical and chemical conditions of the gas in the central
$\sim$100 pc of Circinus at 0.5 arcsec, $\approx$10-pc resolution. The high penetrating
power of X-ray and submm/mm wavelengths, covered by \chandra and ALMA respectively, is
highly suited to study of the nuclear dense gas region with a small bias against the absorption
and dust extinction. The 6.4 keV Fe K$\alpha$ line was used to map the X-ray-irradiated 
gas irrespective of its molecular and atomic phases. The map was compared to that of the
molecular HCO$^{+}$($J$=4--3) emission line (Figure~\ref{fig:iron2hcop43}). 

Our results indicate that molecular gas emission is faint in regions with bright Fe K$\alpha$
line emission and bright in the outer regions. These results imply that the molecular gas
close to the AGN becomes dissociated, and on the far side, atomic-to-molecular gas transition
boundaries form. We have quantitatively discussed this possibility according to the XDR model
developed by \citet{Mal99}. The effective ionization parameter (Equation~\ref{eqn:ion}), which
characterizes the fractional abundance of molecular and atomic gas species, was constrained
in each subregion (Nucleus, \rega and \regb). Those estimated in the \rega and \regb regions
were high enough to support the idea. Furthermore, we have discussed the X-ray irradiation
effect on SF; the inactive SF in X-ray-irradiated regions may indicate that it consequently
suppresses SF, particularly in the proximity of AGN. 

\begin{ack} 
Part of this work was financially supported by the Grant-in-Aid for JSPS 
Fellows for young researchers (T.K.). T.K. was supported by the ALMA
Japan Research Grant of NAOJ Chile Observatory, NAOJ-ALMA-202. 
T.I. and M.I. are supported by JSPS KAKENHI grant numbers 17K14247 and 15K05030, respectively. The scientific
results reported in this article are based on data obtained from the \chandra
Data Archive. This research has made use of software provided by the \chandra
X-ray Center (CXC) in the application packages CIAO. This paper makes use
of the following ALMA data: ADS/JAO.ALMA\#2015.1.01286.S and ADS/JAO.ALMA\#2016.1.01613.S. 
ALMA is a partnership of ESO (representing its member states), NSF (USA) and
NINS (Japan), together with NRC (Canada) and NSC and ASIAA (Taiwan) and KASI
(Republic of Korea), in cooperation with the Republic of Chile. The Joint ALMA
Observatory is operated by ESO, AUI/NRAO and NAOJ. We appreciate the JVO portal
(http://jvo.nao.ac.jp/portal/) operated by ADC/NAOJ for the quick look of the
ALMA archive data. This research made use of APLpy, an open-source plotting package
for Python \citep{Rob12}.
\end{ack}

\appendix
\setcounter{table}{0}
\setcounter{figure}{0}
\def\thesection{\Alph{section}}
\def\thefigure{\Alph{figure}}
\def\thetable{\Alph{table}}

\section{Supplemental \chandra data}\label{sec:app}

\begin{figure}
  \hspace{-8mm}
 \includegraphics[scale=0.31]{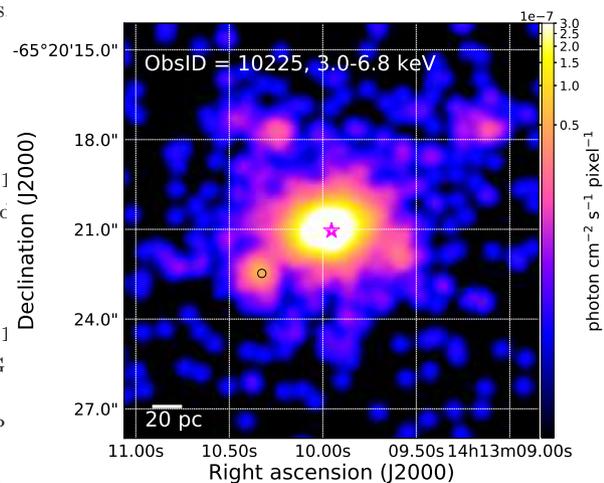}\vspace{15mm}
 \caption{\small{
     Broadband 3.0--6.8 keV image constructed from the ObsID = 10225 data.
     This demonstrates the emergence of a point source (black circle) around the \regb region. 
 }
 }\label{fig:point_src} 
\end{figure}
\begin{figure}
  \hspace{-8mm}
  \includegraphics[scale=0.31]{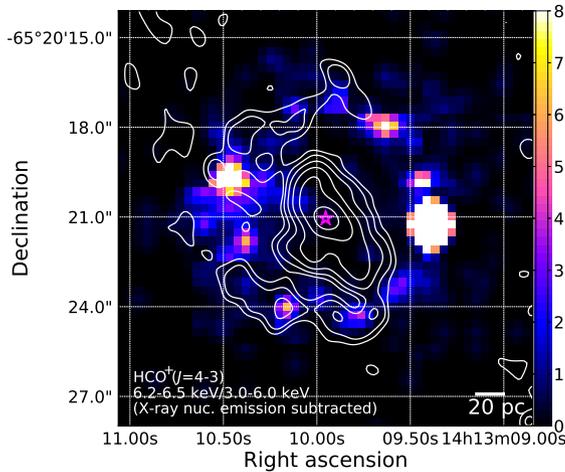}\vspace{15mm}
  \caption{\small{
      Same as Figure~\ref{fig:iron2hcop43}, but the X-ray image ratio is
      created by combining the data in Table~\ref{tab:x_sup_dat_list}. 
 }
 }\label{fig:x_mol_2} 
\end{figure}

  In Section~\ref{sec:xray_map}, we have created the X-ray images from
  the ObsID = 12823 and 12824 data, obtained without the grating. Here, 
  we try to increase the SNR by taking account of 0th order images, 
  and the data obtained for SN 1996cr, $\sim$20 arcsec away from
  Circinus. Table~\ref{tab:x_sup_dat_list} summarizes our data set here. 
  Some of the data are excluded mainly because of the emergence 
  of a point source around the \regb region or very short exposures ($< $
  several ksec). Figure~\ref{fig:point_src}, created from the ObsID = 10225
  data, shows the source, making the analysis and discussion more complex.
  In summary, the total exposure increases by 214 ksec, or effectively
  by $\sim$132 ksec at 6.5 keV given the $\sim$60\% reduced effective
  area\footnote{http://cxc.harvard.edu/proposer/POG/html/chap8.html}. 
  The 6.2--6.5 keV-to-3.0--6.0 keV 
  image ratio is calculated in the same manner as in Section~\ref{sec:xray_map}, and 
  its image is compared with the intensity map of the \molline{HCO$^+$}{4}{3} line in Figure~\ref{fig:x_mol_2}. 
  The result suggests that particularly in the \rega region we can still see the
  anti-correlation, while in the \regb region it seems to be not as clear as 
  in Figure~\ref{fig:iron2hcop43}. The latter might be because we selectively
  exclude the data that have signals in \regb in order to avoid the contamination from
  the point source. It would be however accepted that the molecular line is generally 
  faint in regions with bright Fe-K$\alpha$ emission.

  We finally mention the region around RA, Decl. = 14h13m09.40s, -65d20m22s, 
  where the Fe-K$\alpha$ line with a moderately high EW ($\sim$ 1 keV) can be detected 
  as inferred from Figure~\ref{fig:x_mol_2}. Therein, the molecular lines seem to 
  be very faint (it is clear particularly in the \molline{CO}{3}{2} Moment 0 map),
  consistent with a decreased molecular density. It is however difficult to draw a robust
  conclusion in the same way as in Section~\ref{sec:xdr}, because of the faintness. 
  Hence, we do not discuss this further.


\begin{table}
\caption{\chandra Data for Supplemental Imaging Analysis\label{tab:x_sup_dat_list}} 
\begin{center}
  \begin{tabular}{ccccccc}
  \hline \hline
 ObsID  & Obs. date (UT) & Grating & Exp.    & Tar.    \\
        &                &         & (ksec)  & \\ 
 (1)    &          (2)   &    (3)  & (4)     & (5)  \\ \hline
 4771   & 2004/11/28     & YES     &  52     & C     \\ 
 10223  & 2008/12/15     & YES     &  88     & S     \\
 10844  & 2008/12/24     & YES     &  22     & S     \\
 10873  & 2009/03/01     & YES     &  16     & S     \\
 10850  & 2009/03/03     & YES     &  12     & S     \\
 10872  & 2009/03/04     & YES     &  14     & S     \\
 10937  & 2009/12/28     & NO      &  10     & S     \\
 12823$^\dagger$ & 2010/12/17 & NO &  147    & C     \\ 
 12824$^\dagger$ & 2010/12/24 & NO &  38     & C    \\ 
 &                &      (Total  & 399)   \\
\hline
\multicolumn{1}{@{}l@{}}{\hbox to 0pt{\parbox{78mm}
{\footnotesize
  \textbf{Notes.}\\
  (1) Observation ID. 
  (2) Observation start date.
  (3) Check on the grating observation. 
  (4) Exposure after data reduction.
  (5) Target of each observation (C = Circinus and S = SN 1996cr). 
  ~$^\dagger$ These have been used in the main imaging analysis.
}
\hss}}
  \end{tabular}
\end{center}
\end{table}

\end{document}